\pdfoutput=1
\documentclass[epjH,numbook,smallextended]{svjour}[2012-07-23]
\usepackage{graphicx}
\usepackage{mathptmx}
\begin{document}
\title{Very-high energy gamma-ray astronomy}
\subtitle{A 23-year success story in high-energy astroparticle physics}
\author{Eckart Lorenz\inst{1}\fnmsep\thanks{\email{e.lorenz@mac.com}} 
\and Robert Wagner\inst{1,2}\fnmsep\thanks{\email{robert.wagner@mpp.mpg.de}}}
\titlerunning{Very-high energy gamma-ray astronomy}
\authorrunning{Lorenz and Wagner}
\institute{Max-Planck-Institut f\"ur Physik, F\"ohringer Ring 6, 80805 M\"unchen, Germany \and Excellence Cluster ``Origin and Structure of the Universe'', Boltzmannstra\ss e 2, 85748 Garching b. M\"unchen, Germany}
\abstract{
Very-high energy (VHE) gamma quanta contribute only a minuscule fraction -- below one per million -- to the flux of cosmic rays.  Nevertheless, being neutral particles they are currently the best ``messengers'' of processes from the relativistic/ultra-relativistic Universe because they can be extrapolated back to their origin. The window of VHE gamma rays was opened only in 1989 by the Whipple collaboration, reporting the observation of TeV gamma rays from the Crab nebula. After a slow start, this new field of research is now rapidly expanding with the discovery of more than 150 VHE gamma-ray emitting sources. Progress is intimately related with the steady improvement of detectors and rapidly increasing computing power. We give an overview of the early attempts before and around 1989 and the progress after the pioneering work of the Whipple collaboration.
The main focus of this article is on the development of experimental techniques for Earth-bound gamma-ray detectors; consequently, more emphasis is given to those experiments that made an initial breakthrough rather than to the successors which often had and have a similar (sometimes even higher) scientific output as the pioneering experiments. The considered energy threshold is about 30 GeV. At lower energies, observations can presently only be performed with balloon or satellite-borne detectors. Irrespective of the stormy experimental progress, the success story could not have been called a success story without a broad scientific output. Therefore we conclude this article with a summary of  the scientific rationales and main results achieved over the last two decades.
}
\maketitle
\section{Introduction}
\label{intro}
Since early times, astronomy has been an important part of the human culture. Astronomical observations started way back in the Stone Age. Galileo opened the window of modern astronomy in 1609 AD, by making astronomical observations with optical telescopes. By means of a simple optical telescope, he could observe the four largest moons of Jupiter for the first time. The same year Kepler published the fundamental laws of planetary movements in the {\it Astronomica Nova}. During the following 350 years, astronomers were exploring the Universe in the wavelength range of visible light, successively investigating more and more of the so-called thermal Universe, which comprises all emission coming from systems in thermal equilibrium. In the year 1912, the Austrian physicist Victor Hess showed that some type of high-energy radiation is constantly bombarding the Earth from outer space \cite{hess1912}. These so-called cosmic rays (CR), later identified mostly as charged particles, were a clear evidence of the existence of high-energy processes in our Universe exceeding energies that could be reasonably expected from in thermal emission processes. A fundamental problem of CRs (below some  $10^{18}$~eV) is that these charged particles do not allow their trajectories to be traced back to an astrophysical object, as they are deflected by (unknown) intergalactic magnetic fields and thus lose any directional information. Even today, after a hundred years of CR studies, basic questions about the 
sources of CRs remain unsolved.

Shortly before and after the Second World War, new windows in energy bands below and above visible wavelengths of the electromagnetic spectrum were successfully opened, by observations in radio waves, infrared and ultraviolet light, X-rays, and, eventually, in gamma rays. At around 1980, it was possible to observe cosmic radiation in the entire range of the  electromagnetic spectrum, from $10^{-6}$~eV up to $10^9$~eV. Radio, X-ray and gamma-ray observations  demonstrated that besides the thermal Universe (dominated by stellar production of photons), high-energy  processes based on the acceleration of particles
 are an essential ingredient of the Universe. 
 
In 1989, the window of very-high energy (VHE) gamma-ray astronomy was opened by the detection of TeV gamma rays from the Crab nebula by the Whipple collaboration \cite{Weekes1989}. This seminal detection started a very productive research field in an energy domain which is essentially accessible by ground-based instruments.  In 2012, we are celebrating the 100th anniversary of cosmic ray studies. This article will give an overview of the development of VHE gamma-ray astronomy. The richness of the results achieved over the years necessitated a selection of experiments, that exemplify  the steady progress in VHE gamma-ray astronomy. Obviously, this selection is somewhat personal. Emphasis is put on such experiments that made initial breakthroughs in new detection methods and new results, while less emphasis is put on later experiments, using very similar techniques, although these experiments may have been of same scientific productivity. We also do not describe most of the numerous experiments in the 1970s and 1980s which were optimized for energies above 100 TeV. As we know today, the expectation that these comparatively small arrays could have detected gamma-ray sources was elusive, and up to now no sources in that energy domain have been discovered.

VHE gamma-ray astronomy is part of  high-energy cosmic-ray astrophysics. Many experiments of the past aimed both at the search for VHE gamma-ray emitting sources, as well as at solving fundamental questions concerning the nature of cosmic rays. In this paper, we concentrate on the discussion of gamma-ray studies and refer to \cite{Kampert} for details on CR studies.
 
\subsection{VHE gamma rays, messengers of the relativistic Universe}

Cosmic rays result from distant high-energy processes in our Universe and transmit information about the corresponding phenomena. Besides their energy (and particle type), the most important information they could carry is the location of the astrophysical object of their origin. However, nearly all CRs are charged and therefore suffer deflection from their original trajectories by the weak magnetic fields ($\ll 1$~Gauss) in our Galaxy and, if originating from somewhere in the extragalactic space, also by very weak extragalactic magnetic fields, which are known to exist. Their direction and size is, however, unknown. CRs up to about few times $10^{19}$~eV are nearly completely randomized in direction and cannot be associated with any astrophysical object. Even if the magnetic fields were known, it would currently be impossible to extrapolate observed charged CRs back to their point of origin due to the uncertainty in determining their energy.  With the possible
exception of the very highest energies, the back tracing would result in large correlated areas at the sky. 
Therefore, only neutral particles are currently suited to serve as messenger particles. The two particles types that ideally fall into this category are photons -- gamma ($\gamma$) quanta -- and neutrinos. All other neutral particles are too short-lived to survive over large cosmic distances. The neutron has a lifetime just below 15 minutes in its rest frame. Even at the highest energies of $\approx 10^{19}$~eV, on average it would just travel over a distance from the center of our Galaxy to the Earth. Neutrinos, being weakly interacting particles, are very difficult to detect, and huge volumes of dense material are required to observe the minuscule fraction which makes an interaction. A review of neutrino astronomy and its historical development is given in \cite{Spiering}. 

VHE $\gamma$ rays are therefore currently the best-suited messengers of the relativistic Universe. The challenge to interpret $\gamma$-ray observation is that they can be due to  two fundamentally different production processes (or a combination of these!), namely
{\it leptonic} or {\it hadronic} processes. Neutrinos, on the other hand, can be created only in {\it hadronic} processes, therefore  one could solve this ambiguity. 

The main production processes of $\gamma$ rays are:

\begin{itemize}
\item[a)] Inverse Compton scattering: 

VHE electrons upscatter low energy photons over a broad energy range above the initial one,
$$
{\rm e} +\gamma_{\rm low~energy} \longrightarrow 
{\rm e}_{\rm low~energy} + \gamma_{\rm VHE}.
\label{eq:1}
$$
Normally, there are plenty of low energy photons in the environment of stars due to thermal emission or due to synchrotron emission by the high energy electrons in the ambient magnetic fields. In the lower energy range, the dominant production process of gamma rays from leptons is via synchrotron radiation processes, where electrons lose a fraction of their energy by synchrotron radiation when passing through local magnetic fields. 

\item[b)] Decay of neutral pions produced in hadronic interactions: 

Accelerated protons or heavier nucleons interact with ambient protons, nucleons or photons in stellar environments or cosmic gas clouds. Dominantly, charged and neutral pions are produced. Charged pions decay in a two-step process into electrons and two neutrinos while neutral pions decay with $> 99\%$ probability into two gamma quanta.
For proton-nucleus interactions one has:
$$
\label{eq:2}
{\rm p + nucleus} \rightarrow  {\rm p}' \dots + {\pi^\pm} + {\pi^0} + \dots \hspace{.5cm} {\rm and} \hspace{.5cm} {\pi^0} \rightarrow 2\gamma;\hspace{2mm} \pi \rightarrow \mu \nu_\mu;\hspace{2mm} \mu \rightarrow {\rm e} \nu_\mu \nu_{\rm e}.
$$
Heavier secondary mesons, much rarer, normally decay in a variety of lighter ones and eventually mostly into $\pi^\pm$ and $\pi^0$ and/or $\gamma$. 

\end{itemize}

Spectra and morphology of gamma-ray emissions can provide only circumstantial evidence
for either leptonic or hadronic origin of the gamma rays, while the observation of neutrinos would be an unambiguous proof for hadronic acceleration processes in the source.

\subsection{The long road to the discovery of the first VHE-emitting gamma-ray source}

The main driving force for VHE gamma-ray astronomy was  initially the search for the sources of the charged cosmic rays, while now, after the discovery of many sources, the interests have shifted  to general astrophysics questions. In earlier times the searches were hampered by a few fundamental questions: 
\begin{enumerate}
\item How large is the fraction of cosmic $\gamma$ rays of the total CR flux?
\item What exactly happens when cosmic particles hit the atmosphere?
\item What are the secondary decay products when VHE $\gamma$ rays hit the atmosphere?
\item How can VHE $\gamma$ rays be distinguished from the charged VHE cosmic rays?
\item How transparent is the Universe for $\gamma$ rays of a certain energy, respectively how far one can look with $\gamma$ rays of a certain energy into the Universe?	
\end{enumerate}

It took many years with the detection techniques available in those times to solve these problems step by step -- largely due to  inadequate instruments, slowly developing theories about particle interaction, slowly oncoming additional information from accelerator experiments and the lack of powerful computers.

The exact fraction which $\gamma$ rays contribute to the total CR flux and its dependence on energy is still unknown today. Shortly after the discovery of CRs, Kolh\"orster speculated that CRs originated from cosmic $\gamma$ rays, but the first experiments were too simple to prove or disprove this assumption \cite{Kolhoerster1913}. In 1925, R. Millikan, who introduced the name cosmic rays, was convinced that CRs originally were all $\gamma$ rays \cite{Millikan1925}. In 1930, Millikan and Compton disagreed about the origin of CRs with Millikan pursuing their photonic origin, while Compton was convinced that CRs were originally primary, positively charged particles. This was later proven to be correct when it was possible to observe ionizing particles at the top of the atmosphere or with satellite-borne detectors. In the 1930s and 1940s it was still believed that a significant part of CRs were $\gamma$ rays, while in the early 1980s it was mostly thought that about 1\% of the CRs were $\gamma$ rays. Nowadays this question is still not completely resolved and much smaller flux ratios are assumed. One supposes that at most $10^{-4}$ of all particles coming from the Galactic plane are $\gamma$ rays, while at most only $10^{-5}$ of particles from outside the galactic plane are $\gamma$ rays. 
It took 27 years after Hess' first discovery of CRs until Pierre Auger discovered extended air showers initiated by CRs when hitting the atmosphere \cite{Auger1937}. Furthermore, it took many decades to understand the basics of the showering process; still today only approximate models describe some subtle effects.

%\end{document}

\section{Attempts between 1960 to late 1980s to find the sources of CRs}
\subsection{A short excursion: The basic detection techniques}
The three decades from 1960 to the end of the 1980s saw steady
but  rather slow progress towards discovering sources of VHE gamma rays. Experiments were in a vicious circle: Poor experiments gave doubtful results and the funding agencies were not willing to finance large installations. Many physicists, that started their career in cosmic-ray physics, turned to high energy physics (HEP) experiments at accelerators; this field was and still is in an extremely productive phase. Compared to HEP, technological innovation in very-high energy cosmic-ray physics was rather limited;
little progress in understanding the fine structure of shower developments was due to lack of sophisticated experimental instruments, insufficient computing power and limited theory in high energy hadronic interaction.

Earth-bound gamma-ray detectors make use of particle showers generated by gamma-ray interactions
at high altitude. High-energy gamma rays (as well as hadrons and leptons) enter the Earth's atmosphere and generate a cascade of secondary particles, forming an extended air shower. Initially, in this shower process the number of secondary particles is rapidly increasing. During this multiplication process the energy of the primary is partitioned onto the secondaries until the energy of the secondary particles becomes so low such that the multiplication process stops.  Due to energy loss of the charged particles by ionization, the shower eventually dies out. Depending on the primary energy and nature of the incident particle, the shower might stop at high altitudes or reach ground. 
 
\begin{figure}[h]
\centering
\includegraphics[width=.65\linewidth]{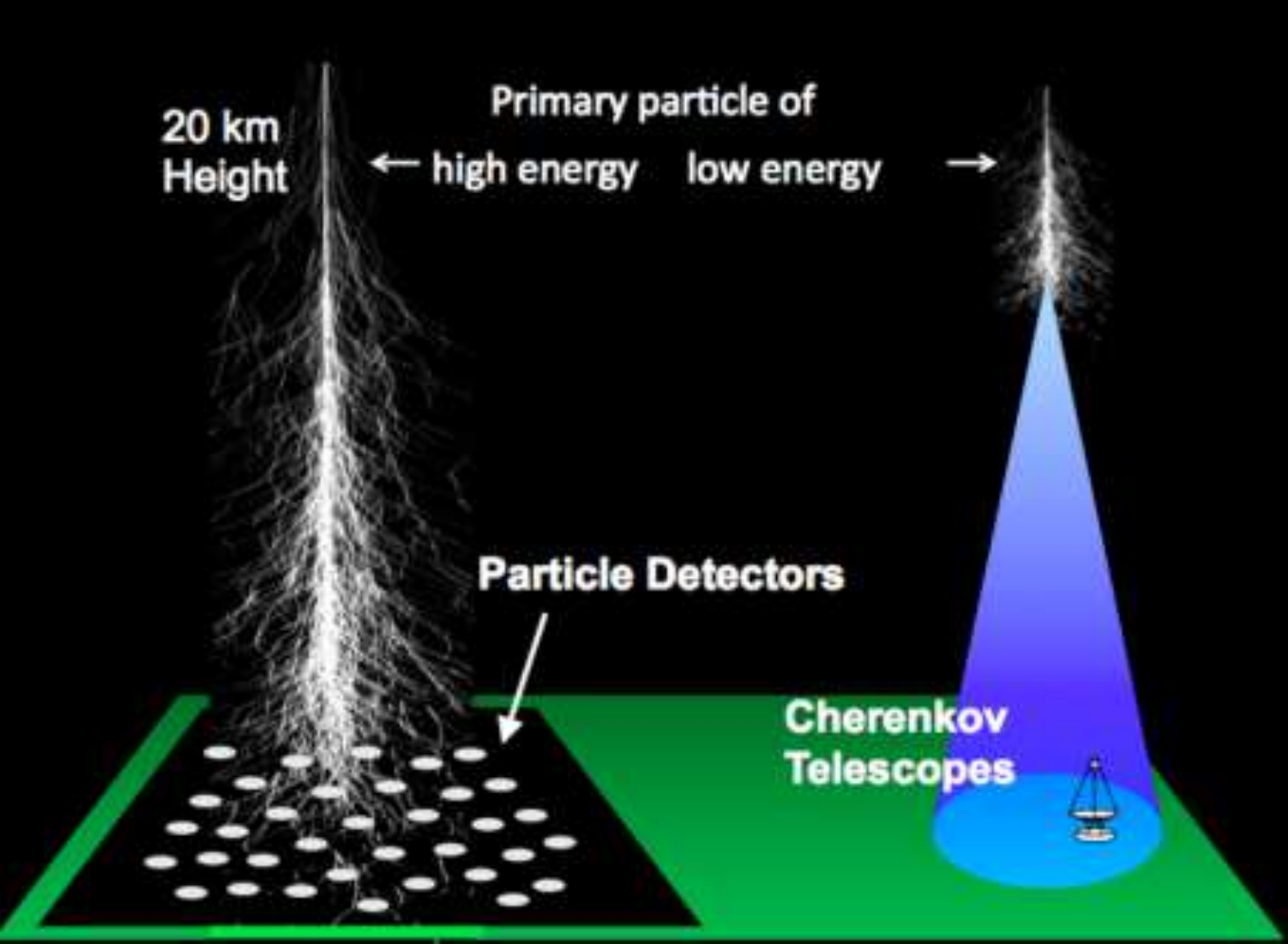}
\caption{Principle of the two commonly used detector techniques for observing cosmic VHE particles. Left: An extended air shower array. Primary particles hitting the Earth's atmosphere initiate an extended air shower. Shower tail particles, which penetrate down to ground level, are detected by an array of particle detectors. Right: Cherenkov light detection of air showers that do not need to penetrate down to ground. The figure illustrates the detection of Cherenkov light by a so-called Imaging Atmospheric Cherenkov Telescope (IACT), comprising  a large mirror focusing the light onto a matrix of high-sensitivity photosensors in the focal plane. Both detector principles are used for the observation of charged cosmic-ray showers as well as gamma-ray induced showers. Courtesy C. Spiering.\label{Fig:cher}
}
\end{figure}

VHE gamma-ray astronomy rests on two basic detector concepts\footnote{Other detection principles like the  fluorescence detectors make use oft the very weak fluorescence light of an air shower; radio detectors detect radio waves emitted by the shower. Both detection principles are currently not used for VHE gamma-ray astronomy, because of their extremely high threshold.} (Fig. \ref{Fig:cher}): 
\begin{itemize}
\item Detectors that measure particles of the shower tail which reaches the ground, the so-called extended air shower arrays (EAS) or, as particle physicists called them,``tail-catcher detectors''. This method provides a snapshot of the shower in the moment it reaches the ground.
\item Cherenkov detectors for observing showers that die out before reaching ground. This method uses the full atmosphere as calorimeter.
\end{itemize}

\begin{figure}[h]
\centering
\includegraphics[width=.24\linewidth]{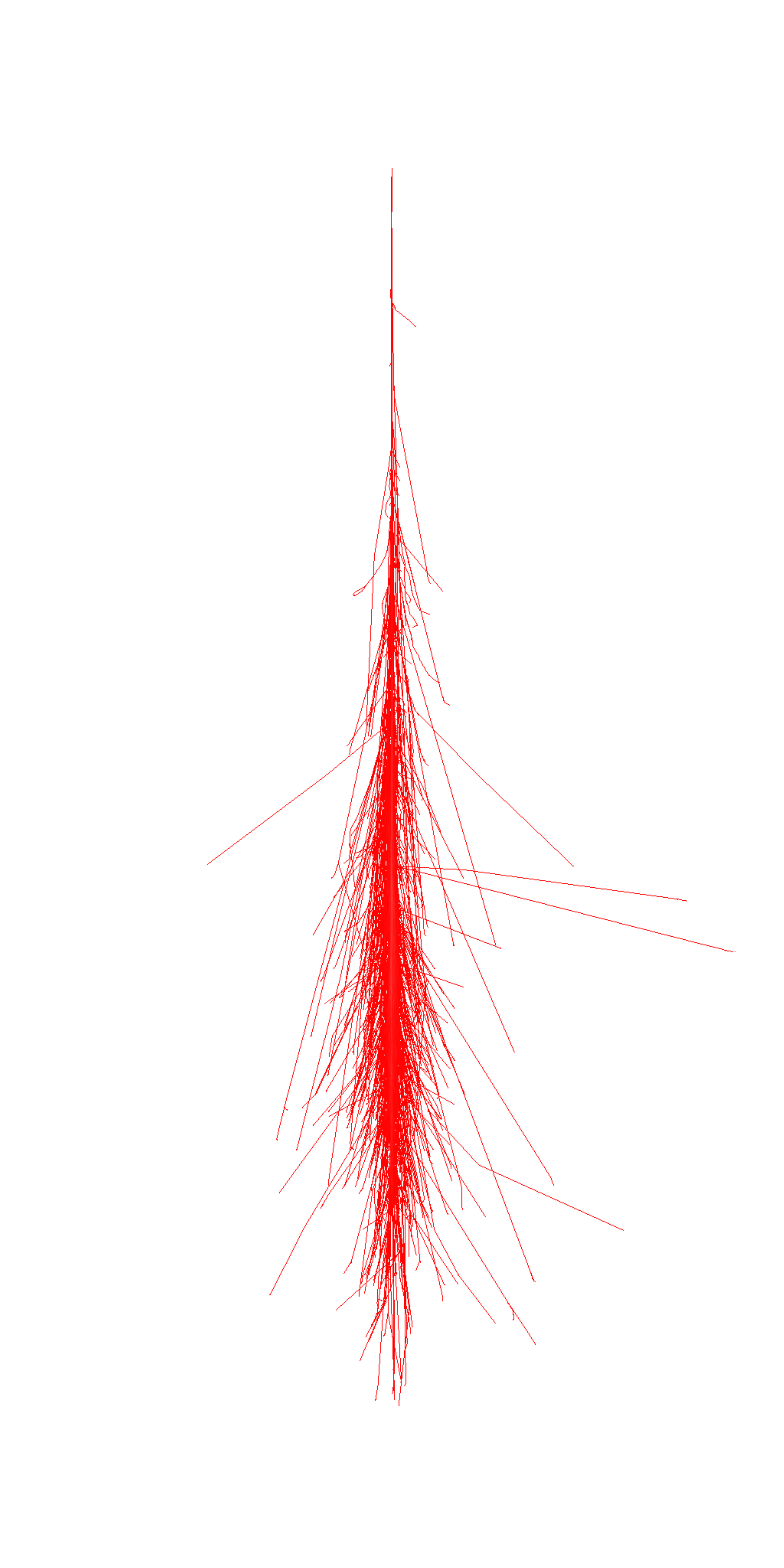}
\includegraphics[width=.24\linewidth]{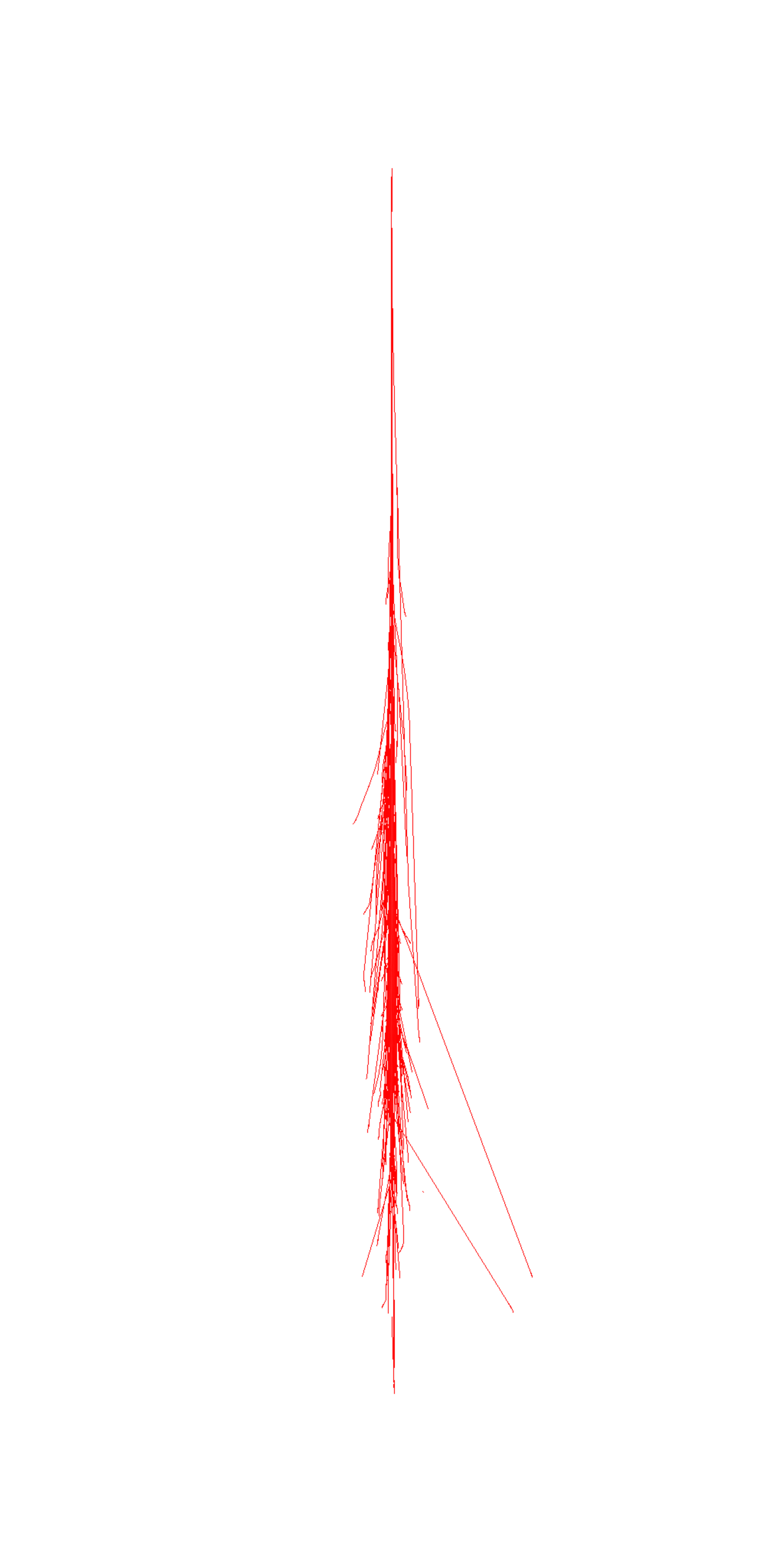}
\includegraphics[width=.24\linewidth]{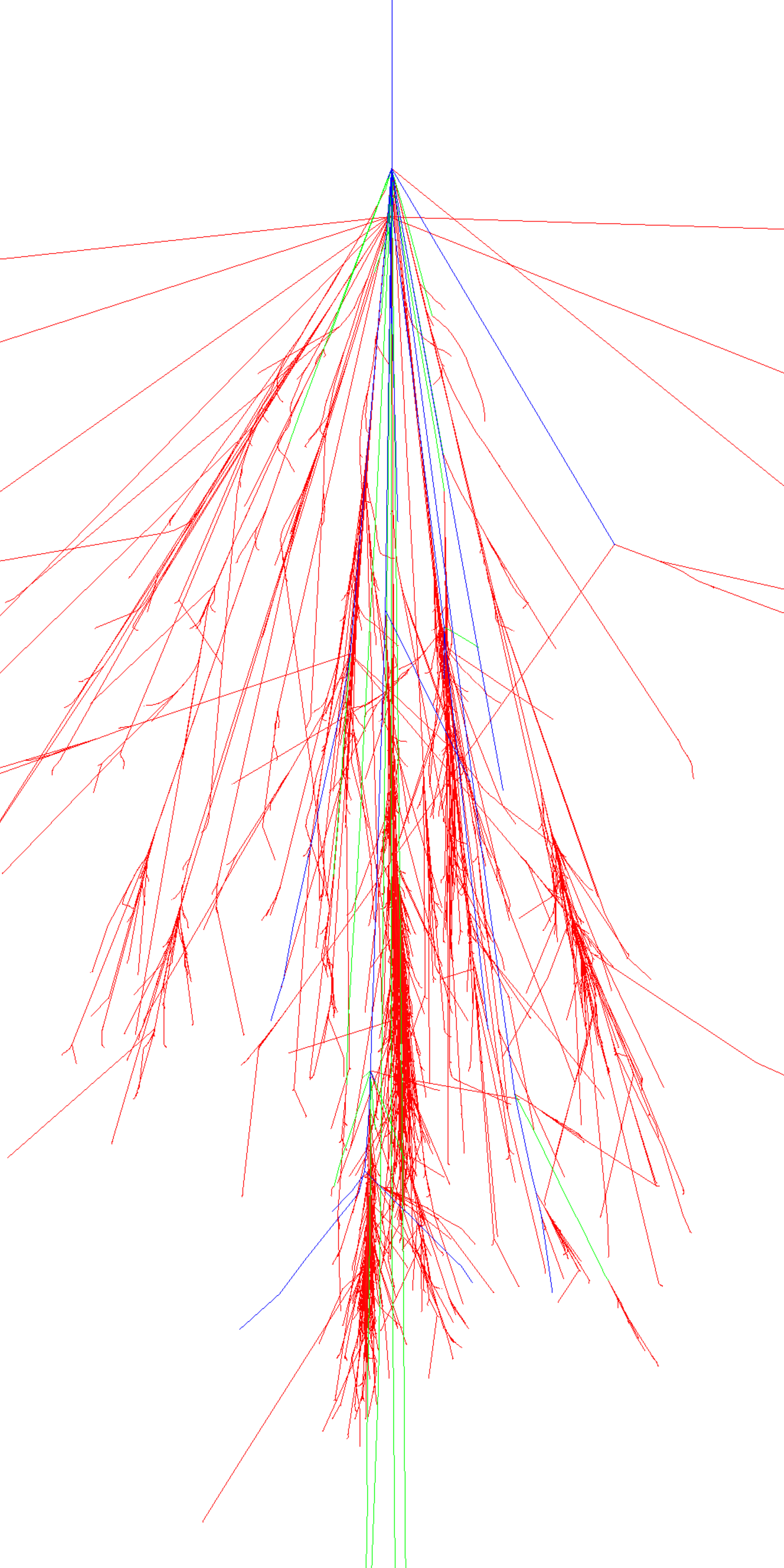}
\includegraphics[width=.24\linewidth]{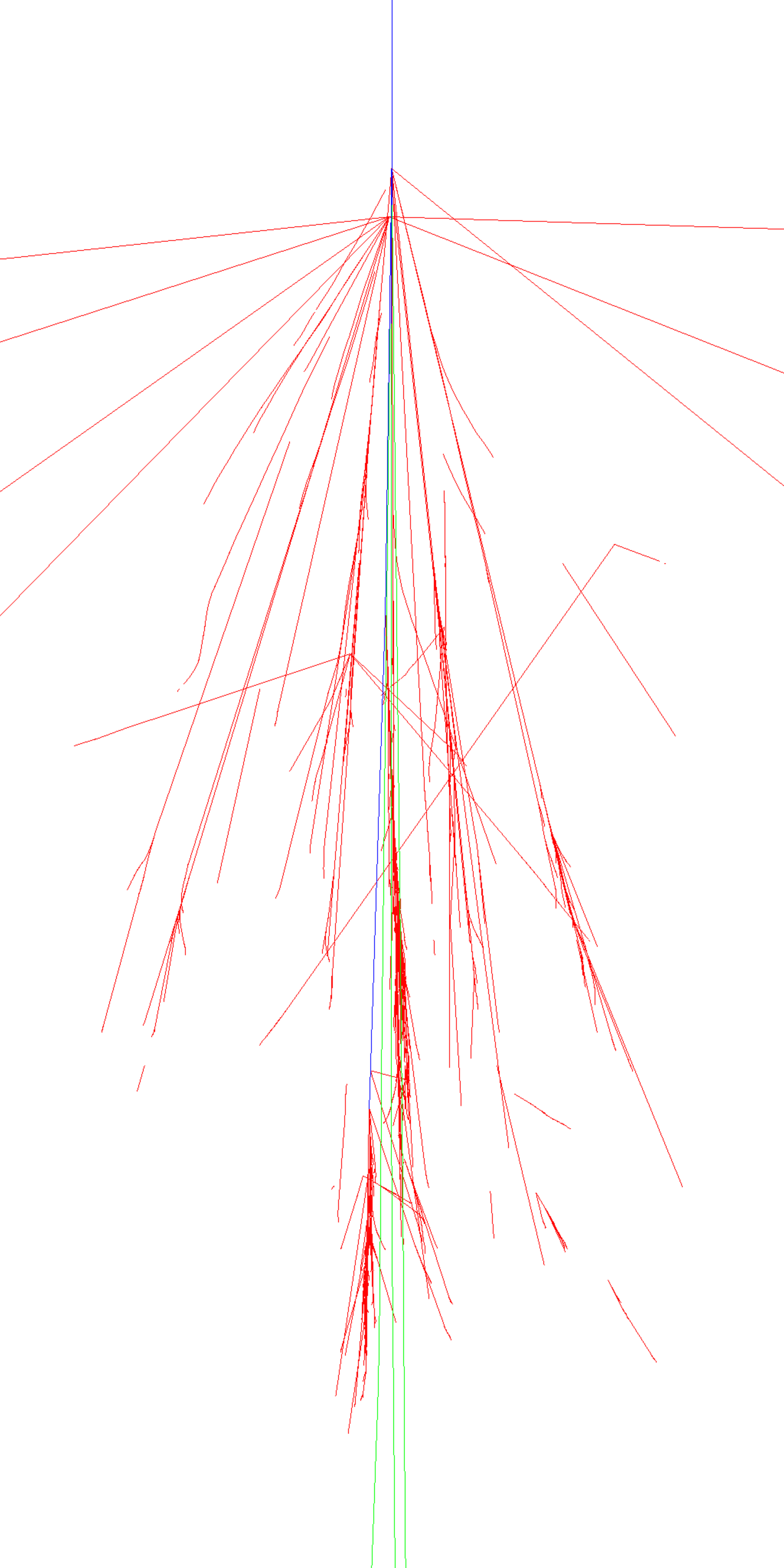}
\caption{Simulations of air showers, from left to right: (a) secondaries of a 50 GeV $\gamma$-ray primary particle, (b) same, but only those secondaries that produce Cherenkov light are plotted. (c) Secondaries of a 200 GeV proton primary particle, (d) same, but only those secondaries that produce Cherenkov light are plotted. 
In all figures, the particle type of the secondaries is encoded in their track color: red = electrons, positrons, gammas; green = muons; 
  blue = hadrons. Figures courtesy Dario Hrupec (Institut Ruder Bo\v{s}kovi\'c, Zagreb), produced using code done by Fabian Schmidt (Leeds University), using CORSIKA.
\label{showers}
}
\end{figure}

Showers originating from hadrons (``hadronic showers'') and  electromagnetic showers, initiated by gamma rays or electrons (positrons) can be discriminated by their development: Fig. \ref{showers} shows examples of showers induced by gamma rays and protons, respectively. If the charged secondary particles are moving faster than the speed of light in the atmosphere, they emit Cherenkov light within a small angle, which depends also on the (altitude-dependent) atmospheric density and particle energy. A hadronic shower starts normally with many secondary pions and a few heavier mesons. Due to the fact that about one third of the secondary particle in each interaction are $\pi^0$ particles,  the electromagnetic component of hadronic showers becomes more and more enriched due to the decay $\pi^0 \rightarrow 2\gamma$.   Most of the charged pions undergo secondary interactions, but some of them decay into muons, which can penetrate deeply into ground. Gamma-ray induced cascades are much narrower. The dominant multiplication processes in electromagnetic showers is electron/positron bremsstrahlung, producing gamma rays  and e$^+$e$^-$ pairs from gamma-ray conversion. In vertical direction, the atmosphere corresponds to 27 radiation lengths and 11 hadronic absorption lengths. Due to the high transverse momentum in hadronic interactions, to multiple scattering and to deflections in the magnetic field of the Earth, the showers are widened, facilitating their detection. In case of the Cherenkov detector principle, the emission angle of the Cherenkov light is small but still illuminates a large area at ground, of typically 200--220 meters in diameter. A telescope anywhere in this area can detect an electromagnetic shower, provided the Cherenkov light intensity is high enough. Further details of the showering process can be found in \cite{Weekes2003} or in numerous publications about calorimetry in high energy physics experiments.

The air-shower array detectors used in most cases are derivatives of the initial Geiger tube counters and nearly all followed the low active-density array concept. Most advanced detectors used large scintillation counters viewed by photomultipliers.
These array detectors sampled the shower tail and measured the arrival signals in each hit counter, thus allowing to determine the energy and direction of the shower. The active area fraction of the array area covered by detectors was normally below 1\% resulting in rather large uncertainties in energy determination and modest angular resolution. A big problem was the precise angular calibration of the detectors, as no reference source was available. Special variants of the air shower array detectors  tracked the charged particles passing the instruments. It was hoped to determine the incident particle direction from the measurement of a few angular measurements of the secondary tracks in the shower tail. These measurements, however, provided only a very poor directional determination because most of the secondary particles at the shower tail were of low momentum where multiple scattering is large. Air-shower arrays have basically a 24 hour up-time and a wide angle acceptance and thus allow the monitoring of a large fraction of the sky. If the gamma fluxes at high energies were higher, these
two features would have made air shower arrays valuable survey devices in searching the sources of 
gamma rays. However, depending on the altitude of the installation, the threshold was very high. At sea level, one achieved a threshold of around $10^{14}$~eV for showers with vertical incidence. For large zenith angle showers, the energy threshold scales with a strong dependence of the zenith angle $\theta$ of around $\cos^{-7} \theta$. The main deficiency of air shower array detectors, however, is their weak gamma/hadron separation power and the poor energy and angular resolution at  their energy threshold
and still quite far above it. Muons might be used as discriminators. Gamma-ray induced showers contain only very few muons (originating from rare photo-production processes), while hadronic showers contain many muons reaching the ground. Muons are normally identified by their passage of substantial amounts of matter. Therefore muon detectors had to be installed a few meters underground, thus making them an expensive component of the detector. Consequently, only a small fraction of the arrays was covered
with muon counters. The identification of cosmic sources of  $\gamma$ rays was expected
from locally increased rates in the sky maps. Hadronic events would be isotropically distributed and should form a smooth background. By means of the muon detectors it was hoped to suppress the hadronic background further.

\begin{figure}[h]
\centering
\includegraphics[width=.46\linewidth]{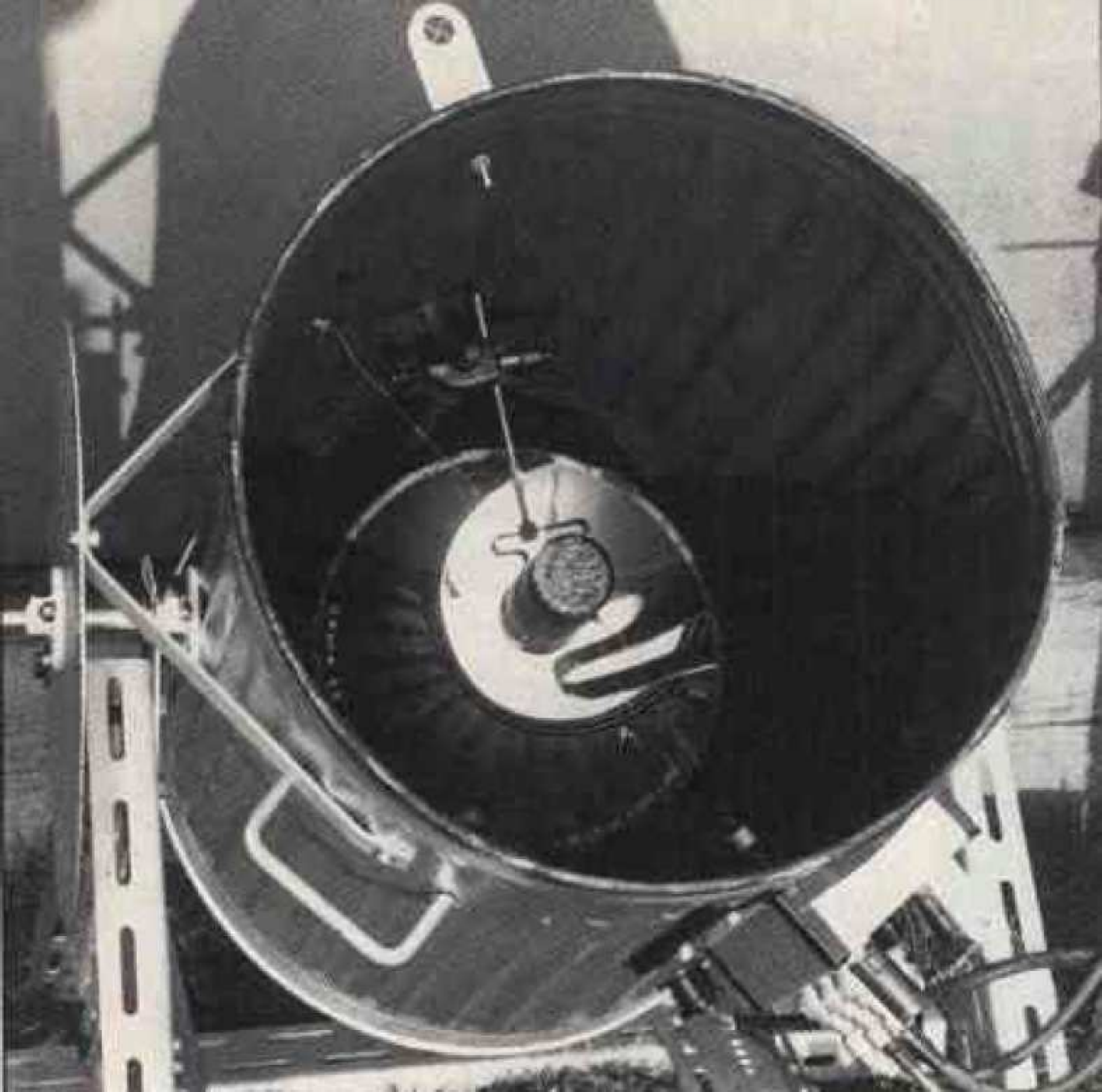}
\includegraphics[width=.53\linewidth]{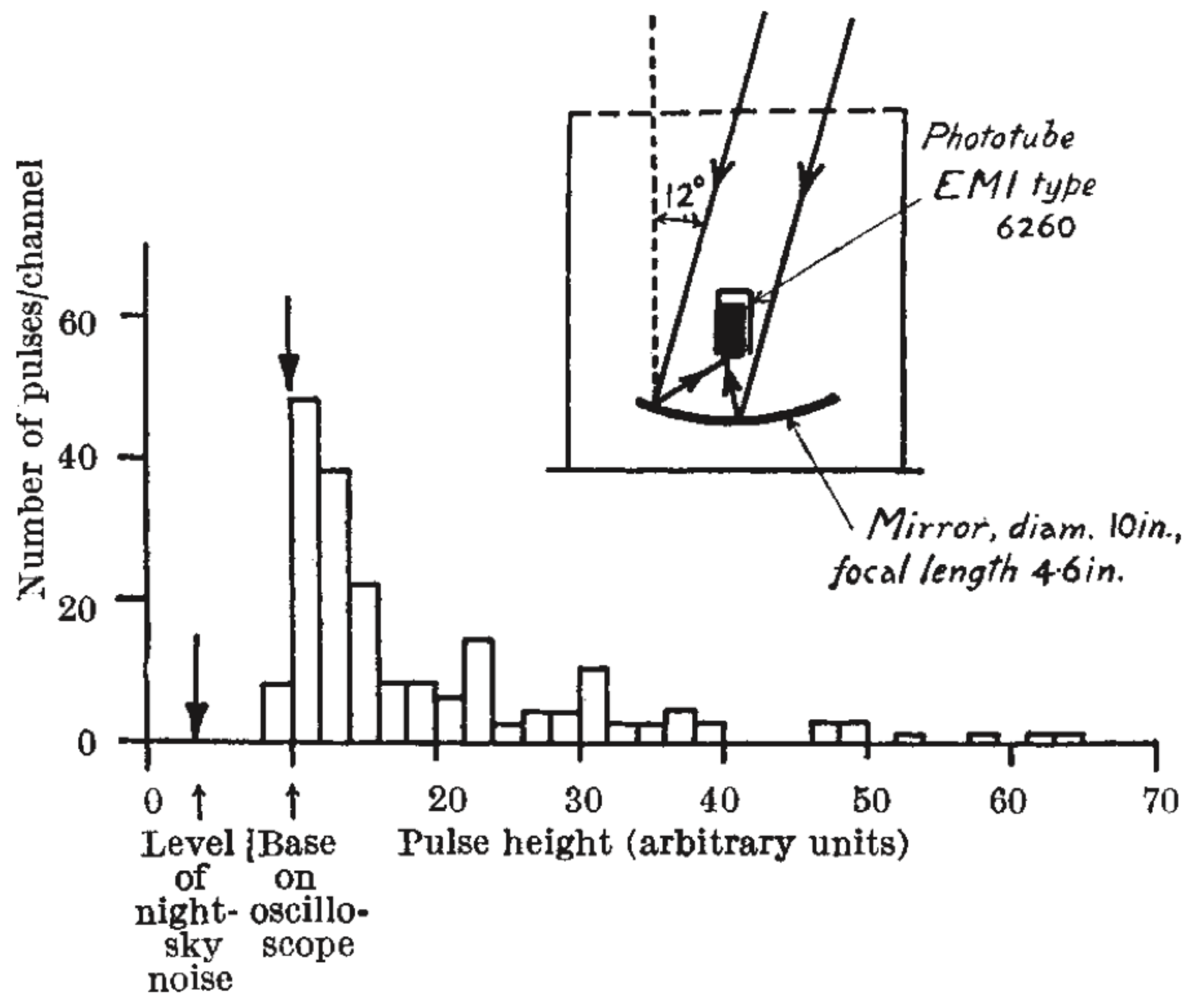}
\caption{Left: The first design of an air Cherenkov counter in a garbage can used by B. Galbraith and J. V. Jelley in 1953 \cite{Galbraith1953} (photograph courtesy T. C. Weekes). Right: Setup and results of the observations of Galbraith and Jelley (figure taken from the original article).
\label{Fig2.1}
}
\end{figure}

The alternative techniques to the air shower arrays  were detectors based on the observation of Cherenkov light from air showers. In 1934, Pavel Cherenkov discovered that charged particles emit some prompt radiation in transparent media when moving faster then the speed of light in that media \cite{Cherenkov1934}. Later, Ilia Frank and Igor Tamm developed the theory for this radiation, dubbed after its discoverer {\it Cherenkov radiation}. All three were awarded the Nobel Prize in 1958. In 1947, the British physicist P. M. S. Blackett predicted that relativistic cosmic particles passing the atmosphere should produce Cherenkov light and even contribute to a small fraction ($\approx 10^{-4}$) of the night sky background light \cite{Blackett1948}. In 1953, B. Galbraith and J. V. Jelley built a simple detector and proved that air showers generate Cherenkov light, which could be detected as a fast light flash during clear dark nights \cite{Galbraith1953}. With a threshold of around four times the night sky noise level they observed signals with a rate of about one event per two to three minutes. This was, by the way, the first demonstration that Cherenkov light was generated also in gases. Later, they could demonstrate that these signals were actually caused by air showers due to coincidences with the nearby Harwell air shower array. The first detectors consisted of a very simple arrangement, i.e., a search-light mirror viewed by a photomultiplier, as shown in Fig.~\ref{Fig2.1}. The first setup was installed in a garbage can for shielding from stray light. In the following years the technique was refined by using larger mirrors, replacing the single photo-multiplier tube (PMT) by a few arranged in the focal plane and even a few of these simple telescopes in coincidences. As in optical astronomy, the air Cherenkov telescopes had to track the source under observation.  Nevertheless, all these many pioneering efforts were not rewarded by any important discovery. The so-called air Cherenkov telescopes had some important advantages compared to the air shower arrays. The telescopes collected light from the entire development of the particle shower and one could, in principle, measure the energy of the initial particle with much higher precision and with a  threshold typically two orders of magnitude lower than that of ``tail catcher'' detectors. The main disadvantages were that one could observe with a field of view of a only few degrees. Thus one could study only a single object at a time. Moreover, observations could only be carried out during clear, moonless nights. Similarly to the air shower arrays, the first-generation Cherenkov telescopes could not discriminate between hadronic and electromagnetic showers. Therefore observers tried to identify sources by just a change in the counting rate when pointing their telescope(s) to the sources and later for the same time slightly off the source. As the gamma-ray flux was very low compared to the CR flux, such a method was prone to secondary effects generating rate changes, for example fluctuations due to atmospheric transmission and the night sky light background from stars.

Because Cherenkov detectors measured the light coming from the entire shower, they had, besides their better energy resolution, also a better angular resolution. Basically, the combination of the atmosphere and the detector forms a fully active calorimeter with some imaging quality due to the directional distribution of the Cherenkov light. Compared to man-made calorimeters for accelerator experiments, the atmosphere is quite a complex absorber, as summarized in Table \ref{Tab2.1}. Fig. \ref{showers} shows simulations of the shower development of typical $\gamma$-ray and proton-induced air showers, particularly illustrating those secondaries that produce Cherenkov light.

\begin{table}[h]
\centering
\caption{Comparison of a typical sampling calorimeter for accelerator experiments and the atmosphere as calorimeter for air Cherenkov counters
}

\label{Tab2.1}
\begin{tabular}{lll}
\hline
Calorimeter type & Sampling calorimeter & Atmosphere\\
\hline
\hline
Extension & few m$^{3}$ & few km$^{\mathrm{3}}$\\
Density & $3-12$ kg dm$^{-3}$ & $0.01-1$ g dm$^{-3}$ (altitude dependent)\\
Uniformity& non-uniform, & smooth, due to fully active material\\
& due to sampling & \\
Density variation & uniform & exponential change with altitude\\
Time dependence of density & none & permanent due to Earth rotation\\
Background light & minimal & open calorimeter, high background,\\
&& light from night sky\\
Signal stability & high & low, because of unpredictable \\
&&changes of atmospheric transmission\\
Differences between lengths & large & similar due to low-$Z$ materials\\
of hadronic and $\gamma $ showers & & \\
\hline
\end{tabular}
\end{table}

One should be aware that only a fraction of less than $10^{-4}$ of the total shower energy is converted into photons, and quite a few of these photons get lost before hitting the ground. Losses are due to absorption by ozone molecules below around 300~nm, Rayleigh scattering (normally well predictable) and Mie scattering due to fine dust or thin clouds or haze in the atmosphere. In the early times of Cherenkov detectors, losses due to Mie scattering were quite unknown -- even until the 1980s; these losses could not be fully explained because no adequate instruments for measuring them were used. Even around 1990, the predictions of the transmission of the atmosphere for Cherenkov light  varied by up to a factor four. Adding to these uncertainties the systematic errors of the instruments, in particular the photon detection efficiency (PDE) of the photomultipliers, provided observers with measurements, which were hardly consistent. Also, as previously mentioned, the first generation Cherenkov detectors did not allow one to discriminate between electromagnetic and hadronic showers. The early Cherenkov telescopes plainly did not have the necessary sensitivity to even observe the strongest sources. Despite of that, several excesses of three standard deviations ($\sigma$) in the rate difference between {\it On} and {\it Off} source observations were reported and claimed as a discovery.

Technical goals in the construction of air Cherenkov telescopes are maximizing light collection and the conversion of Cherenkov photons into measurable photoelectrons. For the conversion of photons into measurable photoelectrons up to now only high-quality photomultipliers are used in order to achieve a very fast readout in the range of the duration of the Cherenkov light flashes of a few nanoseconds. The best overall photon detection efficiency (PDE) of today's Imaging Air Cherenkov Telescopes (IACTs) is around 15\%-18\% when averaged over the spectral range of 300-600nm. This is only a modest improvement in PDE of about a factor of three to four compared to the earliest instruments measuring Cherenkov light.

High-Energy Physics (HEP) has made considerable progress in particle studies in laboratories in the years since 1960. The success could be traced to advances in accelerator developments as well as to the replacement of optical readout techniques for bubble chambers or optical spark chambers by a continuous development of more powerful electronic devices, intense use of computers and the formation of large collaborations. At the same time, cosmic-ray physics progressed slowly. Detectors were small and completely inadequate for the necessary collection of complete shower information. The important discrimination between $\gamma$-ray and hadronic showers was very much hampered by a poor knowledge of the shower development, i.e., by the lack of adequate VHE measurements of especially the high energy hadronic interaction. Progress in technology was also modest due to a lack of resources -- completely different to HEP experiments. Often leftover material from dismantled HEP experiments was used, thus reflecting the state of the art electronics of the 1950s and 1960s. 
Also, the use of computers was rather modest compared to HEP.

In HEP experiments, discoveries were often made as soon as an excess of at least three standard
deviations ($3\sigma$) above background was observed. When cuts based on poor knowledge of shower developments were applied to CR data to find sources, failures were guaranteed because the used selection procedures did not deliver unbiased samples. Thus, detections often were reported when a subset of cuts provided a $3\sigma$ excess, and this was then interpreted as a signal. Heinz V\"olk from MPI f\"ur Kernphysik Heidelberg characterized the situation well: ``The road to VHE gamma-ray astronomy was plastered with many $3\sigma$ corpses''. Particularly the air shower arrays, although simple to operate, suffered from their high and rapidly changing threshold with the zenith angle. The results of that time were often highly  controversial and often disagreed at the level of spectral analyses. These mostly and often contradictory $3\sigma$ observations of claimed sources contributed very much to the low reputation of cosmic-ray physics. Only a few physicists who did not change their focus to HEP in the 1950s and 1960s continued this research. Also the Cherenkov technique, which looked quite promising, was not delivering compelling results.  In retrospective, the lack in finding the sources of the CRs is quite obvious. The reasons were:
\begin{itemize}
\item a very low gamma-ray flux compared to the charged CR flux,
\item interaction of $\gamma$ rays with cosmic low-energy photons suppressed the detection of ultra-high energy $\gamma$-ray sources,
\item shower tails contain very poor information on primary particles,
\item poor energy resolution just above threshold,
\item poor angular resolution,
\item poor understanding of shower processes, in particular hadronic showers, as no precise accelerator measurements existed,
\item the most severe problem: poor $\gamma$/hadron separation power from data recorded with too simple detectors.
\end{itemize}
In summary, the reasons for failure were the use of detectors of insufficient sensitivity, the lack of  information from precision VHE experiments at accelerators, the lack of understanding the details of the dominant hadronic  shower development and the atmospheric response. Paradoxically, the field 
got new momentum from an experiment which retrospectively appears suspicious, or at least
highly controversial: an array detector set up by the University of Kiel.

\subsection{Cygnus X-3 in 1983: New momentum for gamma-ray astronomy from a controversial signal}
\begin{figure}[h]
\centering
\includegraphics[width=.6\linewidth]{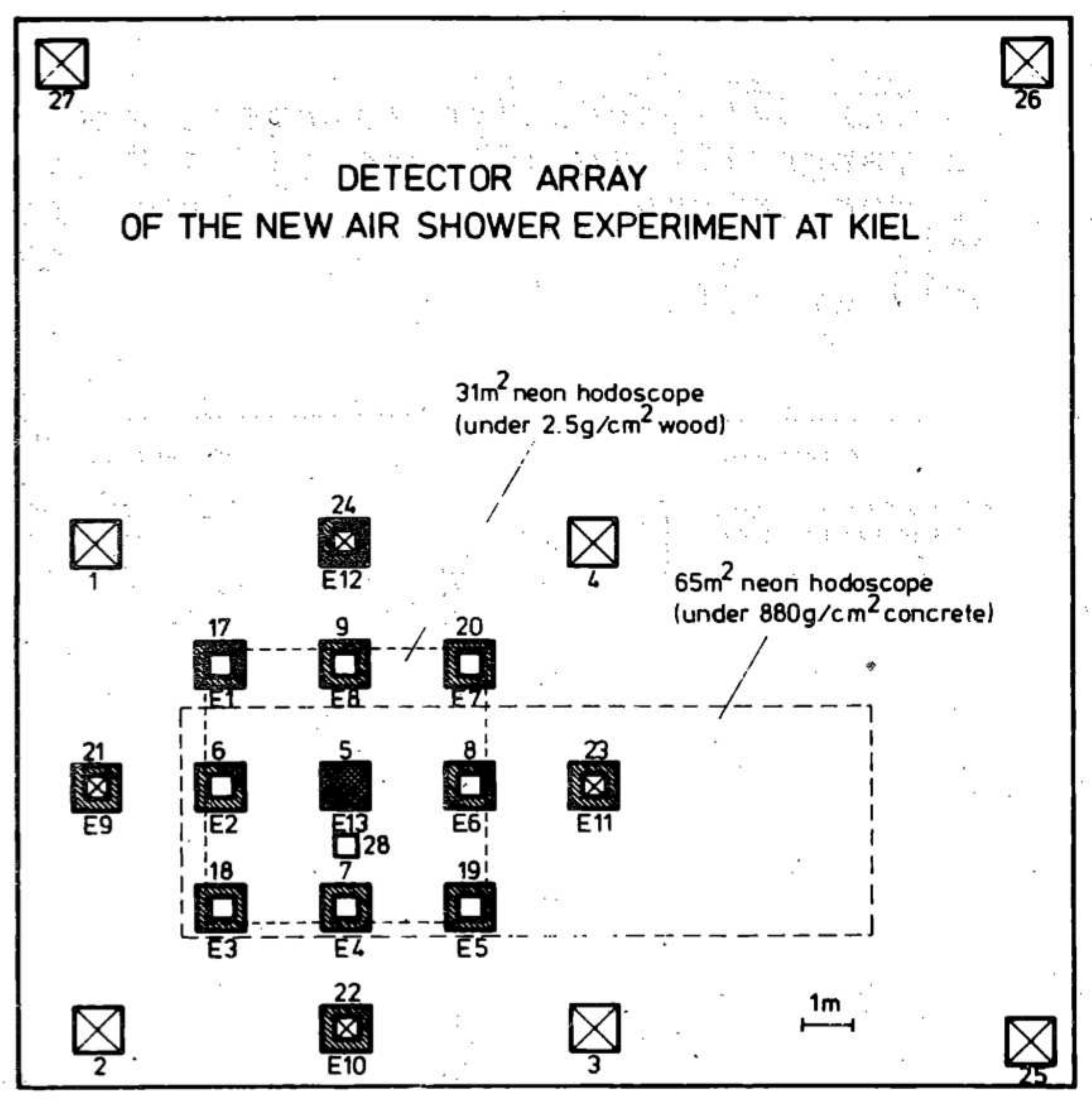}
\caption{Layout of the inner part of the Kiel detector (figure taken from \cite{Bagge1977}): Central part of the EAS detector array. Squares represent scintillation counters of 1(.25)\,m$^2$ area each. Shaded areas indicate a 1.25$\times$1.25\,m shielding of 2\,cm lead plus 0.5\,cm of iron on top of the scintillator. Detectors with additional fast timing photomultipliers are labeled with diagonal crosses.\label{Fig4.1}
} 
\end{figure}

\begin{figure}[h]
\centering
\includegraphics[width=.8\linewidth]{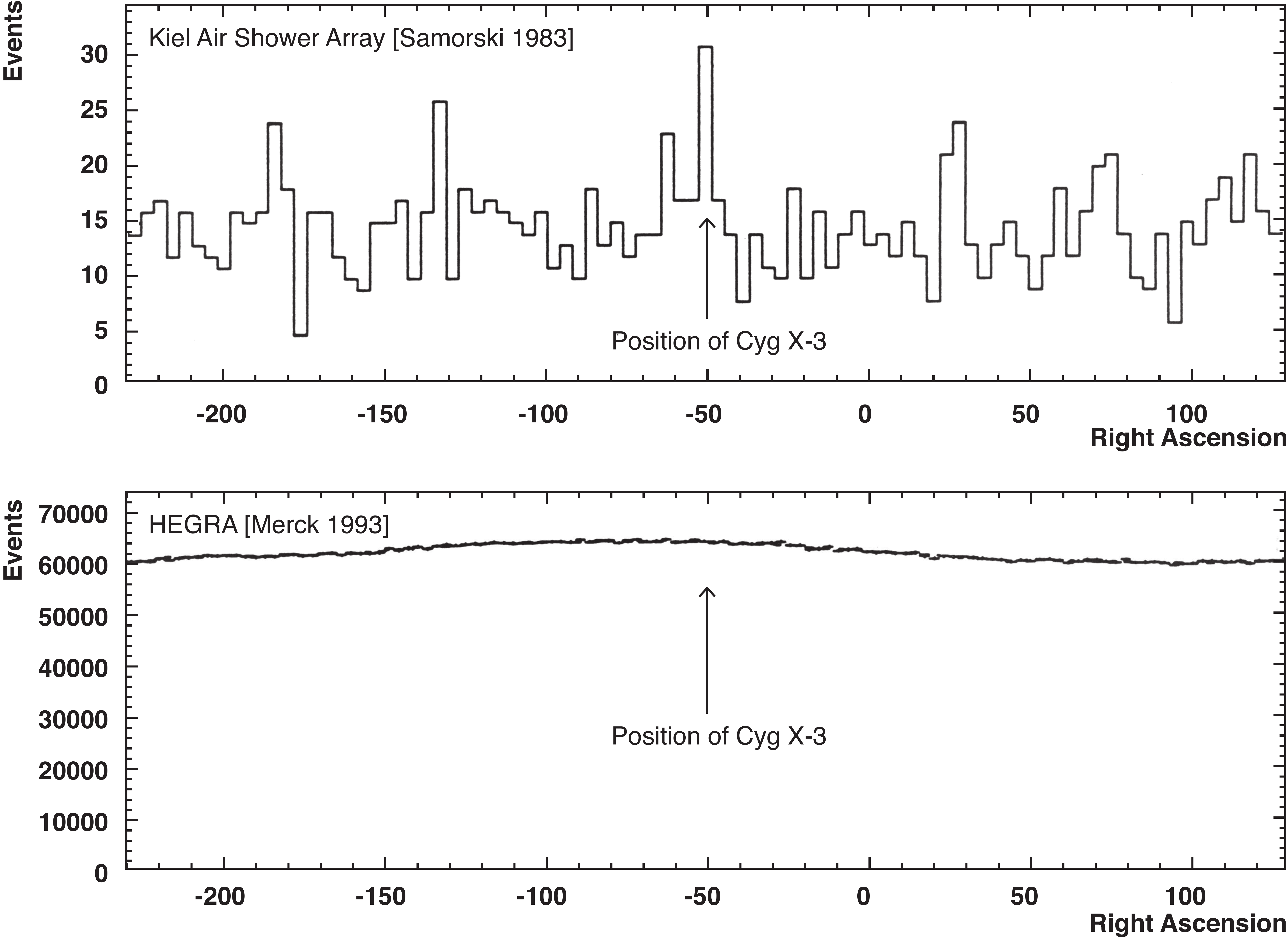}
\caption{Measurement of the cosmic ray flux from the direction of Cygnus X-3 (right ascension band at $40.9^\circ\pm1.5^\circ$ declination) and the surrounding sky region as measured 1983 by \cite{Samorski1983} (upper panel) and 8 years later by HEGRA (lower panel). Figures taken from \cite{Merck1993}.\label{Fig4.2}}
\end{figure}

\begin{figure}[h]
\centering
\includegraphics[width=.35\linewidth]{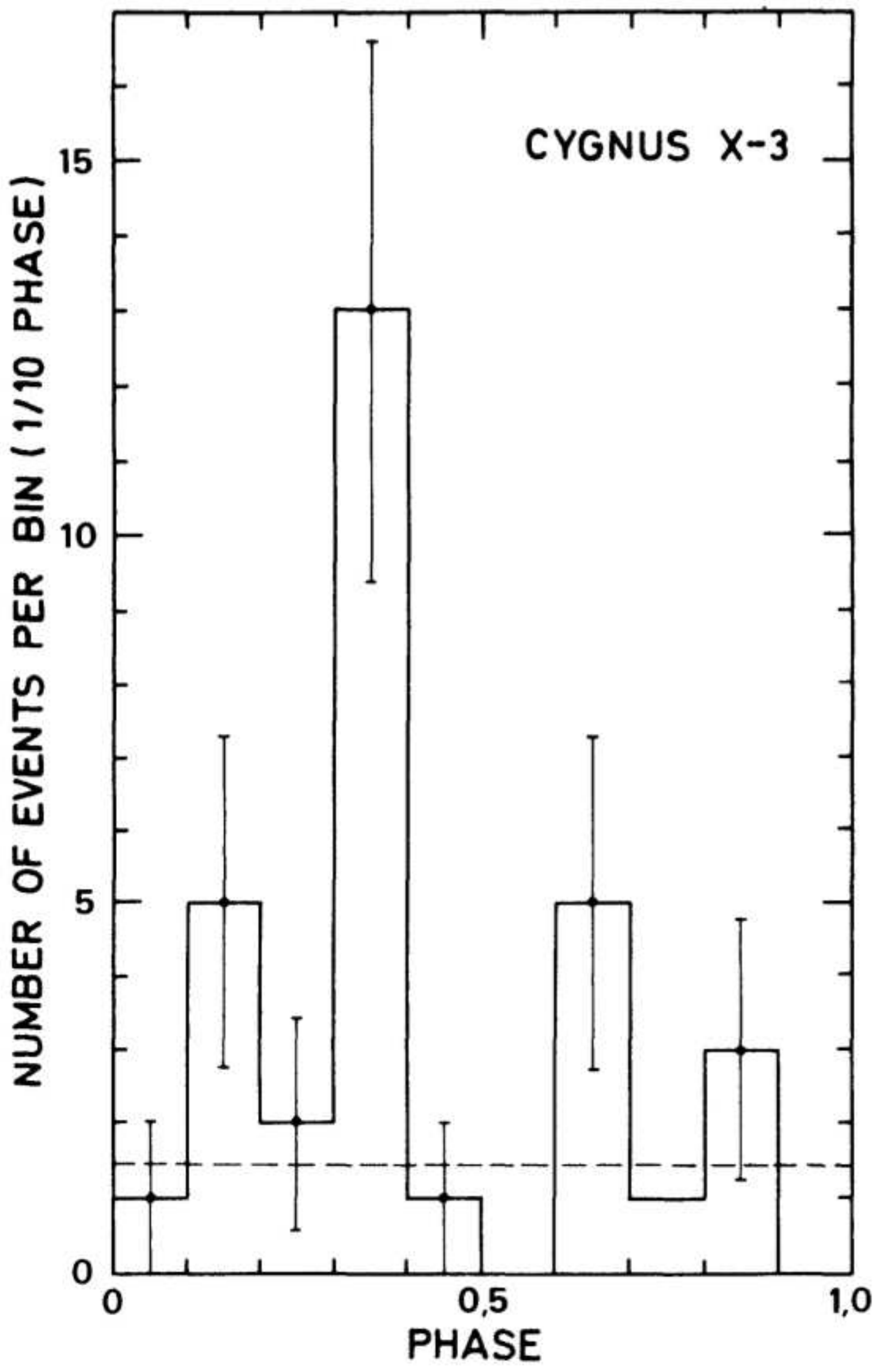}
\caption{Phase diagram of Cygnus X-3, indicating a 4.8-hour periodicity \cite{Samorski1983}.\label{Fig4.3}}
\end{figure}

At the University of Kiel, a small but very active group pursued cosmic ray research. In the mid 1970s, the group improved their cosmic ray experiment by extending the existing scintillator detector array and measuring more parameters of air showers. They added quite a few scintillation counters up to a distance of 100 m from the previous core detector arrangement. Also, they improved the measurement of the different shower components, 
such as a measurement of the electron, the muon and the hadron parameters of individual showers. Fig. \ref{Fig4.1} shows the layout of the inner part of their array. The array comprised the following detectors:
\begin{itemize}
\item 27 unshielded scintillation counters for measuring the shower size and core position,
\item 11 scintillation counters connected to 22 fast timing circuits for determining the arrival directions by means of time of flight measurements,
\item a 31~m$^2$ neon hodoscope with 176,000 flash tubes for measuring the electron core structure. This hodoscope was protected by wood mounted on the ceiling with 2.5\,g\,cm$^{-2}$ density. The flash tubes were recorded on film by four cameras in case of a trigger from a few scintillation counters,
\item 13 scintillation counters shielded by 2\,cm of lead and 0.5\,cm of iron for measuring the energy flow of the electromagnetic shower component,
\item a 65~m$^2$ neon hodoscope of 367,500 flash tubes under a layer of 880\,g\,cm$^{-2}$ concrete. These flash tubes registered shower muons and the hadronic component. Three cameras recorded the flash tube pattern when an event trigger occurred.
\end{itemize}
Operation of this array, one of the most powerful detectors of that time, started in 1976. As the installation was at sea level, the threshold of the detector was quite high, about $1-2\times$10$^{15}$~eV. In 1983, the group published the results from four years of data recording, focused on the search of possible gamma-ray emission from Cygnus X-3 \cite{Samorski1983}. Cygnus X-3 is one of the strongest X-ray emitting binary star systems and was for a long time a prime search candidate for the emission of TeV gamma rays. Quite a few experiments claimed to have seen gamma rays with roughly a $3\sigma$ excess. In their 1983 publication, the Kiel physicists claimed a $4.4\sigma$ excess at the position of Cygnus X-3, at a declination angle of $40.9^\circ\pm1.5^\circ$, and $307.8\pm2.0^\circ$ in right ascension.
The result was based on 3,838 hours of observation time and a sensitive area of 2800\,m$^2$. Fig.~\ref{Fig4.2} shows the published results. What enhanced the belief in the result was the finding of a strong peak in the phase diagram of a 4.8 h periodicity, derived from X-ray data ten orders of magnitude lower in energy (Fig.~\ref{Fig4.3}). 
As early as 1973, the SAS-2 satellite had reported gamma-radiation within a narrow phase interval of the 4.8\,h-phase \cite{Parsignauld1976}. Common belief was that Cygnus X-3 comprises a binary system generating gamma rays and that the eclipsing of the compact star by its companion was most likely causing this periodicity signal.

This result created quite some interest and intense discussion not only in the CR community, but also in the HEP community, and quite some groups started to observe specifically Cygnus X-3. Basically, this result triggered a revival in interest in the search for the sources of CRs. 
In the wake of the Kiel experiment quite a few other experiments 
confirmed the Kiel result, mostly claiming also to observe a 4.8-h periodicity signal. Details go beyond the scope of this article. We refer to the papers \cite{Lloyd-Evans1983}, \cite{Marshak1985}, \cite{Watson1985} which include some general discussion as well
more information on such air shower arrays as Haverah Park, Cygnus, EAS-Top or CASA-MIA. Without these positive observations, in particular that of Haverah Park,
the momentum of gamma-ray astronomy after the Kiel result might have
soon slowed down.

Later, some additional results surfaced, which could have reduced the excitement. It turned out that the excess was about $1.5^\circ$ off the position of Cygnus X-3 (M.\,Samorski, {\it private communication}), but this was considered consistent with the systematic uncertainty of the measurement of the shower arrival direction by means of time of flight measurements. Also, not published in the 1983 article, the muon hodoscope results showed that nearly all showers in the Cygnus X-3 bin had a very similar muon flux as that of hadronic showers, i.e., also the excess showers were consistent with hadronic showers. In the absence of reliable gamma experiments at accelerators it was speculated that electromagnetic showers above 10$^{15}$\,eV had a strong hadronic component, explaining the presence of a strong muon component. Cygnus X-3 is about 12 kiloparsecs away from the Earth, and photons
with an energy $> 10^{15}$\,eV should have been strongly attenuated over this distance
by interaction with the cosmic microwave background , see Sect.~\ref{Sect:EBL}. Again, CR physicists speculated that in the absence of trustworthy accelerator experiments PeV gamma rays behaved quite differently compared to low-energy gamma rays. 

The Kiel result triggered
quite a few $3\sigma$ observations as well a similar number of contradicting results, and a flood of exotic theoretical predictions for an energy range inaccessible to HEP accelerator experiments. In this context it is noteworthy that analysis techniques at that
time did not yet follow the statistical standards of today and did not correctly
consider trial factors. The concept of "blind analysis" was still far from
being accepted. 

It is interesting to note that the Kiel physicists assumed that the gamma-ray flux to be about 1.5\% of the total VHE CR flux \cite{Samorski1983}. 
Eight years later, HEGRA, the High-Energy Gamma-Ray Array, was started by the Kiel group on the Canary island of La Palma, at a height of 2,200 meters above sea level. With much higher precision and higher data statistics, HEGRA did not confirm the signal from Cygnus X-3 \cite{Merck1991}, see the lower panel of Fig.~\ref{Fig4.2}. 
Also the CASA-MIA experiment, at that time the EAS experiment with the highest sensitivity (array size 500$\times$500 m$^2$; median energy of 100 TeV) could not find any signal from Cygnus X-3 \cite{Borione1997}. 
However, as it cannot be excluded that a signal from Cygnus X-3 is variable, the Kiel result is not necessarily disproved by later negative observations.

\section{How far can we ``see'' with VHE gamma rays?}
\label{Sect:EBL}
\label{Sect:5}
An important issue, which was not very much considered by the CR physicists of the 1960s to 1980s, was the question of how far VHE gamma rays would, on average, propagate through the Universe before being lost by absorption or scattering. In 1964, Arno Penzias and Robert Wilson discovered the 2.7-K cosmic microwave background radiation (CMB) \cite{Penzias1965}. Later, it was found that besides the dominant CMB the Universe is also ``filled'' with a wide spectrum of low energy photons, although the 2.7-K CMB photons with about 420 photons per cm$^3$ were by far the most dominant ones. These low energy photon fields are basically a calorimetric measure of all past and present radiating cosmic objects. VHE gamma rays have to pass this soup of low energy photons and might occasionally interact with them forming an e$^+$e$^-$ pair and because of this are lost to the observer. This process is described by quantum electrodynamics (QED). The cross section peaks close to the double electron mass in the center of mass of the two photons.

\begin{figure}[h]
\centering
\includegraphics[width=.65\linewidth]{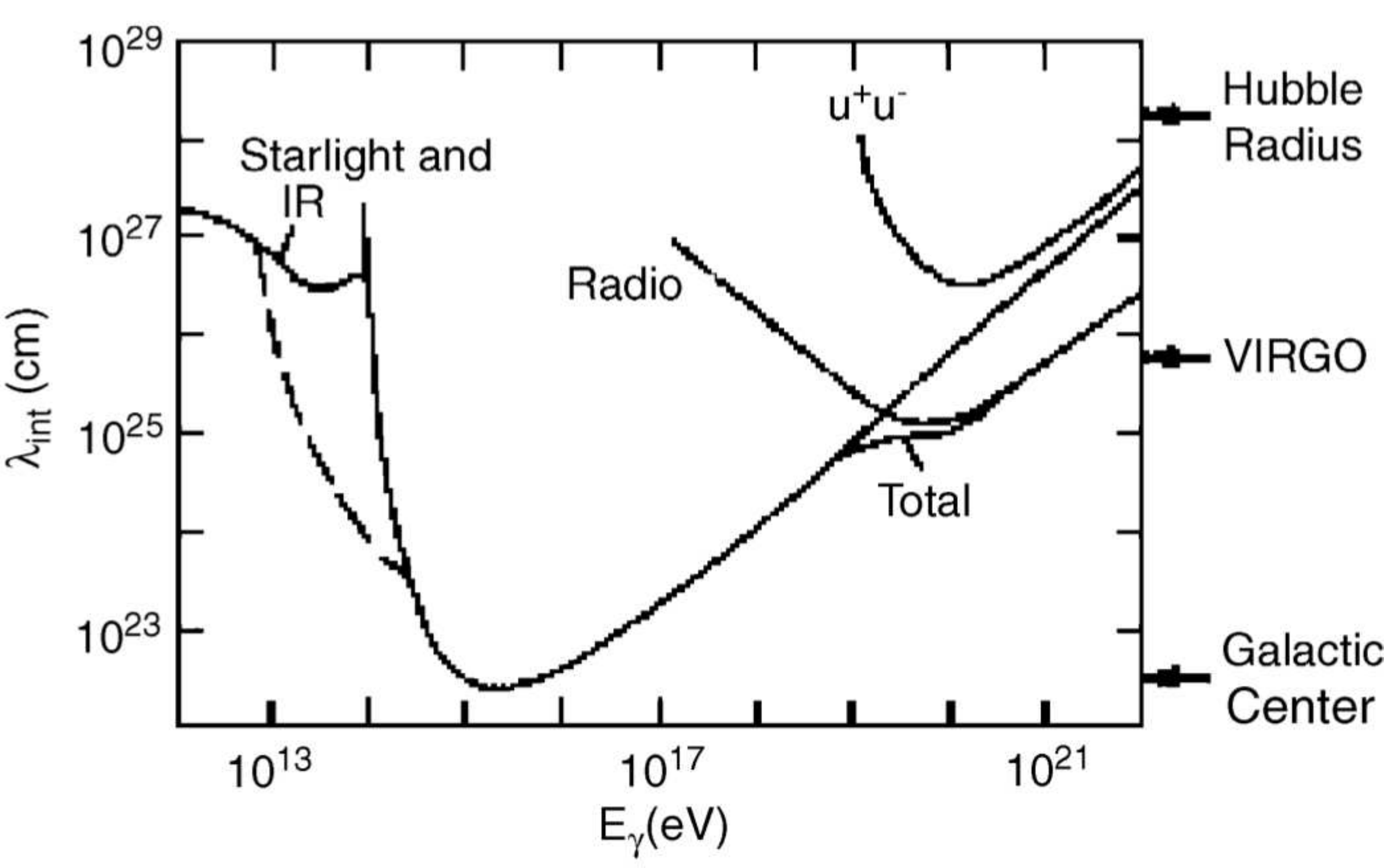}
\caption{Absorption length of VHE gamma rays in the Universe due to interaction with the low energy photon fields. The dominant absorption around 10$^{15}$~eV is caused by interactions with photons of the 2.7-K CMB. From X. Chi, J. Wdowczyk and A. W. Wolfendale \cite{Chi1992}. Figure from \cite{lorenznima}.\label{Fig5.1}}
\end{figure}

\begin{figure}[h]
\centering
\includegraphics[width=.8\linewidth]{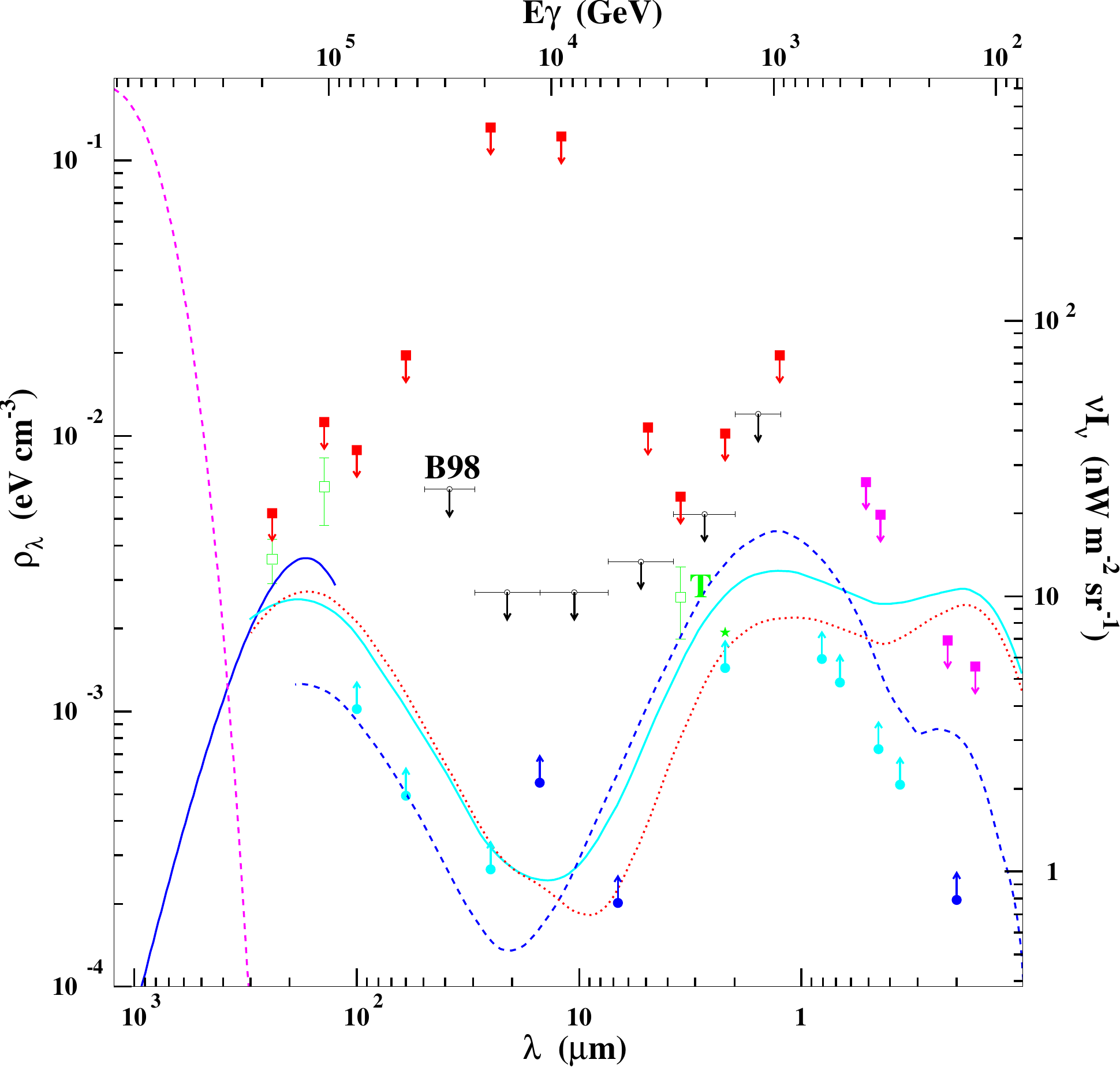}
\caption{The situation of the measurements of the extragalactic photon background around the year 1999. The peak at the very left is caused by the 2.7-K CMB, the right peak around 1 micron wavelength is an estimate of the absorption by stellar radiation and the peak around 100 micron due to recycled photons from reheated dust. The experimental data reflect the measuring precision of that time \cite{Catanese1999}, see reference and \cite{Vassiliev2000} for further references of the individual measurements.\label{Fig5.2}}
\end{figure}

 Depending on density and energy of low energy photons as well as
the energy of the gamma ray, the propagation length of the VHE gamma rays is limited. As early as around 1966, R. J. Gould and G. Schr\'eder \cite{Gould1966} made a first prediction of the opacity of the Universe. At around 10$^{15}$~eV the absorption length is just around 10 kiloparsec, i.e., approximately the distance of the center of our Galaxy to us. Fig.~\ref{Fig5.1} shows the prediction of the absorption length from a later work of X. Chi, J. Wdowczyk and A. W. Wolfendale \cite{Chi1992} for the gamma-ray absorption length due to the different contributions of the density of low energy cosmic photon fields. The density of infrared (IR) photons is more or less a crude estimate. This plot indicates that detections of PeV gamma rays from observing extragalactic sources are impossible, and it is quite obvious that air shower arrays with their high threshold had close to no chance of significantly contributing to the search for gamma-ray emitting sources of CRs, except if these were very close ($<10$~kpc) to the Earth. Also detectors needed to be placed at high altitudes of a few thousand meters. Figure \ref{Fig5.2} reflects very much the uncertainties in experimental data of starlight and IR photons above the 2.7-K CMB \cite{Catanese1999} in the year 2000. Instruments flown on satellites or on the ground carried out most of these measurements. The main experimental difficulties resulted from foreground light: zodiacal light and IR radiation from our Galaxy, which was much stronger than the extragalactic starlight and IR radiation. Since roughly the year 2000, the analysis also included spectral measurements from distant active galactic nuclei (AGNs). These measurements are not influenced by the foreground radiation, but need some model assumptions about the initial spectral shapes. Normally, the high energy part of spectra of distant AGNs is in part suppressed by the interaction with the cosmic low-energy photon fields. These distorted spectra allow to estimate in most cases the density of the low-energy photon fields. It is expected that in the near future sufficient high redshift AGNs will be discovered in VHE gamma rays, and it will be possible to determine a rather precise shape of the cosmic background light densities.

\section{The Whipple collaboration opens the window of VHE gamma-ray astronomy in 1989}
It took over 35 years until the air Cherenkov technique was rewarded with the first discovery of a VHE $\gamma$-ray emitting source since the initial observation of Cherenkov light from air showers by J. V. Jelley. The first-generation Cherenkov telescopes were in general using relatively small mirrors and very simple readouts in the form of a single PMT. In 1968, a large 10-m telescope was completed at the Fred Lawrence Whipple Observatory in Arizona, USA \cite{Fazio1968}. Figure \ref{Fig6.1} shows a photograph of the 10-m Whipple telescope at Mount Hopkins. Again, during the first phase only a single PMT was used as a ``camera'' and thus, $\gamma$/hadron discrimination was impossible. Therefore, no source could be detected although the light-collecting mirror was sufficiently large. Then, under the leadership of Trevor Weekes, both the instrument and the analysis methods were developed further to increase the sensitivity, and a method for the crucial $\gamma$/hadron separation was implemented, enabling the search for sources of much lower $\gamma$-ray fluxes than in other experiments. In 1989, the Whipple collaboration published the first convincing observation of gamma-ray emission from the Crab nebula \cite{Weekes1989}. It was basically a culmination of 10 to 20 years of hard experimental work with many steps of improvements. While quite a number of discoveries in particle physics are just surprise results, like for example the discovery of the $\psi$ particle at the SPEAR storage ring at SLAC \cite{Augustin1974}, the opening of the new window in VHE $\gamma$ astronomy was a long and tediously prepared search for the first VHE $\gamma$ source over many years. In a way, the finding of $\gamma$-ray emission from the Crab nebula was a triple-A discovery:

\paragraph{A1:} The collaboration concentrated on a source that turned out to be the strongest steady state galactic source. Already in 1958, Philip Morrison \cite{Morrison1958} and, independently in 1959, Guiseppe Cocconi \cite{Cocconi1959} had put forward strong arguments for observing VHE gamma rays from the Crab nebula  and made predictions for high $\gamma$-ray fluxes. Two years later, Alexander Chudakov and Georgii Zatsepin published a paper on methods for searching for point sources of high energy photons
\cite{Chudakov1961}, appearing at a time when their detector at Crimea was already under construction. A few years later, this detector started to search for photons from the Crab nebula \cite{Chudakov1965}. Ever since that time, the Crab nebula was a target of VHE $\gamma$-astronomy, but the Whipple collaboration spent a remarkably long observation time of 80~h spread over three years.

\paragraph{A2:} They used a telescope of a large light collection area and for the first time a camera allowing an efficient $\gamma$/hadron separation of the data. The use of an ``imaging camera'' was at first proposed by T. C. Weekes and K. E. Turver \cite{Weekes1977}, but it took another 10 years until the first useful imaging camera was built. This camera with only 37 PMTs covered a field of view (FOV) of 3.5 degrees diameter. It allowed the recording of coarse pictures of air showers and making a simple discrimination of electromagnetic and hadronic showers. This rudimentary camera was nevertheless the start of the design of consecutively improved cameras with finer and finer pixel sampling while the FOV of 3.5$^\circ$ is quite a standard even of today�s telescopes. Fig.~\ref{Fig6.2} shows the first practical Whipple camera with 37 PMTs (pixels) and the pattern of the PMT arrangement.	

\begin{figure}[h]
\centering
\includegraphics[width=.6\linewidth]{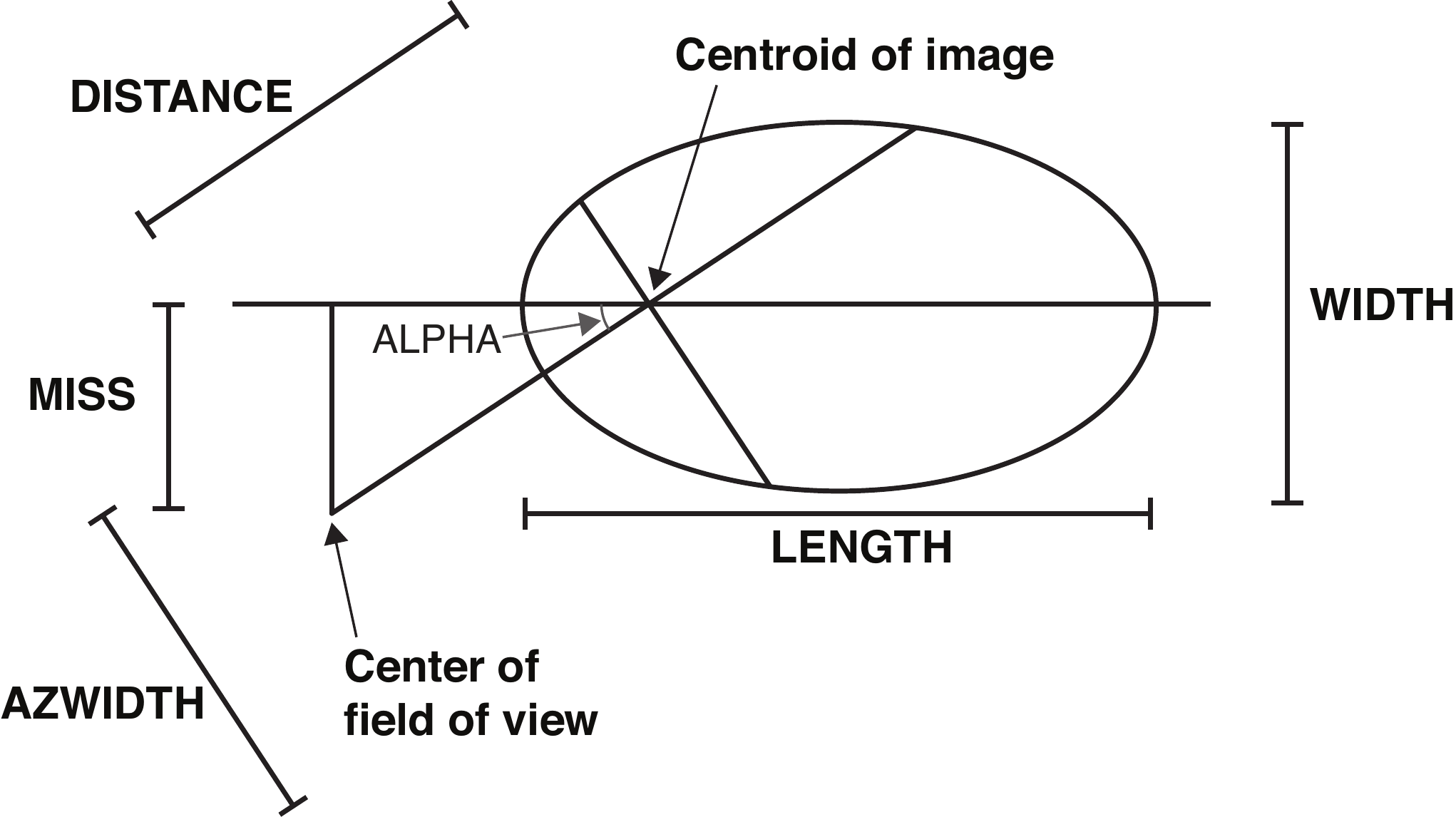}
\caption{The image parameterization employed by \cite{Weekes1989}: The shower image is characterized by the width and length of the shower ellipse along with some parameters describing the position and angle of the shower in the camera plane -- showers originating from the source should point back to the source position, i.e., have a small MISS value. \cite{Weekes1989} showed that there are distinct differences in all parameters given between gamma- and hadron-initiated showers. Today, the original MISS parameter has been superseded by the ALPHA parameter, describing the angle between the weighted center of the shower ellipse and the camera center, and, later still, by the $\theta^2$ parameter, allowing for analyses without assumptions about the source position. \label{Alpha}}
\end{figure}

\begin{figure}[h]
\centering
\includegraphics[width=.6\linewidth]{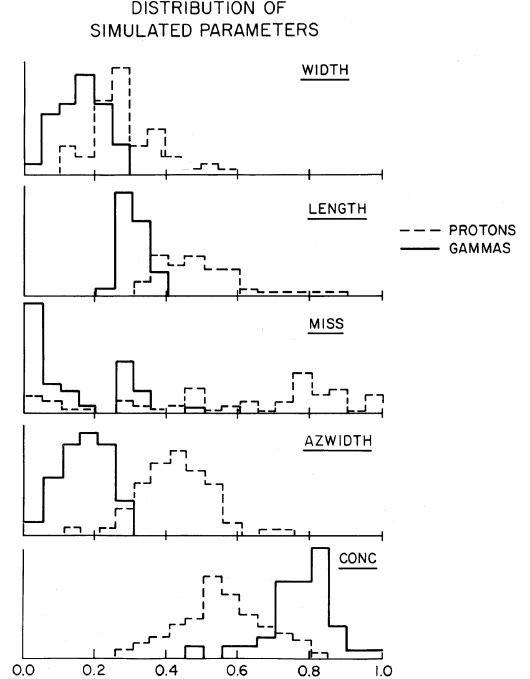}
\caption{Image parameter distributions predicted by simulations for an on-axis gamma-ray source. All parameters can be used for discriminating gamma-ray induced and charged cosmic-ray induced air showers. All horizontal scales (except for CONC) are given in degrees. Figure taken from the original discovery publication \cite{Weekes1989}.\label{Alpha2}}
\end{figure}

\begin{figure}[h]
\centering
\includegraphics[width=.8\linewidth]{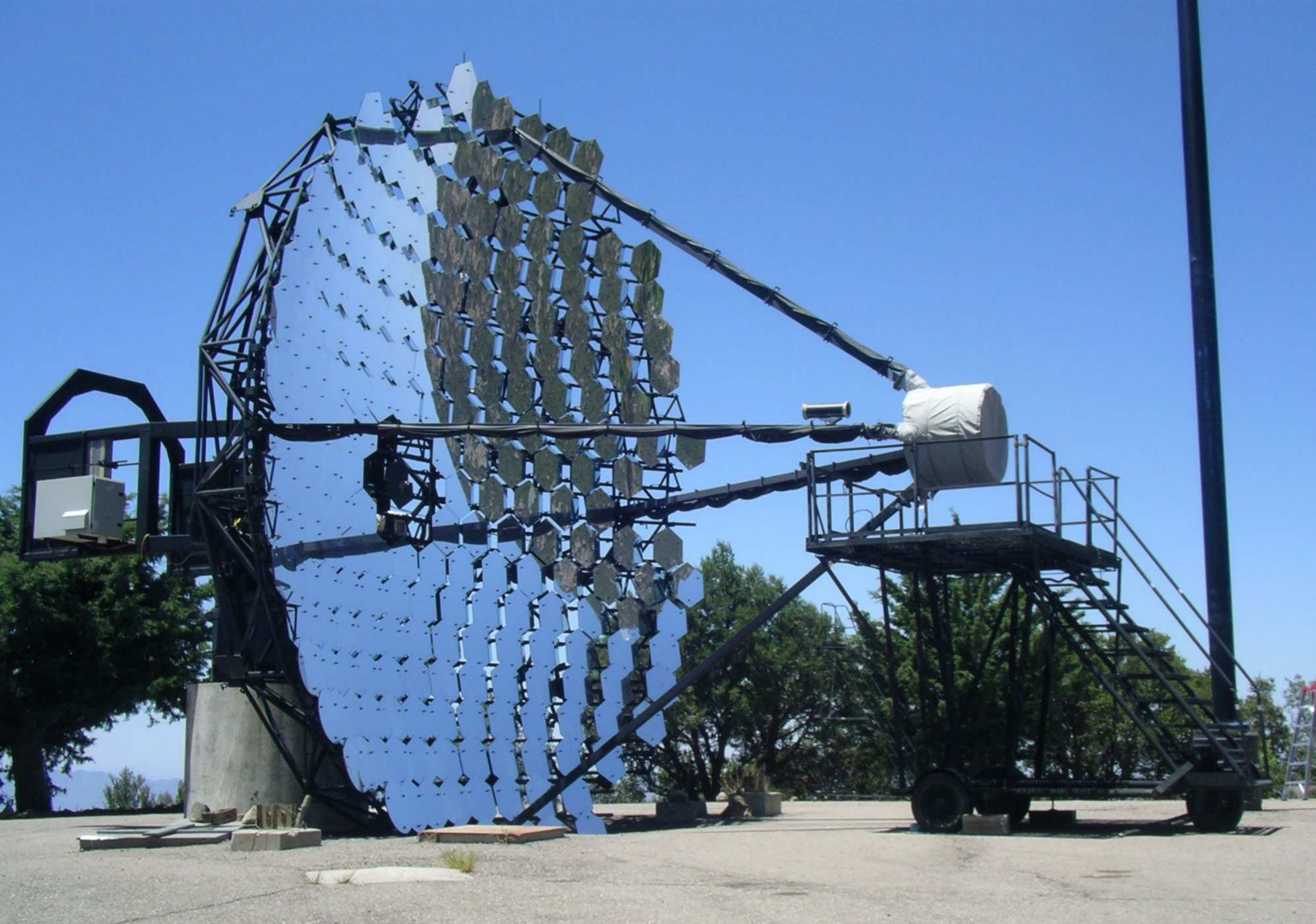}
\caption{Photo of the Whipple 10-m telescope at Mount Hopkins. Courtesy Brian Humensky.\label{Fig6.1}}
\end{figure}

\begin{figure}[h]
\centering
\includegraphics[width=.4\linewidth]{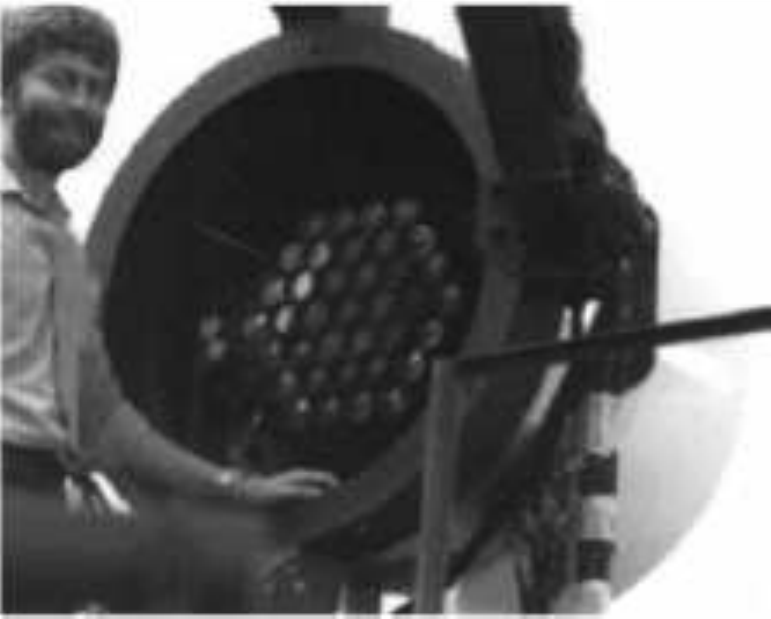} \includegraphics[width=.4\linewidth]{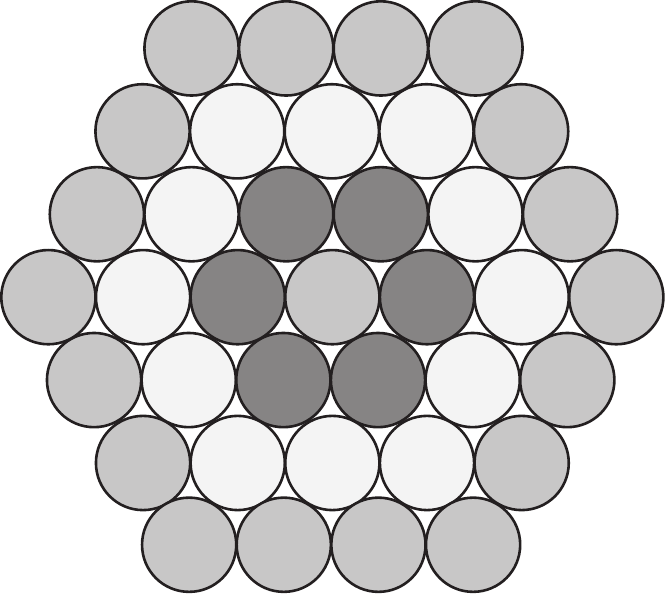}
\caption{The first Whipple telescope camera with 37 PMTs, allowing efficient $\gamma$/hadron separation. Right figure taken from \cite{Weekes1989}.\label{Fig6.2}}
\end{figure}

\paragraph{A3:} The third and most important achievement was the introduction of a refined $\gamma$/hadron separation method based on the calculation of image moments. This analysis developed by the Whipple collaboration in the mid-eighties was based on the combination of both a measurement of the shower image orientation, originally proposed by T. C. Weekes in 1981 \cite{Weekes1981} and an analysis to evaluate the difference in images between gamma-ray showers and hadron showers, originally proposed by A. A. Stepanian, V. P. Fomin and B. M. Vladimirsky, \cite{Stepanian1983}. The shower image should align with the position of the source in the camera (Fig. \ref{Alpha}) and images of gamma and hadron showers should distinctly differ in shape with gamma showers being rather slim and concentrated, while hadron showers are much wider and more irregular. Of course, shower fluctuations could sometimes make discrimination difficult and limit discrimination power. The originally rather simple moment analysis, commonly known as Hillas parameterization analysis \cite{Hillas1985} became the basic concept for $\gamma$/hadron separation in future Cherenkov telescope experiments. It is still in use in most of today�s experiments with some refinements based on additional information retrieved from better cameras with finer resolution and better shower timing data. Fig. \ref{Alpha2} shows image parameter distributions for (simulated) proton and gamma events and how these distributions differ and thus allow for a rejection of hadronic background events.

The original Whipple detection is based on 652,974 {\it On} source and 651,801 {\it Off} source events recorded in equal time spans spread over a total of 80 hours. ({\it On} source means that the telescope is pointed towards the source under study, while {\it Off} source observations are conducted towards a direction close by the target source, at an area where no gamma-ray source is expected; for background normalization). While the classical analysis method gave just a 1-$\sigma$ excess, the $\gamma$/hadron analysis bases on the Hillas moments allowed the hadronic background to be reduced by 98\% (with a loss of around 50\% of $\gamma$ events). Eventually, an excess at a 9-$\sigma$ level was found. This observation was confirmed in the following years by a number of other Cherenkov telescope experiments and opened the window for VHE $\gamma$ astronomy.

A very important, but hardly noticed, byproduct of the detection of gamma rays from the Crab nebula was the first trustworthy measurement of the $\gamma$ flux of $\approx 0.2\%$ of the CR flux in a FOV of 2 degrees around the Crab nebula position and above about 0.7 TeV. This low value explains why past experiments had no chance of finding a real signal due to their low $\gamma$/hadron separation power. Tail-catcher detectors had a double penalty, i.e., a too high threshold and a too poor angular resolution. 

Soon, the analysis of the Whipple data showed that the $\gamma$-shower images were significantly narrower in their width than the used pixel size of 0.4-degree diameter of the first camera. In the following years, with a new camera of much finer pixels and 109 PMTs the significance could be steadily improved. Also, many other experiments followed the concept of the second-generation Cherenkov telescope with a pixelized camera and confirmed the VHE $\gamma$ emission of the Crab nebula (Tab.~\ref{Tab6.1}).

\begin{table}[h]
\centering
\caption{Some Imaging Cherenkov telescopes in the 1990s, similar to the Whipple telescope, which later confirmed the Crab nebula VHE gamma-ray emission and detected also some other gamma-ray sources.}

\begin{tabular}{llll}
\hline
Telescope & \#Cameras/Pixels & Collaboration & Ref.\\
\hline
\hline
Crimean GT48 & 2$\times$37 pixels & Crimean Astronomical Observatory & 1,2\\
Yerevan & 37 pixels & Yerevan & 3\\
Ala-Too & 144 pixels & Lebedev & 4\\
Cangaroo I & 2 Telescopes & Japan/Australia & 5\\
HEGRA & 37$+$5$\times$271 pixels & HEGRA collaboration & 6\\
Granite (Whipple$+$11-m Tel.) & 109$+$37 pixels & extended Whipple collaboration & 7\\
Narrabri & 24 pixels & Durham & 8\\
Telescope array protoype & 8$\times$256 pixels & TA coll. & 9\\
CAT & 600 pixels & CAT collaboration & 10 \\
ASGAT & 7$\times$7 pixels & ASGAT collaboration & 11\\
\multicolumn{4}{l}{(not a telescope array with genuine imaging quality)}\\
\hline
\label{Tab6.1}
\end{tabular}

References--- 1: \cite{Vladimirski1989}, 2: \cite{Fomin1991}, 3: \cite{Aharonian1989}, 4: \cite{Nikolski1989}, 
5: \cite{Kifune1989}, 
6: \cite{Aharonian1991}, 
7: \cite{Akerlof1991}, 
8:  \cite{Bowden1991}, 
9: \cite{Aiso1997}, 
10: \cite{Barrau1998},
11: \cite{Goret1991}
\end{table}

\section{Experiments of the decade 1990 -- 2000}
\subsection{The 1990 VHE sky map} 
\begin{figure}[h]
\centering
\includegraphics[width=.95\linewidth]{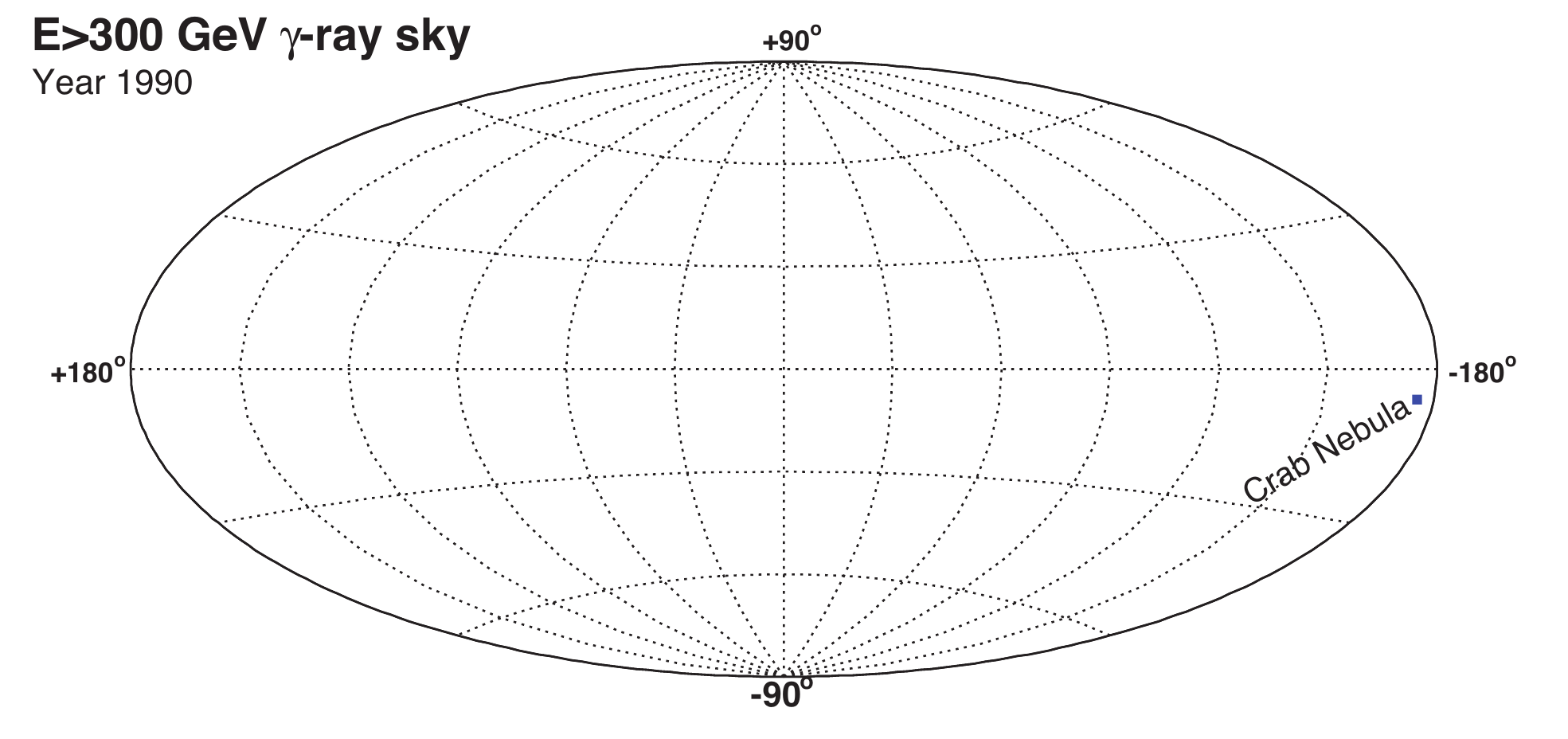}
\caption{The VHE sky map of the year 1990 with only one source -- the Crab nebula. The galactic plane is the equator of the sky map\label{Fig7.1.1}}
\end{figure}

The sky map of 1990 shows just one source, the Crab nebula as detected by the Whipple collaboration \cite{Weekes1989}. Fig.~\ref{Fig7.1.1} shows the sky map with the sole source in the galactic plane.

\subsection{A small sensation: Whipple finds an extragalactic source 5 billion light years away and opens the window for AGN studies}
\begin{figure}[h]
\centering
\includegraphics[width=.45\linewidth]{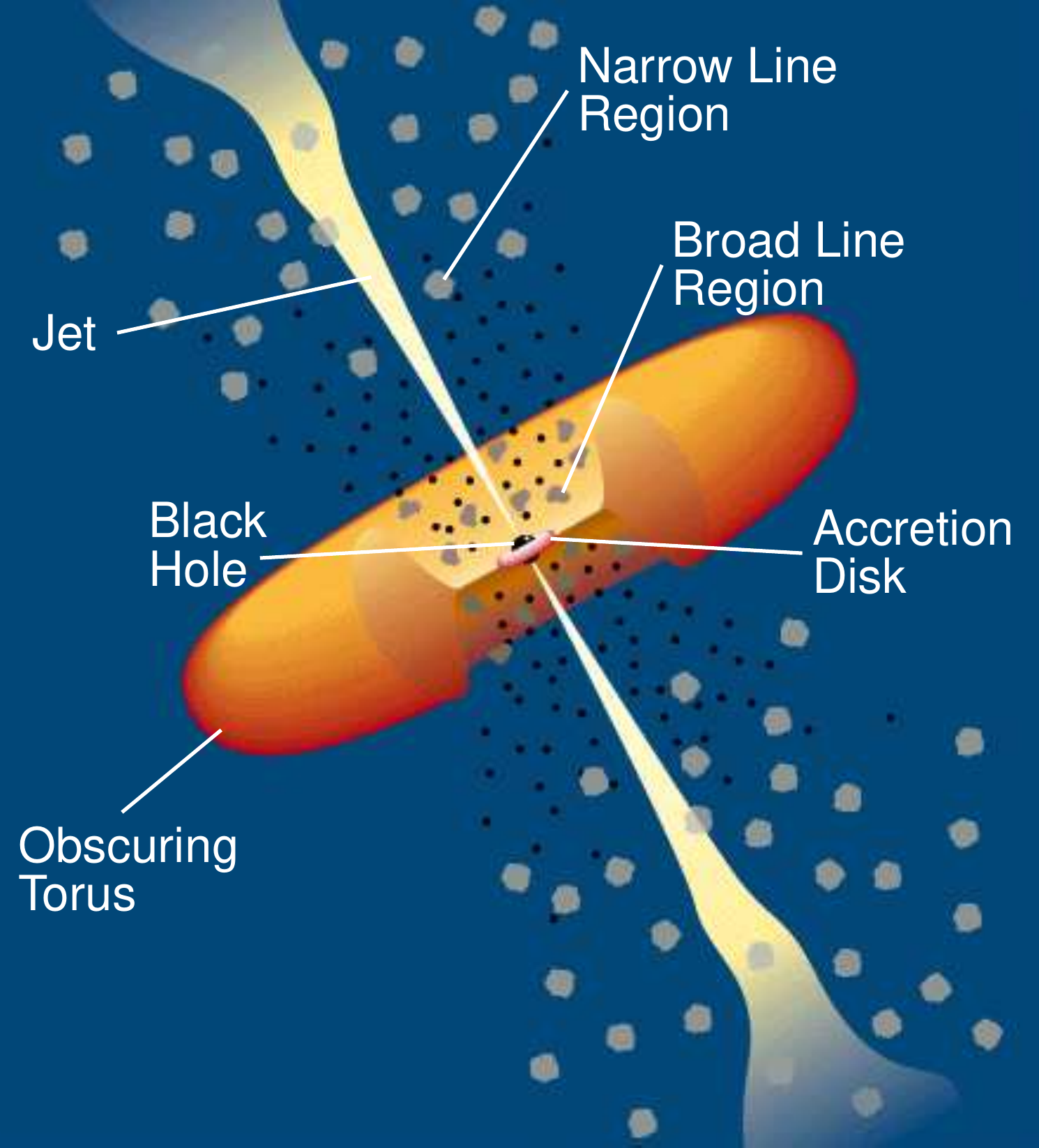}
\caption{Schematic model of an AGN.  Blazars are the subset of AGNs with their jets aligned closely to our line of sight. Picture taken from \cite{UrryPadovani}.\label{Fig7.2.1}}
\end{figure}

Not long after the discovery of VHE $\gamma$-emission from the Crab nebula and the search for some other galactic sources, the Whipple collaboration started a search for $\gamma$ emission from extragalactic sources. Candidates were AGNs of the blazar type, that had been detected in X-rays and low-energy gamma rays in satellite observations. Amongst the five candidate AGNs they selected for their study, only the weakest low energy $\gamma$ emitter, the AGN Markarian (Mkn) 421, showed a strong VHE signal of about 30\% of the Crab nebula flux \cite{Punch1992}.  If converted naively to the intrinsic brightness of the source nearly 5 billion light years away, Mkn 421 must emit over $10^6$ times more VHE gamma rays than the Crab nebula. This observation opened the window of extragalactic $\gamma$-search. Later, quite a few AGNs were detected and now nearly the same number compared to galactic sources are observed. Nearly all of them are blazars, i.e. galaxies with an accreting super-massive black hole in the center, a large accretion disc, and two jets orthogonal to the accretion disc (sometimes only one is seen, presumably due to beaming effects). Most current models assume that $\gamma$-rays are produced in the jets. In case one jet points towards the earth, they are called blazars. Many gamma-detected blazars show rapidly varying $\gamma$-activity, which is called ``flaring''. Intensity variations by a factor ten or more are observed, in extreme occasions up to a factor of $\approx$\,50 with respect to the lowest gamma-ray fluxes seen from the respective blazars. It is likely that most blazars have not yet been detected because they are currently in a ``dormant'' state. Also, the sensitivity of current Cherenkov telescopes might only allow one to see the strongest flaring sources, as up to now nearly all observed blazars have a super-massive black hole of at least 10$^8$ solar masses \cite{Wagner2008}. Fig.~\ref{Fig7.2.1} shows a conceptual view of an AGN.

\subsection{The 1992 Palaiseau conference: Towards a major imaging Cherenkov telescope}
On June 11-12, 1992, Patrick Fleury and Giuseppe Vacanti invited the community to a conference at Palaiseau with the aim of forming a project of a major imaging Cherenkov telescope \cite{Fleury1992}. The Whipple collaboration had already shown the existence of VHE sources and that the imaging Cherenkov technique was a viable method of detecting them. During the meeting many more small projects were discussed but it became obvious that:
\begin{enumerate}
\item	There are VHE sources around that can be detected provided the instruments are as sensitive in the range as the Whipple telescope
\item	Cherenkov telescopes were by far the most promising instruments for VHE $\gamma$-ray astronomy 
\item	One needs very large light collectors to achieve a low threshold
\item	A high $\gamma$/hadron separation was the key to success and methods to enhance the separation power had to be pushed further 
\end{enumerate}
The highlights of the conference were not so much the many talks about a number of different detectors but the many discussions about the fundamentals of the Cherenkov technique, the progress in computing power and electronics as well as a better understanding of the development of air showers and a better understanding of the detector response. Very much to the frustration of Patrick Fleury, no consensus could be achieved on the goal of the conference -- the formation of a large project, although it became obvious that all the above-mentioned conditions needed to be fulfilled together with a large collaborating activity and sufficient financial support. The majority of the participants considered that a very large telescope was too expensive, as it would cost close to the price of a satellite. Nevertheless, the seeds for large telescopes of the third generation were planted.  In the following years, a few more workshops with the same goal were arranged, more and more details on the method of $\gamma$/hadron separation developed and construction elements of large telescopes resolved. 
These meetings paved the way for many ideas of the third generation telescopes. 
Plans for cameras with much finer pixel structures and better readout emerged. Even today's CTA project (cf. Sect. \ref{CTA}) can be traced back to quite a few ideas developed at the Palaiseau conference. 

\begin{figure}[h]
\centering
\includegraphics[width=.95\linewidth]{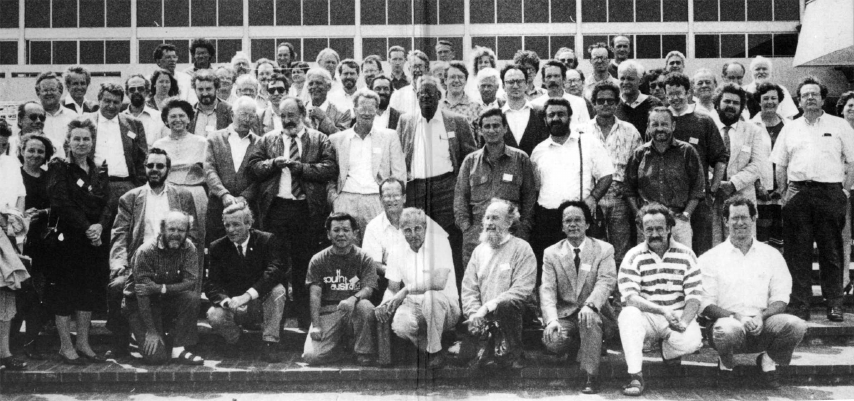}
\caption{Group photo of the participants of the 1992 conference ``Towards a major imaging Cherenkov telescope in Palaiseau'' (taken from the proceedings volume \cite{Fleury1992}).\label{Fig7.3.1}}
\end{figure}

The conference was, in many aspects, the turning point in VHE $\gamma$-astronomy after the Whipple collaboration had opened the window of TeV gamma-astronomy. The participation in the conference was remarkably high. Fig.~\ref{Fig7.3.1} shows the photograph of the conference members. The exodus of cosmic ray physicists to accelerator experiments had ended and already the first physicists from HEP were returning to gamma astronomy (and cosmic-ray studies). Also, the dissolution of the USSR had some influence, as quite a few of their physicists had no chance anymore to pursue cosmic-ray physics in the states of the former USSR and therefore joined Western groups, bringing in particularly their knowledge of the Cherenkov technique. 

In the following sections two important issues, which helped very much to advance the detection technique but are rarely mentioned with sufficient vigor will be shortly discussed: Progress in computing power and electronics and the importance of Monte-Carlo simulations to understand shower structures and as a replacement for the missing test beams. 

\subsection{Progress in electronics and computing power drive the field}
Comparing the development of the high-energy particle physics research with that of the CR research, it becomes clear that the development of high-energy particle physics runs very much in parallel with the progress in electronic and computing power. Initial experiments at accelerators in the 1950s and 1960s dealt with large cross-section reactions, and `simple' reactions of a few secondaries at low energies. Electronic detectors were simple and based mostly on WW-II developments. Computers were of very limited power and slow, and the transition from electronic tubes to transistors had just started. The steady quest to increase the energy and to study smaller and smaller cross-section reactions and more complex interactions was again paralleled by advances in more powerful electronics detectors like spark chambers, proportional -- or drift chambers, superconductive magnets, calorimeters, silicon strip detectors, etc. and the introduction of personal computers changing the computing world. Moore's law about the doubling of the computer speed advanced much faster than the increase in accelerator energy. Better and more powerful computers provided experimentalists with the powerful tools necessary.

It was only around 1990 that the progress of the detector electronics and the affordable computer power reached the necessary level to give sufficient (but still not complete) insight into the entire chain of a cosmic air shower and its interaction with the medium of the atmosphere. In a certain, way CRs were discovered much too early compared with the analysis power of instruments and with the computing power of around 1912. This for many decades very much prevented the understanding of what in detail took place when high energy particles hit the atmosphere and produced rather complex extended air showers. Accelerator experiments of the 1990s provided the first high energy interaction data although not yet in the multi-TeV region. Also, the emerging satellite technique helped to clarify what happens at lower energies of CRs. In the 1980s, electromagnetic showers could already be fairly well simulated although no terrestrial accelerator was able to produce the $\gamma$-energies observed in CRs, and there were still doubts that the shower development could be precisely extrapolated to the TeV/PeV energy range. The simulation of hadronic showers was more complicated and still is, although the first TeV accelerator allowed  interactions in the so-called ``forward direction'' up to one TeV to be studied. Nevertheless, the steady improvement in computing power permitted simulations (with some necessary simplifications) of extended air showers and their Cherenkov photon production such that one gradually got a clearer picture of what took place in the atmosphere. In the mid 1990s the simulation of a simplified PeV hadron shower of an iron nucleus took one week on a IBM 4200 main frame computer while today it takes only a few seconds.

In summary, the power of computing and the advances in performance of electronic detector elements and their affordability in the 1990s started to match the needs for understanding and precise measuring of air showers of very high energy and paved the road to designing instruments -- mainly Cherenkov telescopes -- for detecting the strongest $\gamma$-sources.

\subsection{Monte-Carlo simulations as a replacement for missing test beams}
Detectors for high energy physics experiments need to be calibrated. For accelerator experiments the common procedure is a calibration in a test beam of known parameters, such as energy, direction, particle nature and exact timing. In cosmic ray physics a calibration with an adequate test beam is not possible. This is quite a handicap for taking measurements and contributed very much to many dubious results in the 1960s to 1980s. An alternative method is calibrating individual detector components and deducing the overall performance by huge simulation programs that include physics laws for interactions from accelerator experiments, the Earth's magnetic field, atmospheric parameters from meteorology, all possible detector elements and some extrapolations and assumptions. The Monte-Carlo simulations need quite some computing power and were not able to come close to reality before the 1990s. Still, simplifications have to be used because it is not possible to trace all the low energy particles in a shower as well as all the Cherenkov photons. Also, it is still not proven that the number of muons in hadronic showers is correctly predicted because cross sections in the multi-TeV/PeV energy region are not precisely known and one cannot determine the charge of each individual cosmic-ray particle. 

In summary, Monte-Carlo simulations in the last decades have reached sufficient predictive power for both the Cherenkov telescopes and the air-shower array detectors such that they can be used for reducing systematic uncertainties to acceptable levels. In about 10 to 15 years, the rapid and steady rise in computer power will allow simulating a large number of events and tracing all Cherenkov photons in a VHE shower in acceptable times. 

\subsection{A huge flare from Markarian 421 -- a personal episode}
In spring 1996, one of us (E.~L.) participated in the meeting of the High-Energy Astrophysics Division of the American Astronomical Society in San Diego. Also, Tadashi Kifune was attending. We both asked Trevor Weekes if we could visit the Whipple telescope and see it in operation on our way back after the conference. On May 7 we were at the site. As it was nearly full moon only about one hour of observation time was possible. The students were not very excited to switch on all the necessary instruments for such a short observation time. Nevertheless, Trevor convinced his team to start observations with the telescope, which was then pointed towards Mkn 421. Very surprisingly, the online display events looked mostly like perfect and clean $\gamma$-showers occurring at high rate. Normally one would expect that nearly all events to be from hadronic showers, zipping across the camera in all directions, but nearly never pointing to the reference position of Mkn 421 in the camera. I even suspected that the students had decided to show the visitors just Monte-Carlo events of gamma showers. In order to clarify the situation the data were transmitted via computer link to Ireland and immediately processed. A few minutes later the so-called ALPHA plot (cf. Fig. \ref{Alpha}) was sent back showing a huge signal of more than 10-$\sigma$ excess for 20 minutes of observation time. The next two 20-minute data sets showed a further rise to more than a 14-$\sigma$ excess. Then the moon came up on the horizon and its scattered light resulted in a large increase in the night sky background light and in turn in a steadily increasing background signal in the camera, confirming that these data could not be MC events. To avoid damage to the photomultipliers, the observation had to be terminated. I have never before seen such a fast flare of such a huge intensity online.

\begin{figure}[h]
\centering
\includegraphics[width=.6\linewidth]{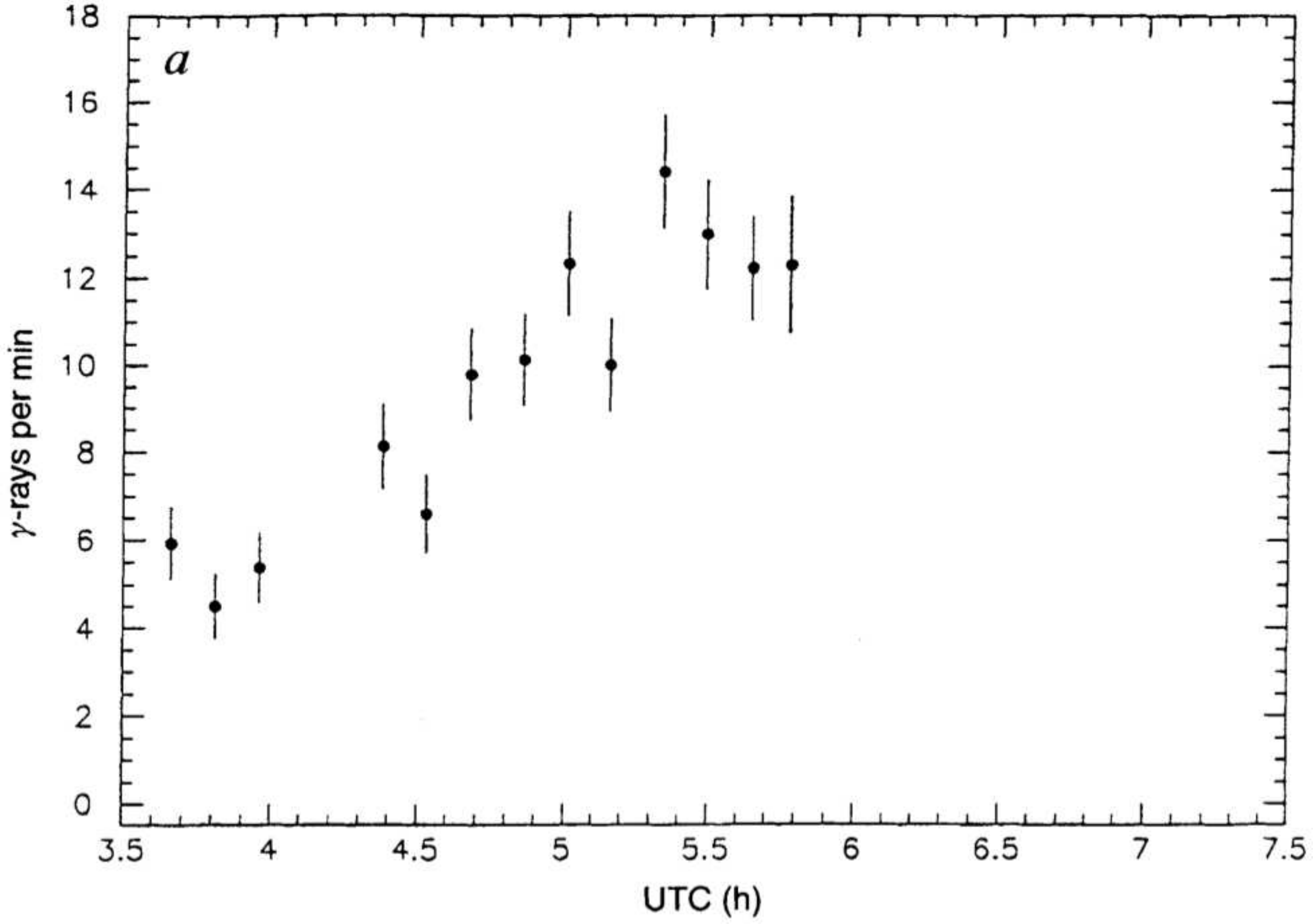}
\includegraphics[width=.6\linewidth]{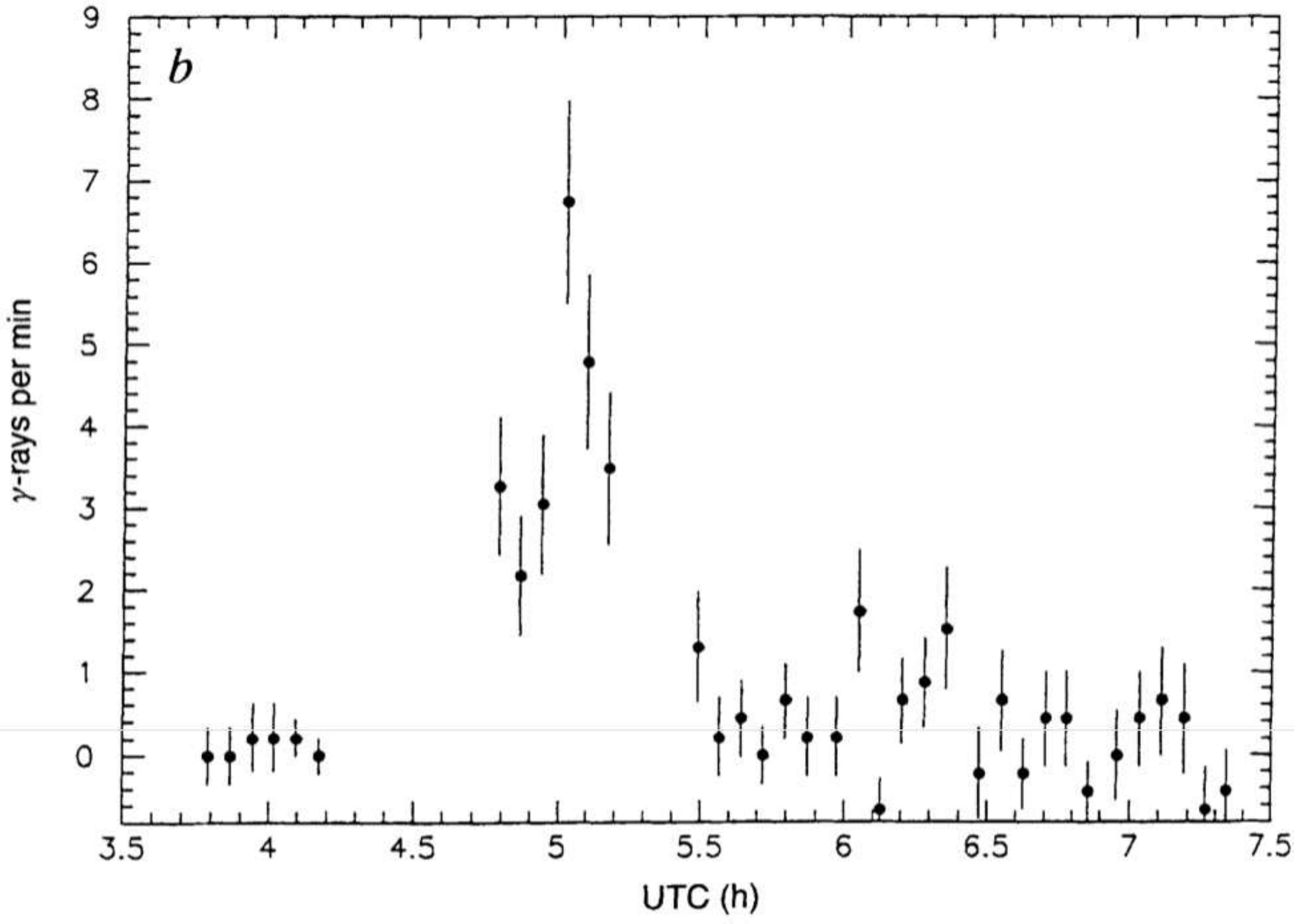}
\caption{Temporal histories of the two short but large flare events of Mkn 421. Rates are determined from the excess events after background subtraction on the intervals of $a<15^\circ$. The horizontal axes show coordinated universal time (UTC) in hours. For the 1996 May 7 flare (panel a), each point is a 9-min integration of the flux and for the 1996 May 15 flare (panel b), the integration time is 4.5 min per data point. The error bars are statistical standard deviations. The cutoff in the data in plot (a) is due to the rising full moon. Figure taken from the Nature publication \cite{Gaidos1996}.\label{Fig7.6.1}}
\end{figure}

On May 15th, Whipple observed another intense short-term flare from Mkn 421. 
While already in the past flaring had been occasionally observed from AGNs, 
and in particular Mkn 421 had flared both years before this episode,
very short flares had not been expected. The data were published in Nature \cite{Gaidos1996}, the flare curves are shown in Fig.~\ref{Fig7.6.1}. After that event, quite a few fast, high-intensity  flares were observed also from other AGNs, the most remarkable flaring activity so far in PKS 2155-304 in 2006 \cite{2155high2}. Particularly the fast evolution of these flares, that point directly to very compact emission regions, remain difficult to be explained in acceleration models.

\subsection{A persistently flaring blazar: Mkn 501 flares for over six months}
Soon after the discovery of Mkn 421, the Whipple collaboration discovered another blazar, Mkn 501 \cite{Quinn1996} at a redshift of $z = 0.034$, at nearly the same distance as Mkn 421 and of very similar performance. The VHE $\gamma$-emission of Mkn 501 was soon afterwards confirmed by the HEGRA collaboration \cite{Bradbury1997}. 

\begin{figure}[h]
\centering
\includegraphics[width=.8\linewidth]{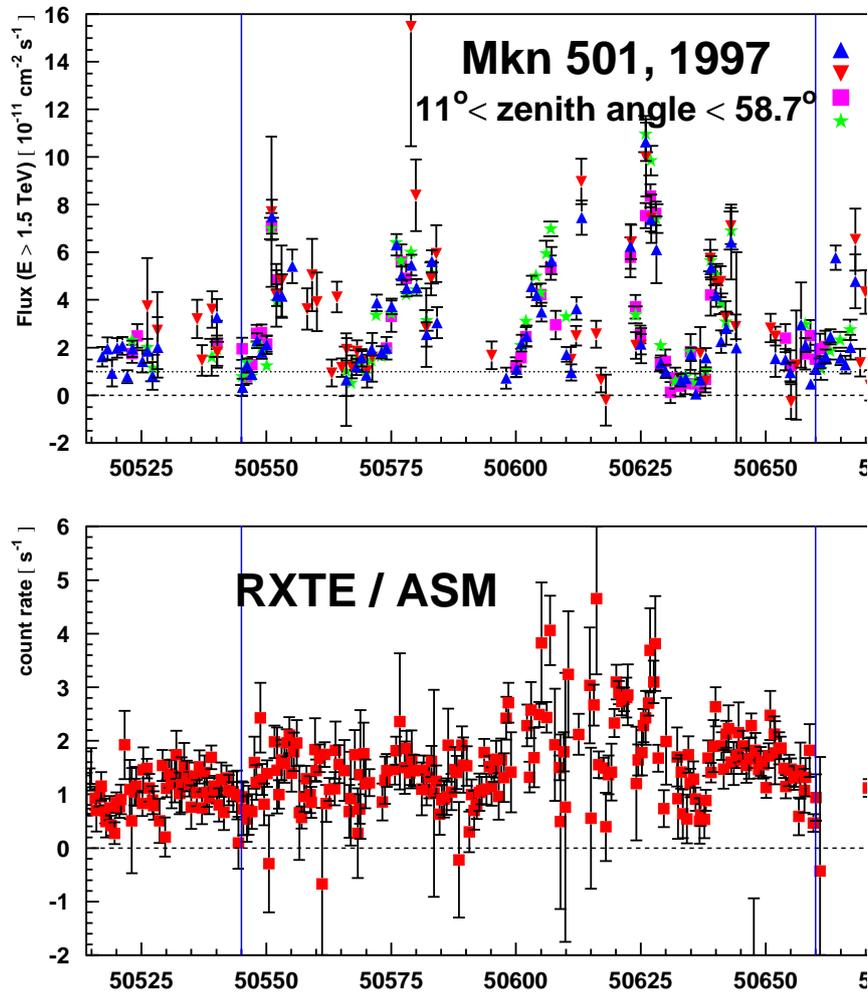}
\caption{The light curve of Mkn 501 in summer 1997. Flux variations in the range of 20 were observed in the VHE domain. The flaring activity extended over the entire observation period of 6.7 months. Due to a new observation method introduced by HEGRA it was possible to observe such a strong source also during partial moonlight. Data from  D. Kranich's Ph.D. thesis \cite{Kranich2002}. The TeV data show much larger fluctuations than the X-ray data recorded by RXTE \cite{Remillard1997}. \label{Fig7.7.1}}
\end{figure}

In 1997, Mkn 501 showed a series of extremely large outbursts extending in time over the entire observation period in 1997 and, up to now, never seen from any other AGN. The flare intensities reached peak values exceeding the low state by up to approximately a factor 20. 
The flaring activity was observed by HEGRA stereoscopic system \cite{hegra501}, TACTIC \cite{tactic501}, and the Whipple telescope \cite{whipple501}. At that time, the HEGRA collaboration introduced a new method for observing strong sources also during partial moonlight, thus
HEGRA was able to collect a nearly continuous light curve over nights during nearly 6 months.
Fig. \ref{Fig7.7.1} shows this flux measurement above 1.5 TeV from the HEGRA collaboration during the observation period in 1997. The data are compared with the X-ray data from the RXTE satellite between 2\,keV$ < E <10$\,keV \cite{Remillard1997}. 
Fig. \ref{Fig7.7.1} highlights the enormous variation in the highest energy domain while at lower energies also a change in the X-ray flux was observed but with a much smaller and smoother flux variation. 

\subsection{Stereo observations improve the sensitivity of Cherenkov telescopes}
\begin{figure}[h]
\centering
\includegraphics[width=.95\linewidth]{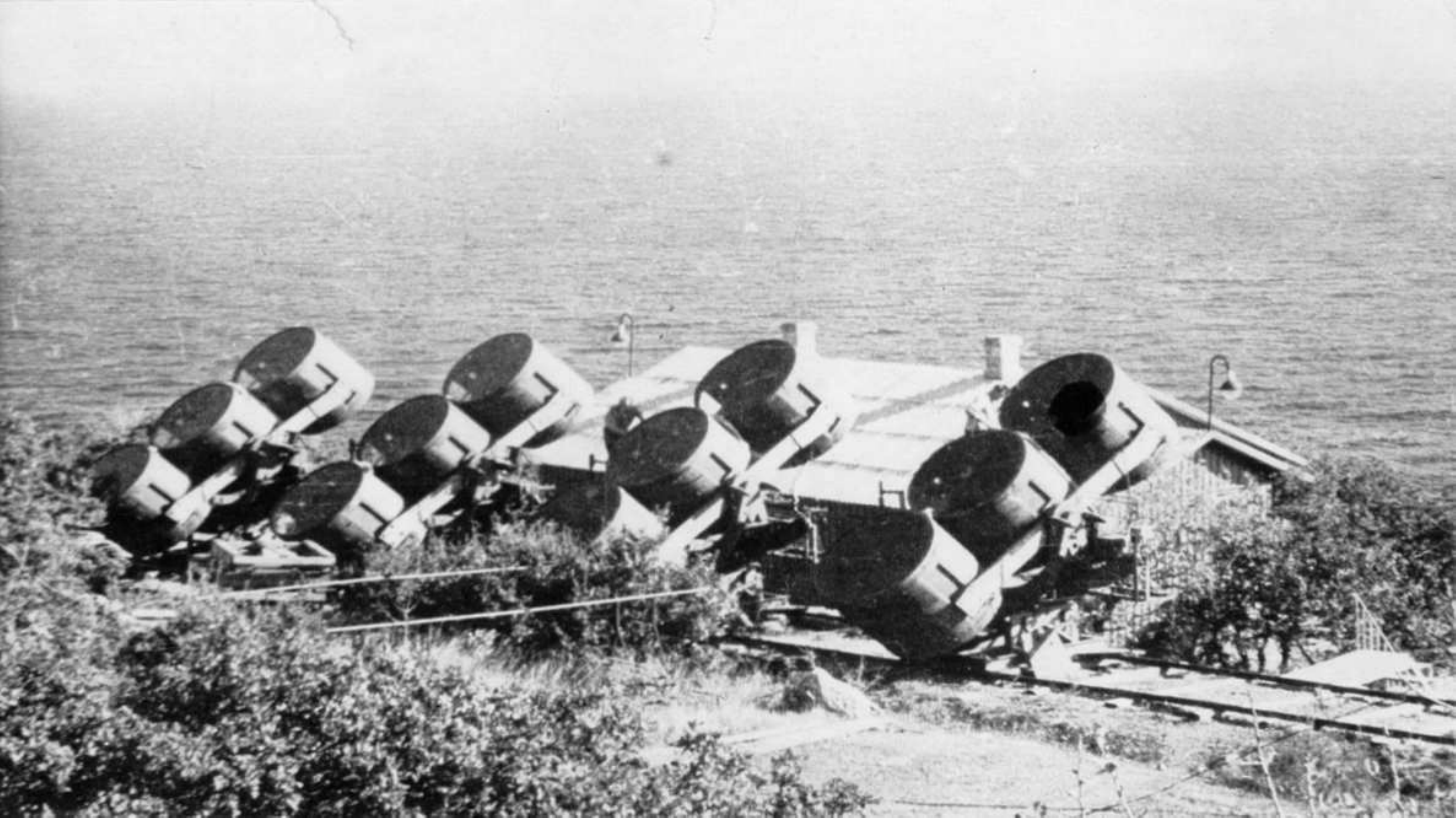}
\caption{Photograph of the Crimean multi-telescope setup. Each barrel contains one Cherenkov telescope. Three telescopes form a unit. Each of those units can be positioned along a railway system and view the air showers under slightly different angles. This arrangement allowed both coincidence measurements and simple multi-telescope observations. The system was used from 1960 until 1963. Figure courtesy T. C. Weekes.
\label{Fig7.8.1}
}
\end{figure}

\begin{figure}[h]
\centering
\includegraphics[width=.65\linewidth]{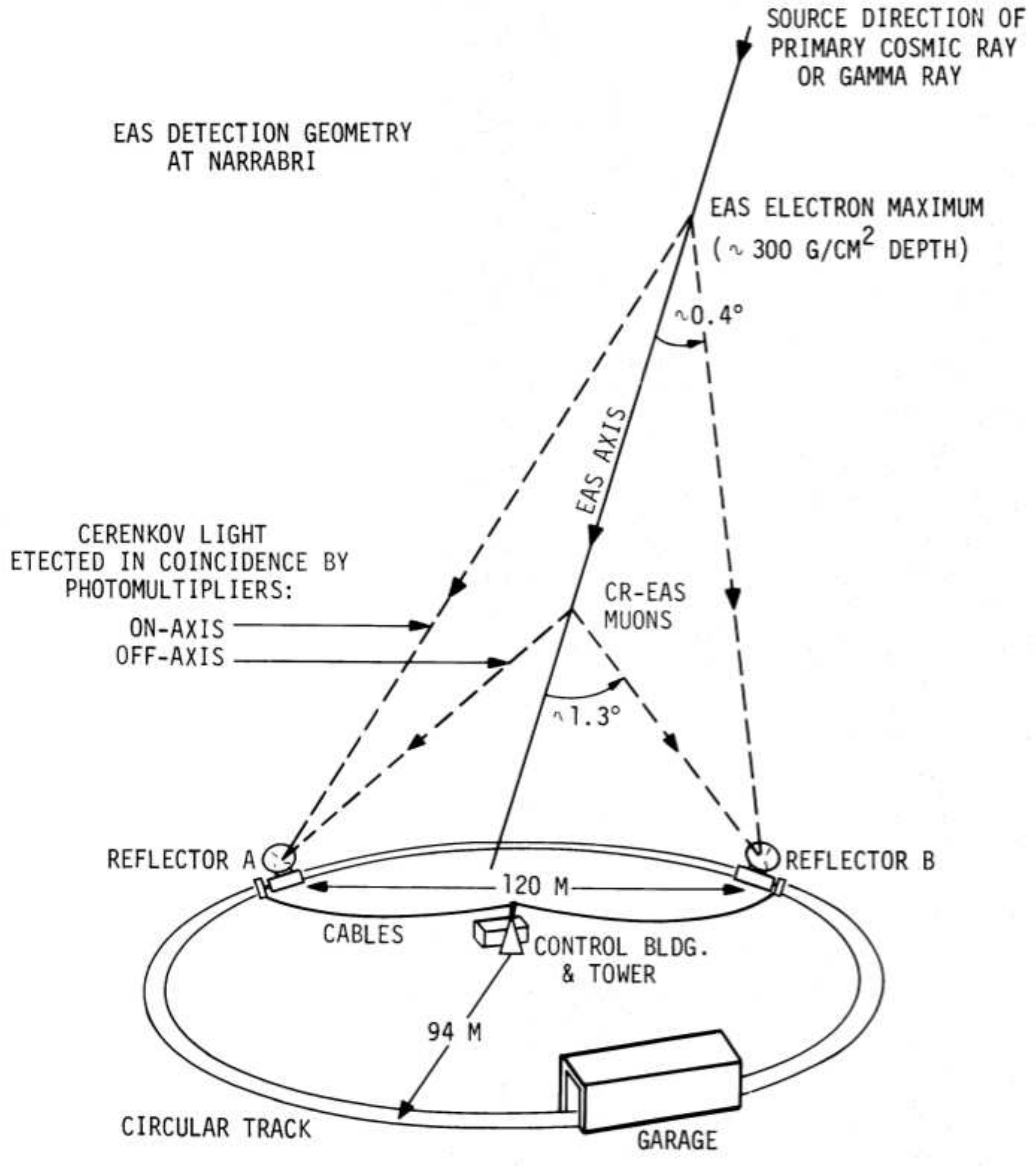}
\caption{Another stereo telescope configuration by J. Grindlay and coworkers \cite{Grindlay1975} used between 1972 and 1976 at Mount Hopkins and Narrabri. The two single-PMT equipped telescopes run on rails and can be operated at different distances.
\label{Fig7.8.2}
}
\end{figure}

Soon after the first Cherenkov telescopes were used to look for the sources of cosmic rays one tried to improve the sensitivity by means of the stereo technique, i.e. by viewing the showers from spaced telescopes. Chudakov and coworkers at the Catsiveli site in Crimea were the first to attempt designing a multi-telescope stereo system \cite{Chudakov1963}, which also facilitated simple stereo observations. They used 12 detectors each comprised of a large mirror and only one photomultiplier per telescope. Units of three detectors each were installed on a simple mount, which could be separated on rails. Fig. \ref{Fig7.8.1} shows a photo of their arrangement. With normally only 20\,m separation and a single large-diameter photomultiplier/telescope, the stereo quality was rather poor and more a coincidence measurement for reducing accidental triggers. Some time later, J. Grindlay \cite{Grindlay1975} tried another stereo approach (Fig.~\ref{Fig7.8.2}) with only two similar telescopes mounted on a circular rail system allowing a separation of up to 180\,m. Later, some other similar attempts were made, but again none of them, however, led to a high-significance source detection. The lack of any discovery can be traced back to the missing $\gamma$/hadron separation power. After the breakthrough discovery of the Whipple collaboration using a pixelized camera, part of the extended Whipple collaboration converted an 11-m solar telescope, originally located in New Mexico, into a 37-pixel camera Cherenkov telescope, dubbed Granite, and genuine stereo observations were pursued. Unfortunately, the sensitivity of the stereo system was worse than the Whipple telescope alone. Reasons were mirrors of poorer optical quality and a tendency of icing due to radiation cooling caused by low heat conductivity of the foam backing of the mirror. 
Additionally, the  spacing between the two telescopes of $\approx 120$\,m did not yield enough events simultaneously detected in both telescopes and thus was far from optimal.
The first successfully operating stereo system with significantly improved sensitivity was build by the HEGRA collaboration.

\subsection{The first high-sensitivity stereo imaging Cherenkov telescope system as part of the HEGRA observatory}
After the publication of a 4.4-$\sigma$ excess from the direction of Cygnus X-3, the Kiel physicists in 1989 started to build an improved scintillation counter array, the HEGRA experiment, at the Roque de los Muchachos (2200 m asl) observatory on the Canary island of La Palma. Four groups coming from particle physics joined and helped to increase the number of counters by a factor eight, spread over an area of 200$\times$200\,m, and to add a few muon tracking detectors (concrete absorbers interspaced with proportional chambers). 
It was hoped that a much larger array, complemented by muon detectors for $\gamma$/hadron separation, at 2200\,m altitude would very much increase the sensitivity, confirm the Cygnus X-3 results with much higher significance, and allow more sources to be observed, but these hopes were not rewarded by a positive result. 

Already in the early 1990s, the Kiel institute leader, the late Otto Claus Allkofer, had discussed with Felix Aharonian from the Armenian group in Yerevan about the possibility of adding five Cherenkov telescopes because of the excellent optical conditions at the La Palma site. The Armenian group had already built a small imaging Cherenkov telescope on Mount Aragats and had plans for a stereo system. Heinz V\"olk from the Max-Planck-Institut f\"ur Kernphysik Heidelberg later took up this idea and joined the HEGRA collaboration. Also, quite a few of the leading physicists of the Yerevan group joined  the HEGRA experiment as, due to the dissolution of the USSR, the research conditions in Armenia were very limited. This group brought the know-how of how to build imaging Cherenkov telescopes.

\begin{figure}[h]
\centering
\includegraphics[width=.5\linewidth]{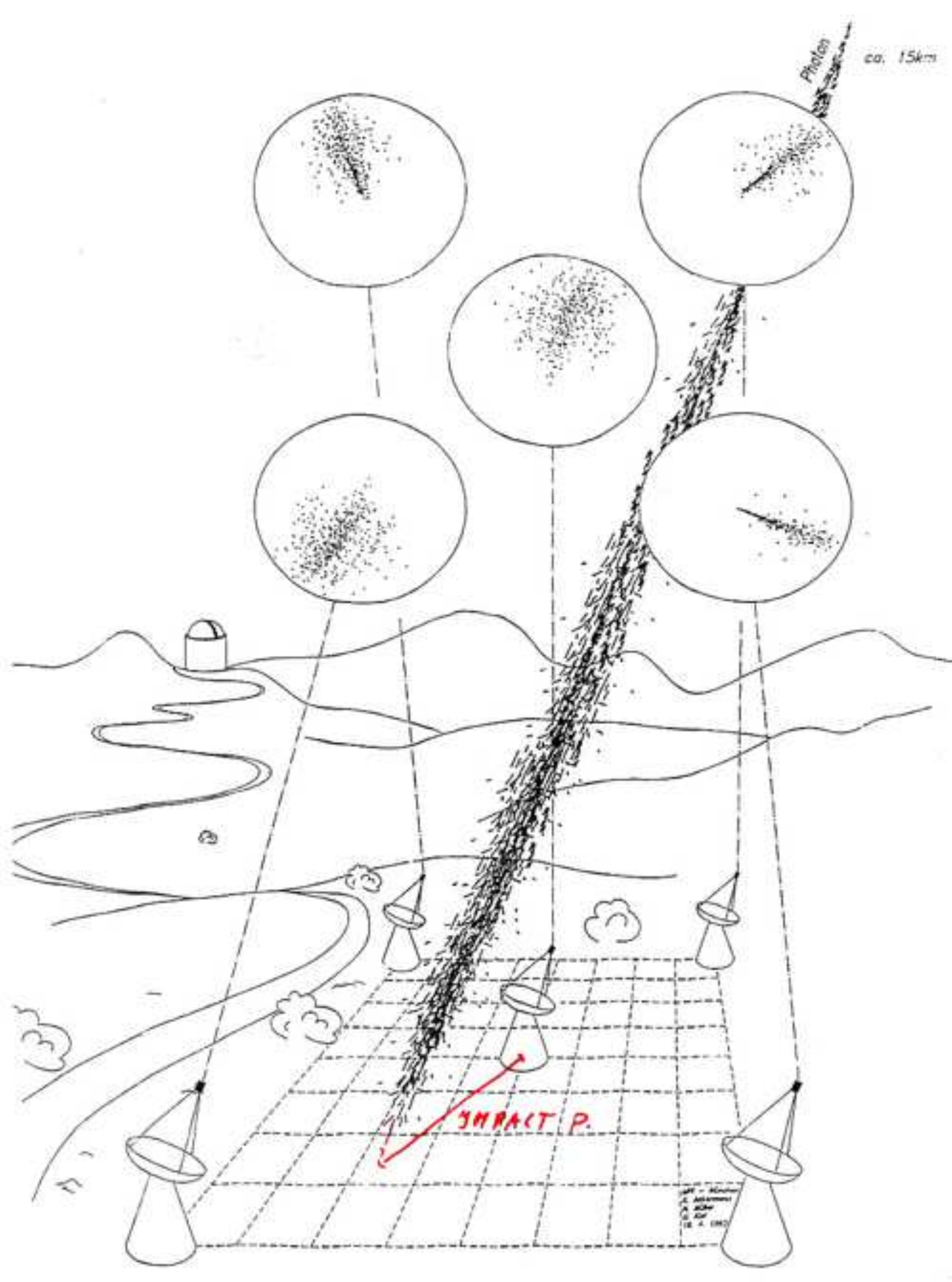}
\caption{A sketch of the principle of stereo observations. In most cases one or two telescopes record a shower image of excellent quality from different angles. Altitude not to scale.\label{Fig7.9.1}}
\end{figure}

Eventually, a prototype Cherenkov telescope and 5 telescopes, operating in a stereo system, were built. The system was very successful with an increase in sensitivity of about a factor 10 compared to a single telescope of the same size. The reasons were manifold and are shown in a sketch in Fig. \ref{Fig7.9.1}. With a stereo system, showers are observed from different directions. This can improve the $\gamma$/hadron separation by means of viewing the shower in part under optimal condition and by suppressing the so-called head-tail ambiguity of single telescopes. In single-telescope pictures recorded by a classical gated analog-to-digital converter (ADC) readout, there is an ambiguity about the shower direction pointing either towards or away from the potential source location. In stereo systems one can cut the background by a factor two by solving this ambiguity. 
Stereo observations also provide a much better shower energy determination and a better angular resolution allowing the study of extended sources. The HEGRA stereo system was the first one that used regularly a readout with flash ADCs, now common in all Cherenkov stereo systems. 

In the last decade of the last century a few other stereo systems were built (Table \ref{Tab6.1}), but none reached the sensitivity of the HEGRA experiment. Nowadays stereo telescope systems are the main tool in VHE $\gamma$-astronomy.

\subsection{Alternative Cherenkov detectors that did not reach high sensitivity}
The success of the Whipple telescope triggered a number of similar but smaller imaging Cherenkov telescopes (Table \ref{Tab6.1}) and also some other designs of Cherenkov counters, which will be briefly mentioned here. One design direction makes use of the ozone layer high up in the atmosphere, acting as a blocker of UV-light. The detectors use photosensors that are sensitive to photons only below 290 nm. The disadvantage of this principle is, that the Cherenkov light from the upper part of the air showers is lost due to the ozone absorption. As an example we discuss briefly the CLUE detector. Another alternative to the standard imaging Cherenkov telescopes is the class of the so-called wave front samplers. These detectors consist of an extended matrix of photosensors with some light collectors that measure the Cherenkov light front with high precision at many positions. This configuration is particularly suited for point-like sources but nearly always lacks a good $\gamma$/hadron separation. Detectors of this type often make use of existing very large mirror arrays of solar energy generators or use small, simple light collectors around large diameter PMTs in order to achieve a large angular acceptance.

\paragraph{CLUE}
Already the Whipple collaboration modified the camera of their telescope by using so-called solar blind PMTs to operate the telescope during moonlight \cite{artemis}. These first tests were not very successful and were soon abandoned because the solar-blind PMTs had a rather low quantum efficiency ($\approx 10\%$ between 220 and 300 nm) and furthermore the camera had to be changed every month because the solar blind PMTs were only superior to normal PMTs during moonlight. The concept was then taken up and modified by the CLUE (Cherenkov Light Ultraviolet Experiment) collaboration that installed at the HEGRA site eight small telescopes with 180\,cm mirror diameter and a novel light sensor in the focal plane \cite{Bastieri1999}. Besides VHE gamma-astronomy, the other aim of CLUE was to measure the ratio of matter/antimatter in the cosmic radiation at energies around the TeV by using the moon as an absorber and analyzing the deflection of the halo particles. This measurement can help to distinguish the different evolution models of the Universe. To prevent any background light from the moon, the light sensors had to operate in a spectral domain below 290\,nm where the ozone layer is impermeable for the UV light.

The light sensor was a multi-wire UV light sensitive proportional chamber (MWPC) with the anode backplane subdivided in a 24$\times$24 matrix of square electrodes read out by capacitive coupling. The front window and the front anode being formed by wires were transparent to UV photons. The CLUE detector was filled with TMAE (Tetrakis-Methyl-Amino-Ethylene, C$_4$H$_{11}$NO). This substance has a good quantum efficiency of converting radiation in the range 180-240 nm into photoelectrons. The ionization potential is around 4.5\,eV. TMAE was admixed with (Ethane-Isobutane, 3:1), which is a good operating gas for MWPCs. The detector was operated for only two years (1997-1999) and could observe the Crab nebula, Mkn 421 and Mkn 501, as well as the moon shadow. The sensitivity was too limited to detect an antiproton flux.

\paragraph{THEMISTOCLE}
The first ``wave front'' Cherenkov light detector (using some mounts of a solar array) was THEMISTOCLE at the partly decommissioned Themis solar power site in the French Pyr\'en\'ees \cite{Baillon1992}. The 18 telescopes of the THEMISTOCLE setup used old mirror stands, which supported parabolic mirrors of 80\,cm in diameter focusing the light onto a single PMT. The mirror focal length was 40.4\,cm and the image size for a point source of the order of 1\,mm. The light was detected by a single photomultiplier normally operated with a 1.6-cm diameter diaphragm covering the 5 cm diameter cathode. Relative timing calibration was provided using a pulsed nitrogen laser whose light pulse had a 2\,ns risetime. The arrangement worked in the \textit{On-Off}-mode and sources were detected by an excess signal in the \textit{On} mode. No $\gamma$/hadron separation mode was possible. THEMISTOCLE just detected the Crab nebula.

\begin{figure}[h]
\centering
\includegraphics[width=.4\linewidth]{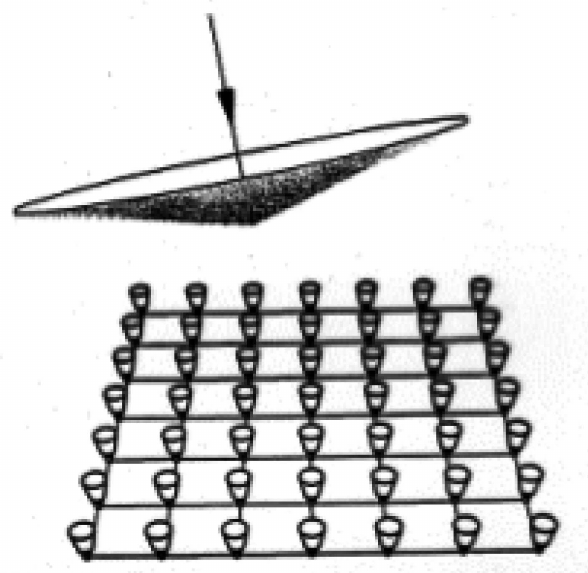}
\caption{The principle of the wide-angle Cherenkov detector AIROBICC: An array of open PMTs look into the night sky. Their light-collection efficiency is enhanced by 40-cm diameter Winston cones \cite{Winston}, restricting the FOV per pixel to 1 sterad. The array is able to detect the wave front of the Cherenkov light: A Cherenkov light disk from an air shower, as indicated, typically is 250 meters in diameter and very sharp in time.
\label{Fig7.10.3.1}
}
\end{figure}

\paragraph{AIROBICC}
Another wave-front detector was the wide angle detector AIROBICC, installed in 1994 inside the HEGRA scintillator array \cite{Karle1995}. Basically, it was an array of 169 large diameter PMTs, viewing the night sky directly (Fig. \ref{Fig7.10.3.1}). In order to enhance light collection the PMTs were inserted into parabolic light collectors of 40\,cm entrance diameter, limiting the FOV to 1 steradian.  The detector did not track a source but recorded any signal within 1 steradian of the sky. The angle was determined by time of flight measurements and the shower energy from the PMT amplitude signals. As the detectors integrated the night sky light background over 1 steradian angle, the threshold was rather high: 14\,TeV for gamma-ray showers and 22\,TeV for hadronic showers. Again, the performance was rather limited due to a poor $\gamma$/hadron separation power (by comparing the light signal with the signal of the scintillator array). Only the signal from the Crab nebula could just be detected with 4.5 $\sigma$ at a gamma-ray energy of 70 TeV. As in scintillator arrays, one of the fundamental problems was the determination of the exact orientation of the detector plane for seeing the showers. Therefore a rather large systematic angular resolution had to be used.

Due to a forest fire, the array was largely destroyed in 2000 and the remaining stations subsequently decommissioned. A similar detector, dubbed BLANCA \cite{Cassidi1997}, was later installed in the CASA-MIA detector \cite{Cassidi1997}, but used mainly for CR studies. Another similar detector, dubbed TUNKA \cite{Budnev2005} near lake Baikal, is currently in operation. Like BLANCA, TUNKA also is more suited for the study of the chemical composition of CRs because of the high threshold.

\subsection{Heliostats as Cherenkov light detectors: CELESTE, STACEE, GRAAL, Solar II}
In the 1990s, solar heliostat fields attracted quite a few physicists as possible detectors. The basic idea was converting theses installations into gamma-ray detectors during night time. The use of many large-area heliostats collecting Cherenkov light and focusing it onto a detector mounted on a tower had quite some appeal to save costs and achieve a low energy threshold. Four projects reached a level of operation, namely the CELESTE detector using the Themis heliostats in the French Pyr\'en\'ees, the STACEE detector using a prototype solar power station near Albuquerque, New Mexico, the Solar II heliostats in Barstow, California, and the GRAAL detector using the heliostats of the Plataforma Solar in Almer\'\i a, Spain.

\begin{figure}[h]
\centering
\includegraphics[width=.35\linewidth]{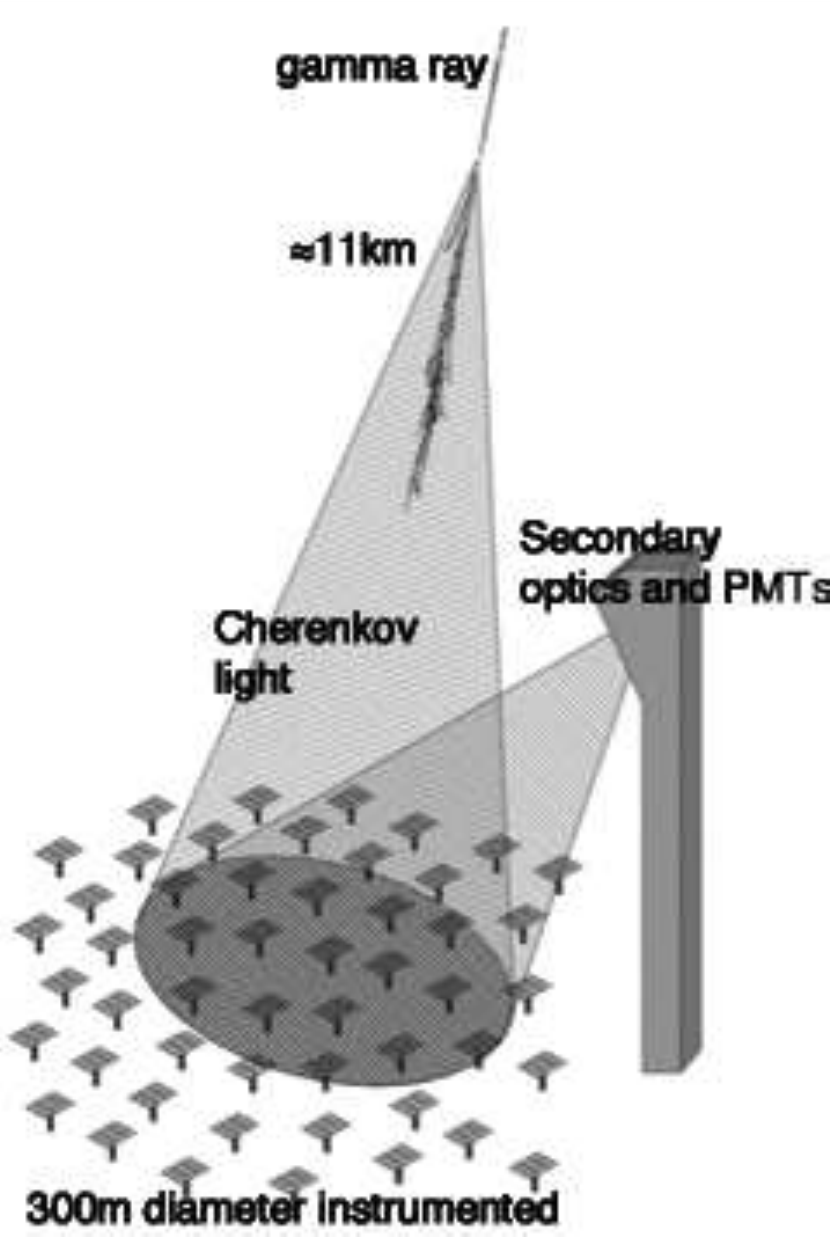}
\caption{The CELESTE detector in the French Pyr\'en\'ees. The heliostats were used to focus the Cherenkov light from an air shower onto the tower where previously the power generator was located. The power generator was replaced by the secondary optics and a matrix of photomultipliers. Although CELESTE had only a rather limited $\gamma$/hadron separation power, the Crab nebula, Mkn 421, Mkn 501, and 1ES 1426+428 could be detected \cite{Smith2006}.\label{Fig7.10.4.1}}
\end{figure}

The CELESTE experiment \cite{Pare2002} used the heliostats of a decommissioned solar farm Themis (42.50$^\circ$ N, 1.97$^\circ$ E, 1650 m asl) in the French Pyr\'en\'ees to detect gamma-ray induced air showers. CELESTE was used from 1997 to 2004. A large (2000\,m$^2$) mirror surface area from 40 independent heliostats projected the Cherenkov light onto a secondary optic system followed by an array of PMTs. The secondary optics and the PMTs were mounted on the former power generator location on a tower. A trigger system using analog summing techniques and signal digitization with 1-GHz flash ADCs made possible the detection of cosmic gamma rays down to 30 GeV. The 40 used stations were spread over an area of 200$\times$300 m. Fig. \ref{Fig7.10.4.1} shows a sketch of the arrangement. 

CELESTE observed VHE gamma rays from the Crab nebula, Mkn 421, Mkn 501, and 1ES 1426+428 \cite{Smith2006}. Again, modest $\gamma$/hadron separation limited the sensitivity.

Later, some nearly identical air Cherenkov telescopes based on preexisting solar power plants had been set up using very similar arrangements. The Solar Tower Atmospheric Che\-ren\-kov Effect Experiment (STACEE) was implemented at a solar power pilot plant near Albuquerque and went online in 2001 \cite{Gingrich2005}.

Another very similar experiment was the GRAAL detector \cite{Plaga2000}, installed at the Plataforma Solar de Almer\'\i a, Southern Spain. The experiment started in 1999 and ran until 2003. Although light was collected by means of 63 heliostats of 39.6 m$^2$ each, only the Crab nebula could be detected with high significance. The main problem was, as in all detectors using solar heliostats, the modest $\gamma$/h separation power and the variation of the brightness of the night sky and the uncontrollable reflectivity of the mirrors due to dew formation, resulting in fluctuations in the counting rates in the \textit{On-Off} method often resulting in fake signals.

The fourth try of the concept was based on the solar heliostat field Solar II near Barstow, California \cite{Tumer1999}. The experiment, dubbed Keck Solar Two , eventually used 160 heliostats spread over a field of in total 2000 heliostats and operated only for two years (2002-2004).

All detectors based on solar heliostat plants were able to detect the strongest sources like the Crab nebula and the flaring Mkn 421 and Mkn 501 but did not reach the sensitivity of the third-generation imaging telescopes and therefore the method was abandoned around 2005. 

\section{Progress in the first decade of the new millennium}

\begin{figure}[h]
\centering
\includegraphics[width=.95\linewidth]{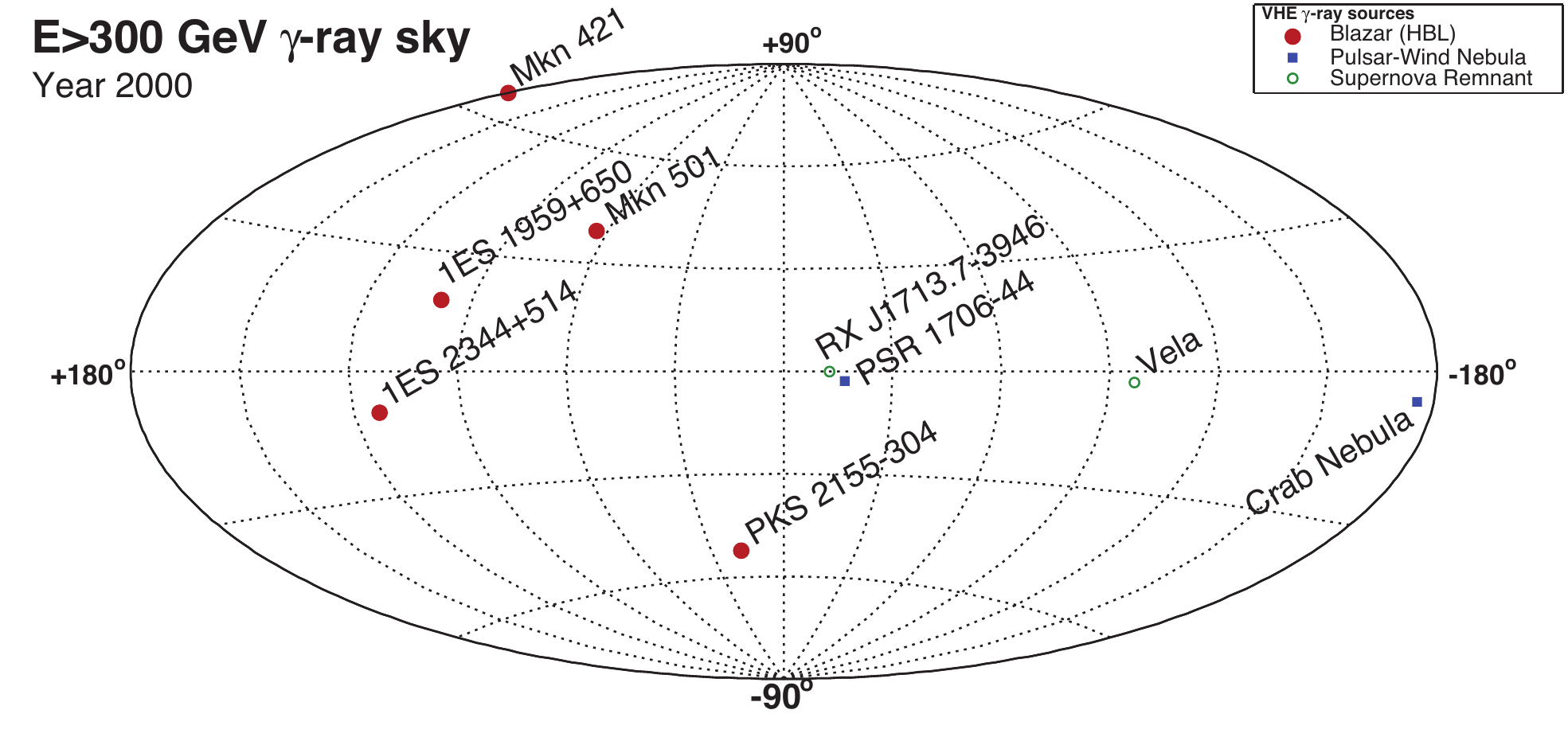}
\caption{
The VHE ($E>300$ GeV) sky map at the year 2000.
\label{Fig8.1.1}
}
\end{figure}

The progress in discovering new VHE gamma-ray emitting sources after the discovery of the Crab nebula was initially rather slow. Figure \ref{Fig8.1.1} shows the VHE sky map in the year 2000.  Only 8 more sources were discovered, all of them by �imaging� Cherenkov telescopes, which became the ``workhorse'' for the searches. These second-generation Cherenkov telescopes were simply not sensitive enough to observe sources that emit VHE gamma rays below 10\% of the Crab nebula flux. Nevertheless, confidence in the observation techniques and analysis methods developed. For nearly every group observing on the northern half of the Earth the Crab nebula was the test bench. The number of extragalactic sources found was equal to that of galactic ones detected. All extragalactic sources were blazars, while two galactic sources were pulsar wind nebulae and two supernova remnants (SNR). The community followed a suggestion of Trevor Weekes that observed sources were accepted as discoveries only if their significance exceeded 5 $\sigma$ and all sources on the sky map were at least confirmed by one other experiment.

\subsection{The large third-generation imaging Cherenkov telescopes}
As in any emerging area of scientific research, the financing of large detectors is the issue of hard negotiations. On the whole, the majority of the astrophysicists and astronomers were still not convinced that the new field would really contribute to the fundamental understanding of the relativistic Universe and the meager results of the past times did not justify the diversion of funding from other areas. Nevertheless, the results from mainly the last decade of the last century made it obvious that new, better telescopes would lead to a breakthrough in the field. Also, the stereo-observation technique was generally accepted as the approach that would reach sensitivities around 1\% of the Crab nebula flux within 50 h observation time for achieving a 5-$\sigma$ excess signal. Eventually, four large projects materialized: Cangaroo III, H.E.S.S., MAGIC and VERITAS. 

It was particularly Germany that was willing to push the development and also to finance it. 
Heinz V\"olk convinced the BMBF (\textit{Bundesministerium f\"ur Bildung und Forschung}, the German federal ministry of education and research) to support the construction of the first H.E.S.S. telescope. The Max Planck Society then provided major funds for the Heidelberg MPI group of Werner Hofmann and Heinz V\"olk for the construction of the four H.E.S.S. telescopes in Namibia, while some time later the BMBF provided major funds for the construction of the 17-m diameter MAGIC telescope on La Palma. The BMBF that time played a very supportive role for the revived activity in high energy astroparticle physics. Regrettably, the competing plans for a stereoscopic system of seven telescopes, proposed by the VERITAS collaboration, were very much delayed by both the lack of funding and problems with the site approval. While these three projects were all pushed by large international collaborations, the Japanese and Australian colleagues tried a somewhat more modest project, dubbed Cangaroo III, and located in Australia. 

The plans for these third-generation improved telescopes started to evolve around the year 1994 onwards. The construction of the first Cangaroo III telescope started already in 1997, the main activities of H.E.S.S. basically around 2000, MAGIC in 2002, and VERITAS in 2003. Table \ref{Tab8.2.1} lists some essential information about the third-generation observatories.

\begin{table}
\centering
\caption{Table of the third-generation observatories with large mirror telescopes. The overview lists location and altitude of the observatories, the diameter and number (``\#'') of the individual telescopes, and the start dates of operations.\label{Tab8.2.1}}

\begin{tabular}{llllll}
\hline
Name  & Location & Diameter &	\# &	Altitude	& Start \\
\hline
\hline
Cangaroo III & 	Australia	&10 m &	4	&\phantom{0}160 m asl&	1999 (4 telescopes in 2003)\\
H.E.S.S.&	Namibia&	12 m &	4&	1800 m asl 	&2002 (4 telescopes in 2003)\\
MAGIC&	La Palma&	17 m &	2&	2225 m asl&	2004 (2 telescopes in 2009)\\
VERITAS&	Arizona&	12 m &	4	&1390 m asl	&2006 (2 telescopes in 2006, 4 in 2008)\\
\hline
\end{tabular}
\end{table}

\subsection{Cangaroo III}
\begin{figure}[h]
\centering
\includegraphics[width=.95\linewidth]{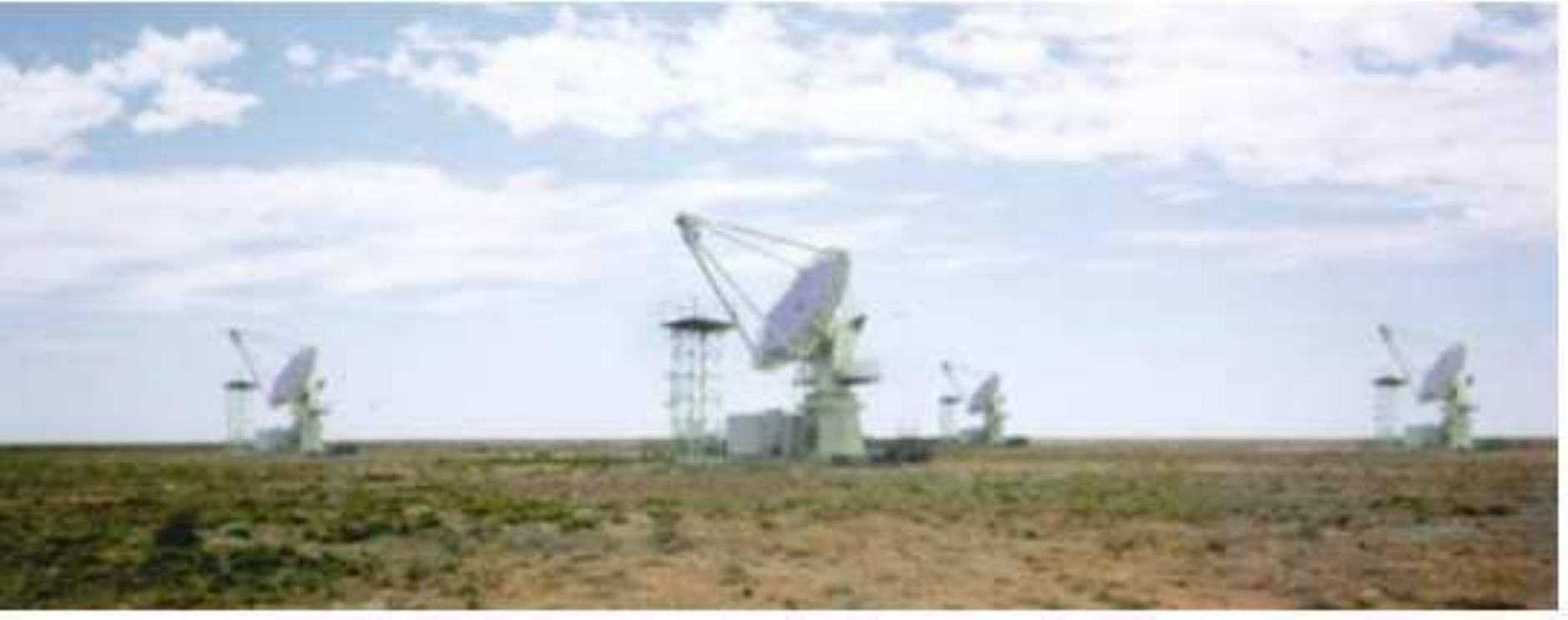}
\caption{
Photograph of the four Cangaroo III Cherenkov telescopes in the Australian outback. The telescopes were arranged in a diamond pattern. Each telescope had a nearby access tower for servicing the camera. The mirrors had a diameter of 10 m.
\label{Fig8.3.1}
}
\end{figure}

Cangaroo III (Collaboration of Australia and Nippon for Gamma-Ray Observation in the Outback) was built by a Japanese-Australian collaboration at low altitude near Woomera in Australia, at 31$^\circ$06' S, 136$^\circ$47' E and 160 m asl \cite{Enomoto2006} (Fig. \ref{Fig8.3.1}). The telescopes had 57 m$^2$ mirror area each, i.e., about half that of one of the H.E.S.S. and VERITAS telescopes. The Cangaroo III telescopes used plastic mirrors with rather modest focusing quality and significant aging (both for reflectivity and focusing), while the PMTs with a square cathode had a relatively low photon detection efficiency (PDE). The telescopes were located at a low altitude of $\approx$160\,m above sea level, where usually significant Mie scattering from fine dust leads to significant light losses. In summary, the Cangaroo III telescopes were not very competitive with H.E.S.S., MAGIC and VERITAS. Eventually, the activities in Australia were stopped in 2011. 

\subsection{H.E.S.S.}
\begin{figure}[h]
\centering
\includegraphics[width=.95\linewidth]{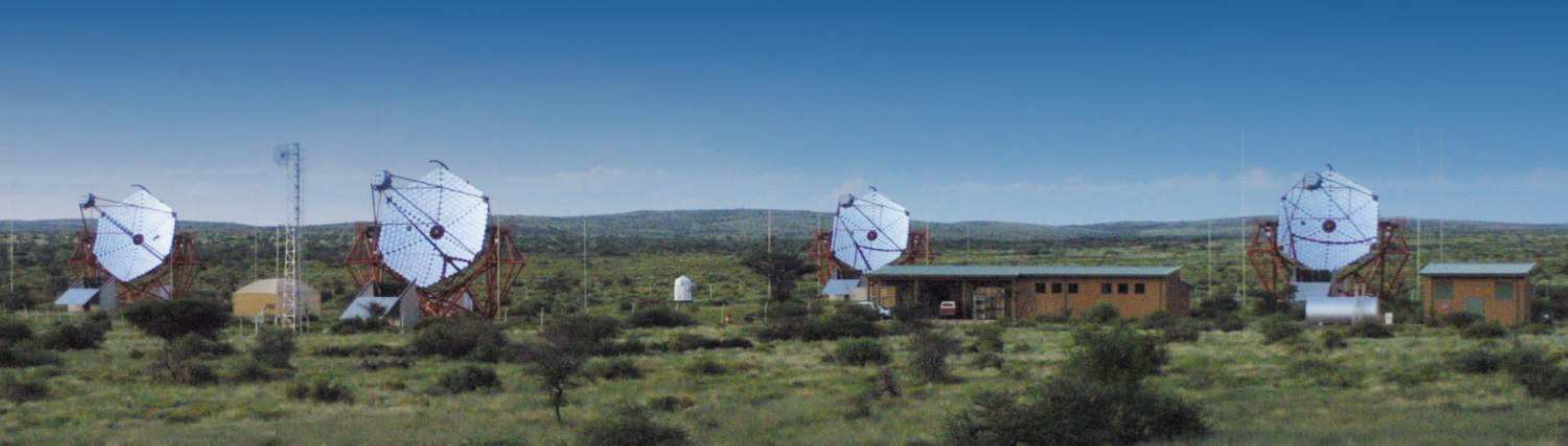}
\caption{Photograph of the four H.E.S.S. 12-m diameter Cherenkov telescopes in the Khomas Highland of Namibia. The future 28-m diameter H.E.S.S. II telescope will be located in the center of the four telescopes.\label{Fig8.4.1}}
\end{figure}

\begin{figure}[h]
\centering
\includegraphics[width=.65\linewidth]{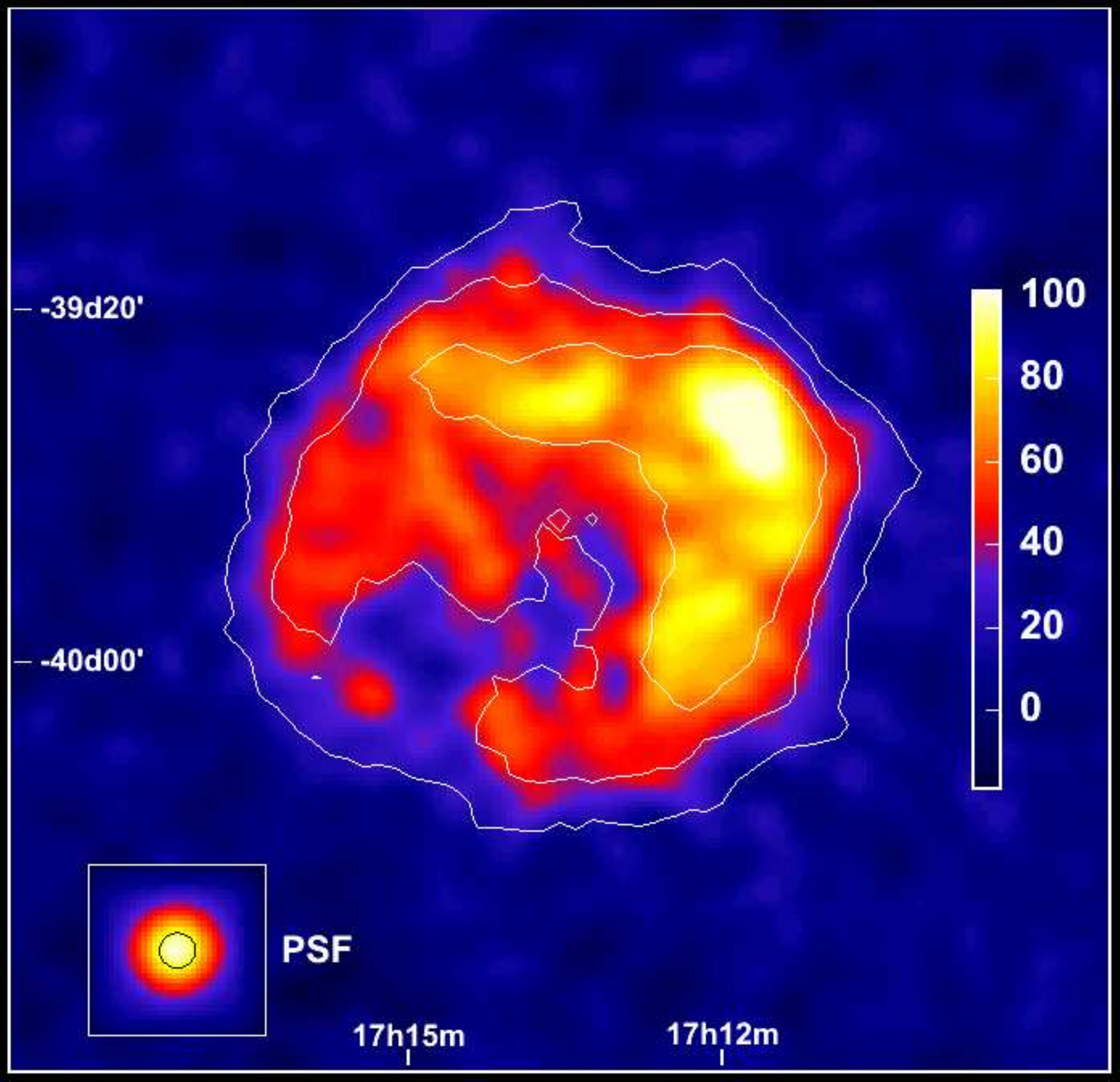}
\caption{Scan of the extended source RX J1713.7-3946, overlaid with a radio scan (black lines) from the satellite-borne $\gamma$-detector ASCA in the 1-3 keV energy range. The insert in the left lower corner shows the resolution of a point-like source \cite{Aharonian2006}.\label{Fig8.4.2}}
\end{figure}

H.E.S.S. (High Energy Stereoscopic System) was built by a large international collaboration in the years 2000-2003 in Namibia at 23$^\circ$16' S, 16$^\circ$30' E, at 1800 m asl \cite{Hofmann2001}. Figure \ref{Fig8.4.1} shows a photograph of the observatory. H.E.S.S. comprises four 12-m diameter imaging Cherenkov telescopes with a 110-m$^2$ mirror and a multi-pixel camera of 960 PMTs each. The observatory is suited for the study of gamma-ray sources in the energy range between 100 GeV and 100 TeV. The stereoscopic system has a sensitivity of 0.7\% of the Crab nebula flux within 25 hours of observation time when pointing to zenith. Like Cangaroo III, H.E.S.S. is located in the Southern hemisphere and is particularly suited for the observation of sources in the central region of the galactic plane. H.E.S.S. is currently the most successful observatory as it has discovered more than half of all known VHE sources. Due to their large diameter cameras of 5$^\circ$ FOV, H.E.S.S. has studied quite a number of extended sources. For example, a scan of the supernova remnant RX J1713.7-3946 in the Galactic plane (discovered in X-rays by \textit{ROSAT} \cite{Pfeffermann1996}) highlights the detection power for extended sources and is shown in Fig. \ref{Fig8.4.2} \cite{Aharonian2006}. In 2012/13, H.E.S.S. will be extended by a central fifth telescope with a 28-m diameter reflector and an energy threshold of 30-40 GeV.

\subsection{MAGIC}
\begin{figure}[h]
\centering
\includegraphics[width=.95\linewidth]{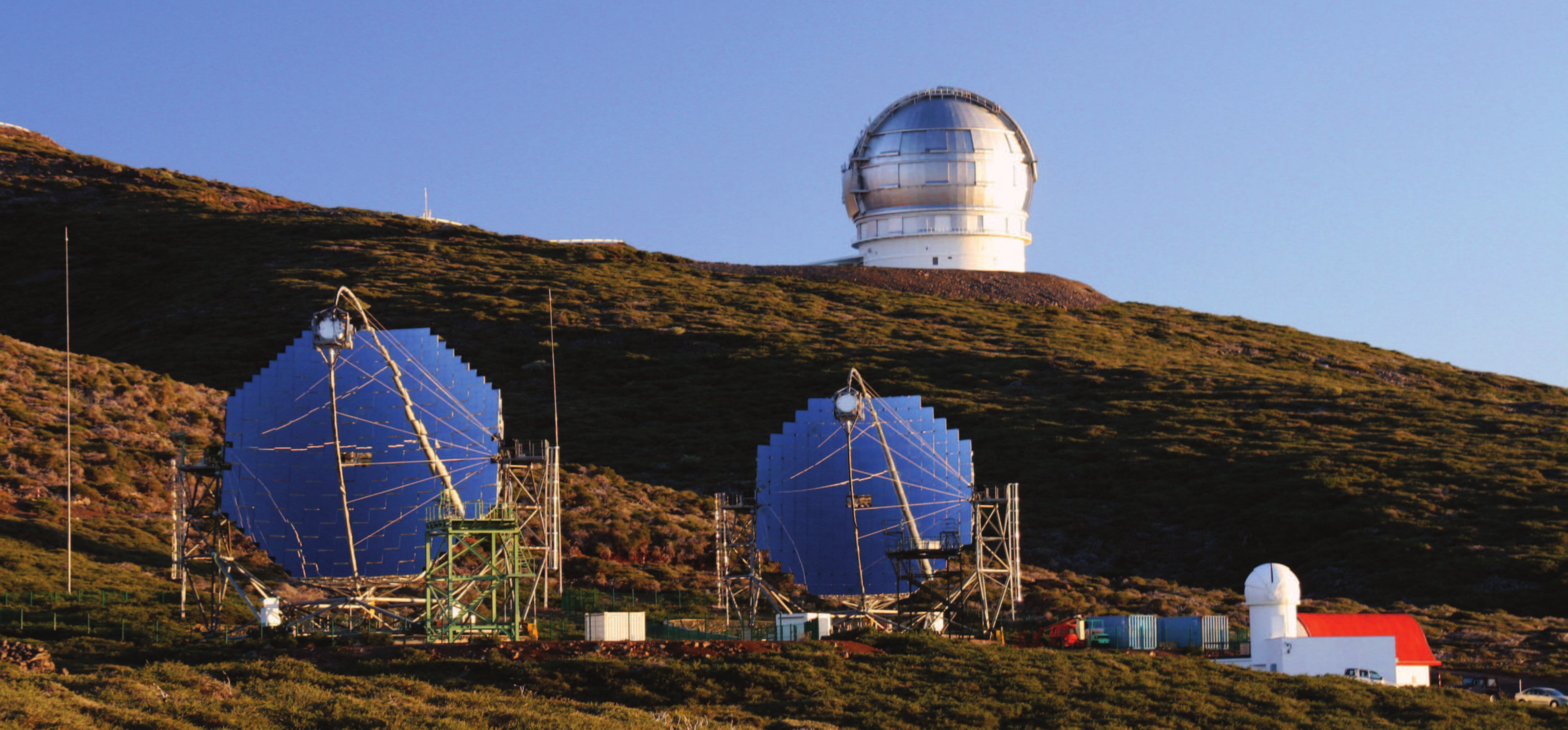}
\caption{Photograph of the two 17-m diameter MAGIC telescopes on La Palma. The two MAGIC telescopes are located in the foreground,  85\,m from each other. The red-roofed house is the control house. In the background the 10-m optical Gran Telescopio Canarias as well as the 2.5-m Nordic Optical Telescope (upper left) are visible.\label{Fig8.5.1}}
\end{figure}

The MAGIC collaboration pursued another path in the development. They designed an ultra-large Cherenkov telescope with a 17-m diameter mirror \cite{Baixeras2003}. Fig. \ref{Fig8.5.1} shows a photograph of the MAGIC telescope on La Palma (28.8$^\circ$ N, 17.8$^\circ$ W, 2225 m above sea level) with second one, which was constructed later. The telescope is based on numerous novel concepts, such as a low-weight carbon-fiber reinforced plastic space frame, supporting the diamond-turned, low-weight, sandwich aluminum mirrors.  To counteract small deformations during tracking, the matrix of small mirror elements, approximating a parabolic mirror profile, was corrected by an active mirror control system. Two curved masts held in position by steel cables to minimize obscuration supported the camera. To minimize the weight of the camera, the PMT signals are transferred to a counting house by optical fibers operated in analog mode. The total moving part of the telescope has a weight of only $\approx 70$ tons and could be repositioned to any point on the sky within 20 seconds in order to observe at least part of gamma-ray bursts (GRB). As it was obvious that a stereo system of such telescopes with so many new features would never been funded in the first round, a second telescope was built only after the new items of the first one proved to work. The first telescope started to take data in 2004, and stereo observations with both telescopes commenced in 2009. The first telescope has a threshold of 60\,GeV and initially a sensitivity of $\approx 1.5\%$ of the Crab nebula flux while the stereo system has a threshold of 50 GeV and a sensitivity of $0.8\%$ of the Crab nebula flux. 

\subsection{VERITAS}
\begin{figure}[h]
\centering
\includegraphics[width=.95\linewidth]{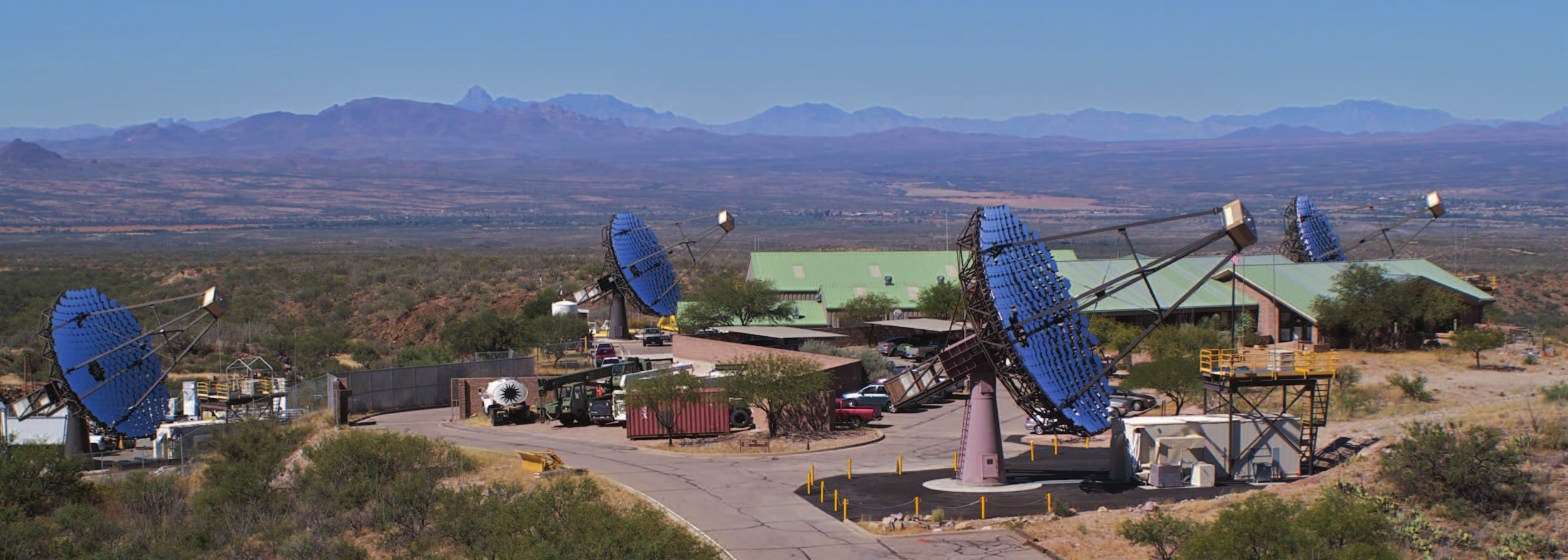}
\caption{Photograph of the four VERITAS 12-m diameter Cherenkov telescopes at the Mount Hopkins base camp in Arizona.\label{Fig8.6.1}}
\end{figure}

The fourth of the third-generation imaging Cherenkov telescopes is the VERITAS telescope complex \cite{Holder2006}. VERITAS stands for Very Energetic Radiation Imaging Telescope Array System (for gamma-ray astronomy). VERITAS comprises four 12-m telescopes and is located in Arizona (31.75$^\circ$ N; 110.95$^\circ$ W, 1268\,m asl). The four telescopes at the base camp of the Mount Hopkins telescope site (Fig. \ref{Fig8.6.1}) are quite similar to the H.E.S.S. telescopes in mirror size, but the cameras have a smaller FOV. As for H.E.S.S., the threshold is $\approx 100$ GeV; the sensitivity is also better than $1\%$ of the Crab nebula flux. The first telescope started operation in late 2005, while the full system saw first light in 2007. 
The project could have been completed a few years earlier, but construction at another site was halted after a Native American tribe filed a federal lawsuit. The VERITAS telescopes have already undergone a major upgrade, in which cameras with new electronics, photomultipliers with increased quantum efficiency, and a new trigger were installed.

\subsection{Milagro}
\begin{figure}[h]
\centering
\includegraphics[width=.6\linewidth]{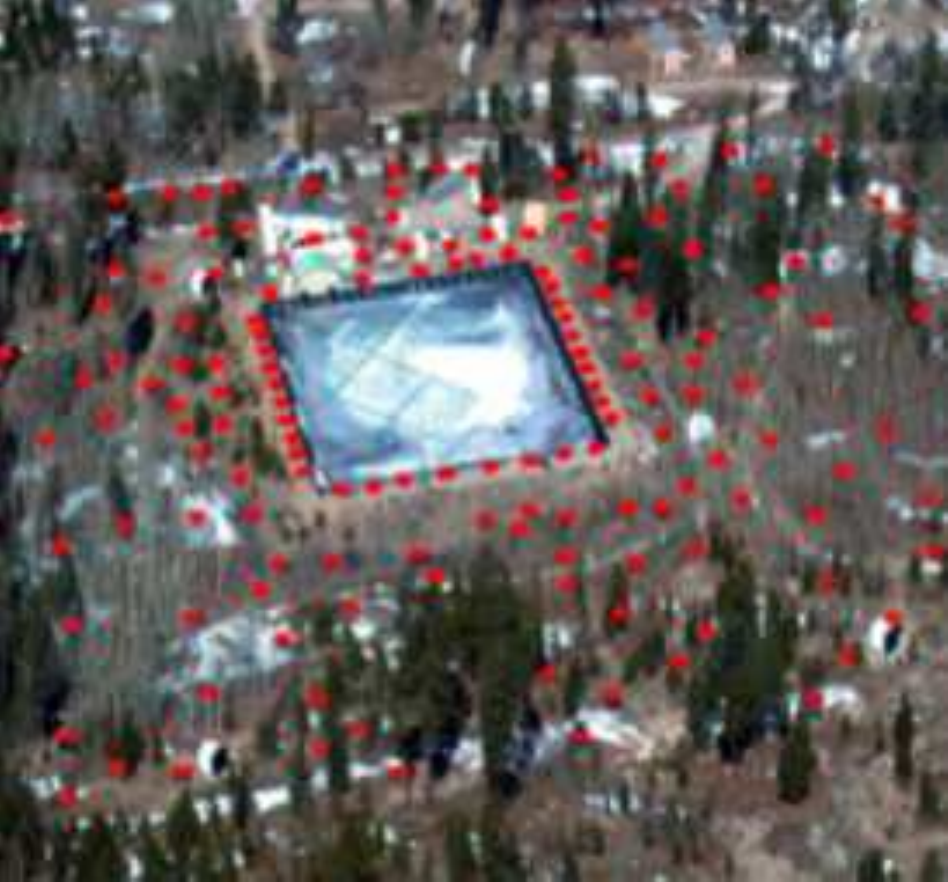}
\caption{Areal view of the Milagro detector (without the black cover). The red dots mark the outrigger water tanks.\label{Fig8.7.1}}
\end{figure}

\begin{figure}[h]
\centering
\includegraphics[width=.65\linewidth]{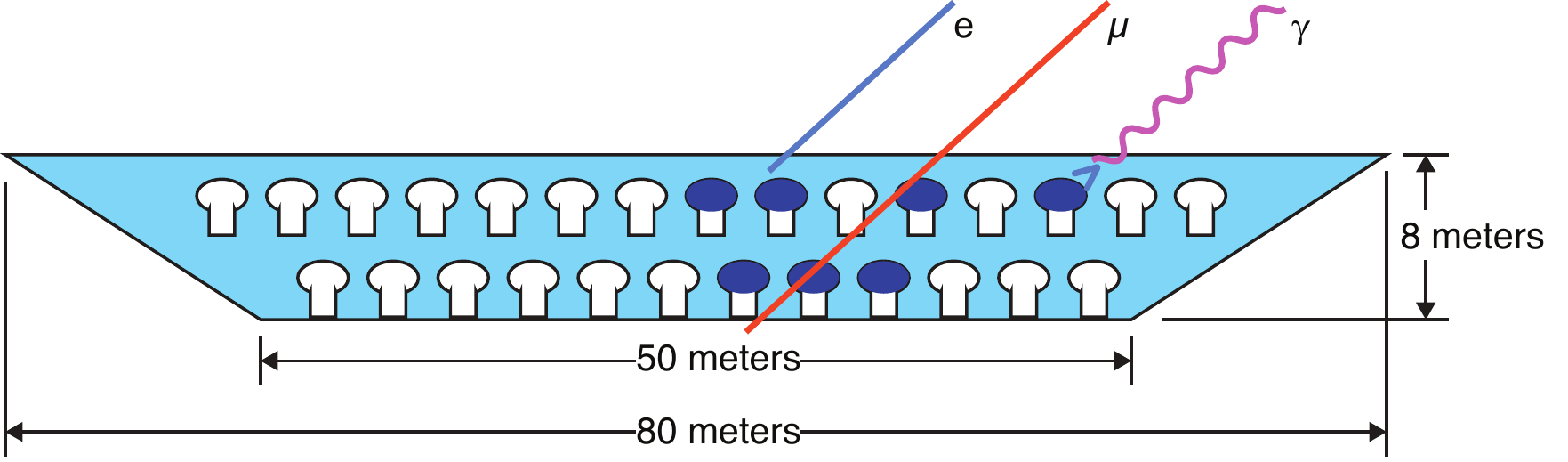}
\caption{Schematic cross section of the water pond of the Milagro detector. Depending on the type of incident particle, PMTs in the upper and lower region of the pond would detect light, as illustrated. The gamma/hadron separation of Milagro was based on these different penetrating powers. 
\label{Fig8.7.2}}
\end{figure}

Milagro was the first really successful tail-catcher detector. Progress in understanding the shower development at its tail and using a detector with 100\% active area around the shower core axis finally produced the first convincing detection of some VHE gamma-ray sources. This detector, dubbed Milagro \cite{Sinnis2009}, made use of a large water pond of $80 \times 50$~m with a depth of 8\,m. The detector was located near Los Alamos at an altitude of 2630\,m above sea level. 175 small water tanks surrounded the water pond (Fig. \ref{Fig8.7.1}) to collect information about the radial shower extension. The charged shower tail particles generated Cherenkov light when passing the water. Electrons from $\gamma$-showers stop normally in the first 2 meters while hadronic showers contain some particles that penetrate deeply into the water pond. The water pond was subdivided into two layers of $2.8\times2.8$ m cells. Each cell was viewed by one large PMT. The top layer of 450 PMTs was under 1.4 meters of water and the bottom layer of 273 PMTs was under 6 m of water, as illustrated in Fig. \ref{Fig8.7.2}. The PMTs in the top water layer and the outrigger array were used to reconstruct the direction of the primary gamma rays (or cosmic rays) to an accuracy of $\approx 0.5$ degrees by means of time of flight measurements. The bottom water layer was used to discriminate against the background cosmic radiation. A black foil sealed the pond against external light.

Milagro had some considerable $\gamma$/hadron separation power. Air showers induced by hadrons contain a penetrating component (muons and hadrons that penetrate deeply into the reservoir). This component resulted in a compact bright region in the bottom layer of PMTs. A cut based on the distribution of light in the bottom layer removed 92\% of the background cosmic rays while retaining 50\% of the gamma-ray events. The detector was suited for the observation of showers above 2\,TeV (from showers coming close from the zenith) and had an uptime of 24\,h. At 45$^\circ$ zenith angle the threshold was 20\,TeV. The collaboration operated the detector from 2002 to 2006.

\begin{figure}[h]
\centering
\includegraphics[width=.95\linewidth]{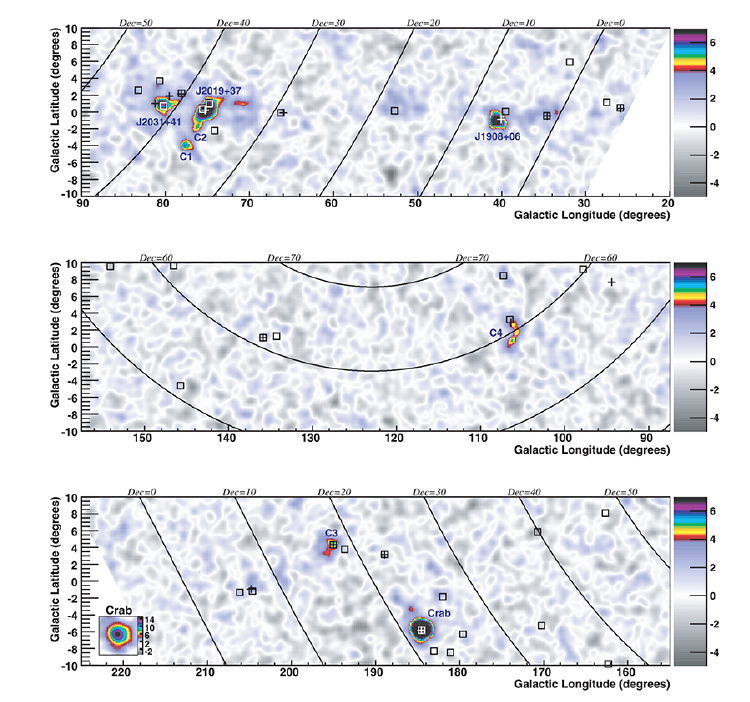}
\caption{Milagro galactic plane survey with 3 newly discovered sources and 5 locations of interest in the Galactic plane. 5 out of 8 of these TeV sources have GeV counterparts. The chance coincidence probability is $6\times10^{-6}$ \cite{Abdo2007}.\label{Fig8.7.3}}
\end{figure}

Milagro with its rather high threshold was best suited for the search for galactic sources in the outer part of the galactic plane. During a survey of the galactic plane \cite{Abdo2007} three new, in part quite extended sources were discovered and a few already known sources confirmed (Fig. \ref{Fig8.7.3}). Milagro stopped operation in 2007 and the enlarged team is currently preparing a new air-shower array dubbed HAWC, to be installed at 4100 m altitude in Mexico. 
Another successful air-shower array is Tibet AS operated by a Japanese collaboration \cite{Huang2009}. This detector at 4300 m asl comprises a large number of scintillation counters but still has only a fractional sampling of the surface and has therefore a threshold of 3\,TeV. Air-shower detectors have a 24\,h up-time and should in principle be well suited for the detection of gamma-ray bursts, but their currently high threshold has prevented any detection up to now.

\subsection{A bonanza of galactic sources: H.E.S.S. scans the galactic plane}
\begin{figure}[h]
\centering
\includegraphics[width=.8\linewidth]{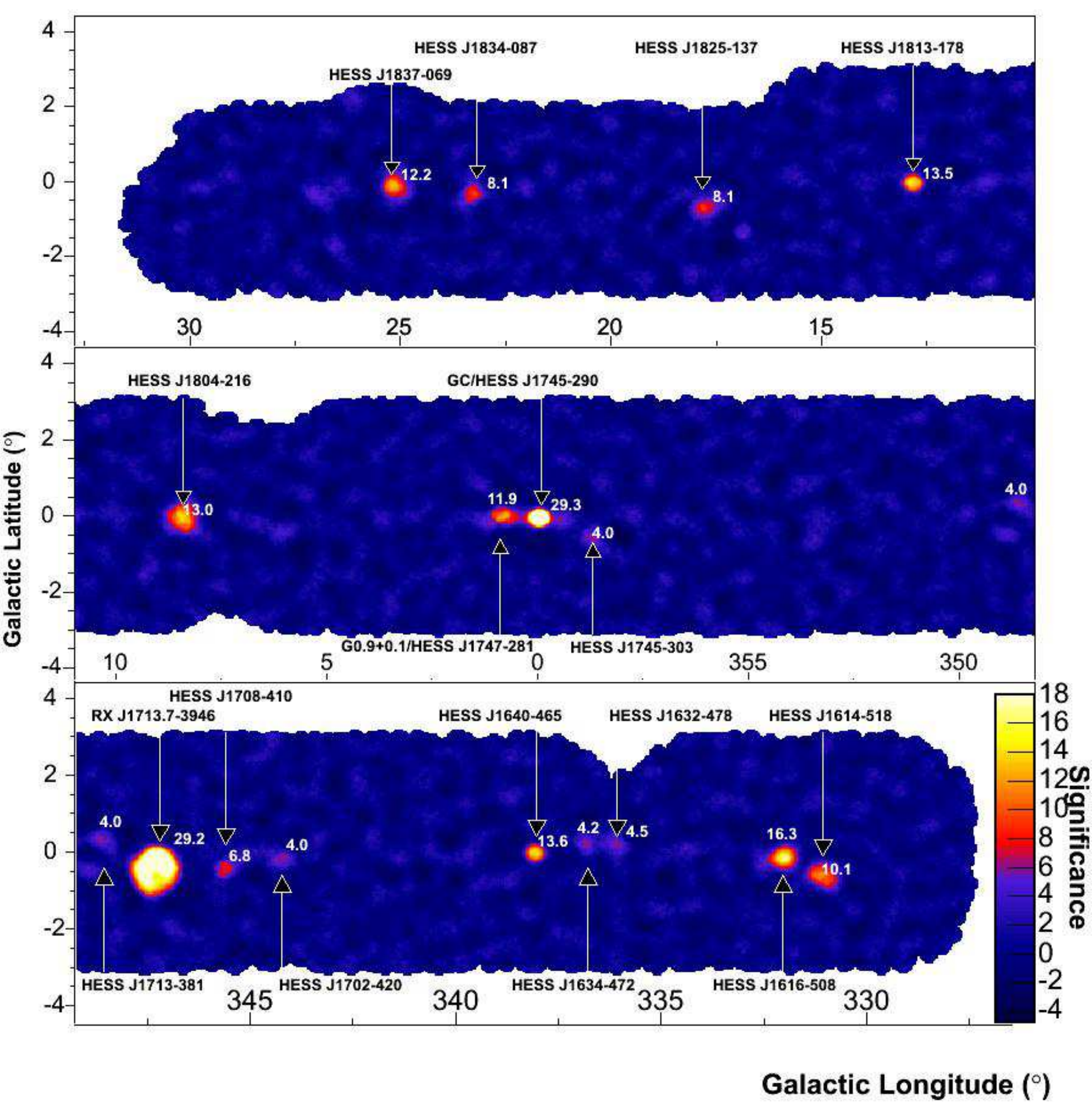}
\caption{The H.E.S.S. scan of the inner region of the Galactic plane with 13 newly discovered sources \cite{Aharonian2006}.\label{Fig8.8.1}}
\end{figure}

Shortly after completion of the four H.E.S.S. telescopes, the collaboration started scanning the inner part of the galactic disk with a sensitivity of 2\% of the Crab nebula flux above 200 GeV. In order to achieve a nearly uniform sensitivity across the galactic disk, the four telescopes were slightly re-adjusted to cover a strip of $\pm3^\circ$ latitude relative to the Galactic plane. The scan extended from $-30^\circ$ to $+30^\circ$ in longitude, covered by 500 pointings in a total of 230 hours.  In total, 14 new sources were discovered (Fig. \ref{Fig8.8.1}), about half of them unidentified sources and the other half in part pulsar-wind nebulae (candidates for the sources of CRs) and SNR with $\geq 4 \sigma$ significance after all trials \cite{Aharonian2006}. Later, a partial rescan with higher sensitivity, respectively with an improved analysis method, increased the number of detected sources to over 30. Also, a few binary objects were found to be gamma-ray emitters. This scan made H.E.S.S. the most successful observatory for the detection of galactic sources. Quite a few sources could not be classified. The richness of sources found in the galactic plane tells us that one could expect a significantly larger number with the next generation higher sensitivity telescopes.

\subsection{H.E.S.S. and MAGIC discover the first binaries}
\begin{figure}[h]
\centering
\includegraphics[width=.99\linewidth]{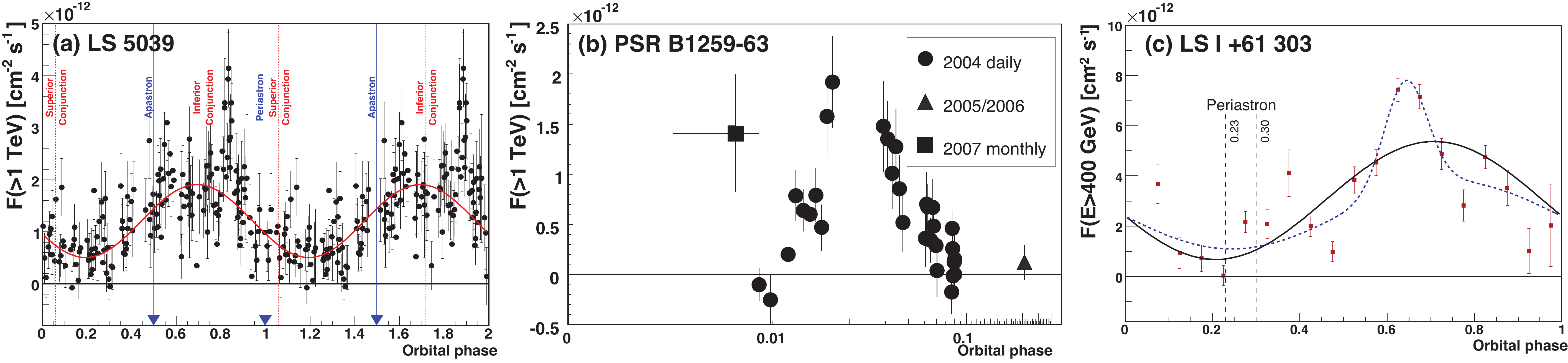}
\caption{Light curves of the binaries (a) LS 5039 (periodicity 4 d) \cite{ls5039}, (b) PSR\,B1259-63 (periodicity 3.9 y) \cite{b1259} and (c) LSI +61 303 \cite{lsi} (periodicity 26 d)\label{Fig8.8.12}.}
\end{figure}

\begin{figure}[h]
\centering
\includegraphics[width=.85\linewidth]{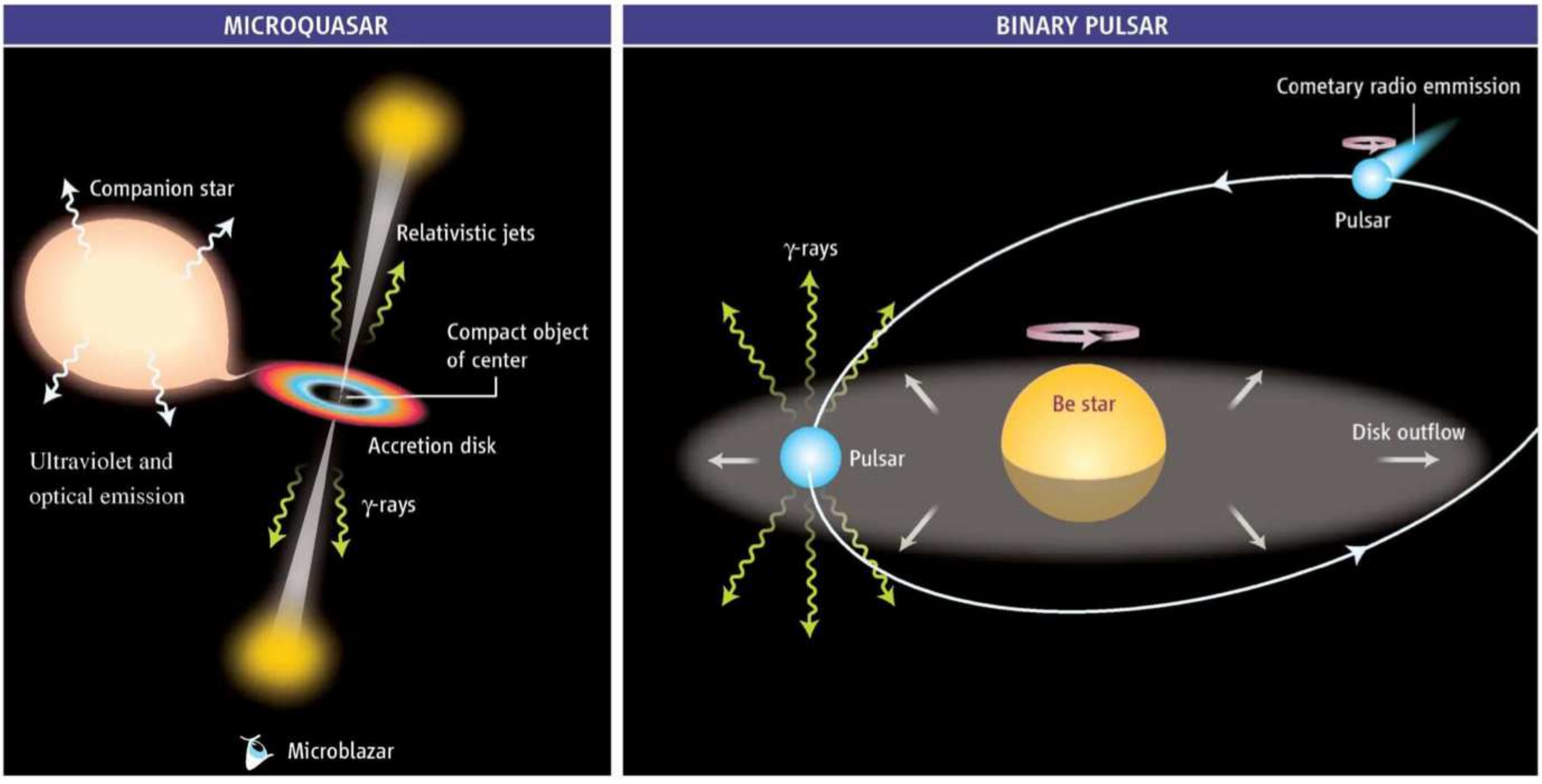}
\caption{The two preferred models of binary system emitting gamma rays. Left: the so-called microquasar model with a small black hole accreting mass from the companion star. Gamma rays are produced in the jets. Right: a binary system proposed by Felix Mirabel \cite{Mirabel2006}. A pulsar circulates around a Be star.\label{Fig8.8.2}}
\end{figure}

About one third of all stars are arranged in binary systems. Already during the Cygnus X-3 studies by the Kiel and other groups, the mostly accredited model for the VHE gamma-ray production was assumed to be a binary system with a periodicity of 4.8 hours. In the 1980s, binaries were considered as \textit{the sources} of cosmic gamma rays. Later, after quite a few VHE gamma-ray sources were discovered and none of them could be explained as binary systems, the question after the discovery of the Crab nebula was raised at nearly every International Cosmic Ray Conference before 2005: Where are the binaries? Eventually, both H.E.S.S. and MAGIC detected binaries in the Galactic plane. H.E.S.S. published the first discovery of a VHE binary, PSR\,B1259-63 \cite{Aharonian2005a} and LS\,5039 on the Southern sky \cite{Aharonian2005b}. Soon afterwards MAGIC discovered the first binary on the Northern sky, LS\,I\,+61\,303 \cite{Albert2006}. Fig. \ref{Fig8.8.12} shows the light curves of the three binaries. The composition of the binaries is not evident; Fig. \ref{Fig8.8.2} shows the two preferred models.

\subsection{MAGIC discovers the first VHE pulsar}
\begin{figure}[h]
\centering
\includegraphics[width=.65\linewidth]{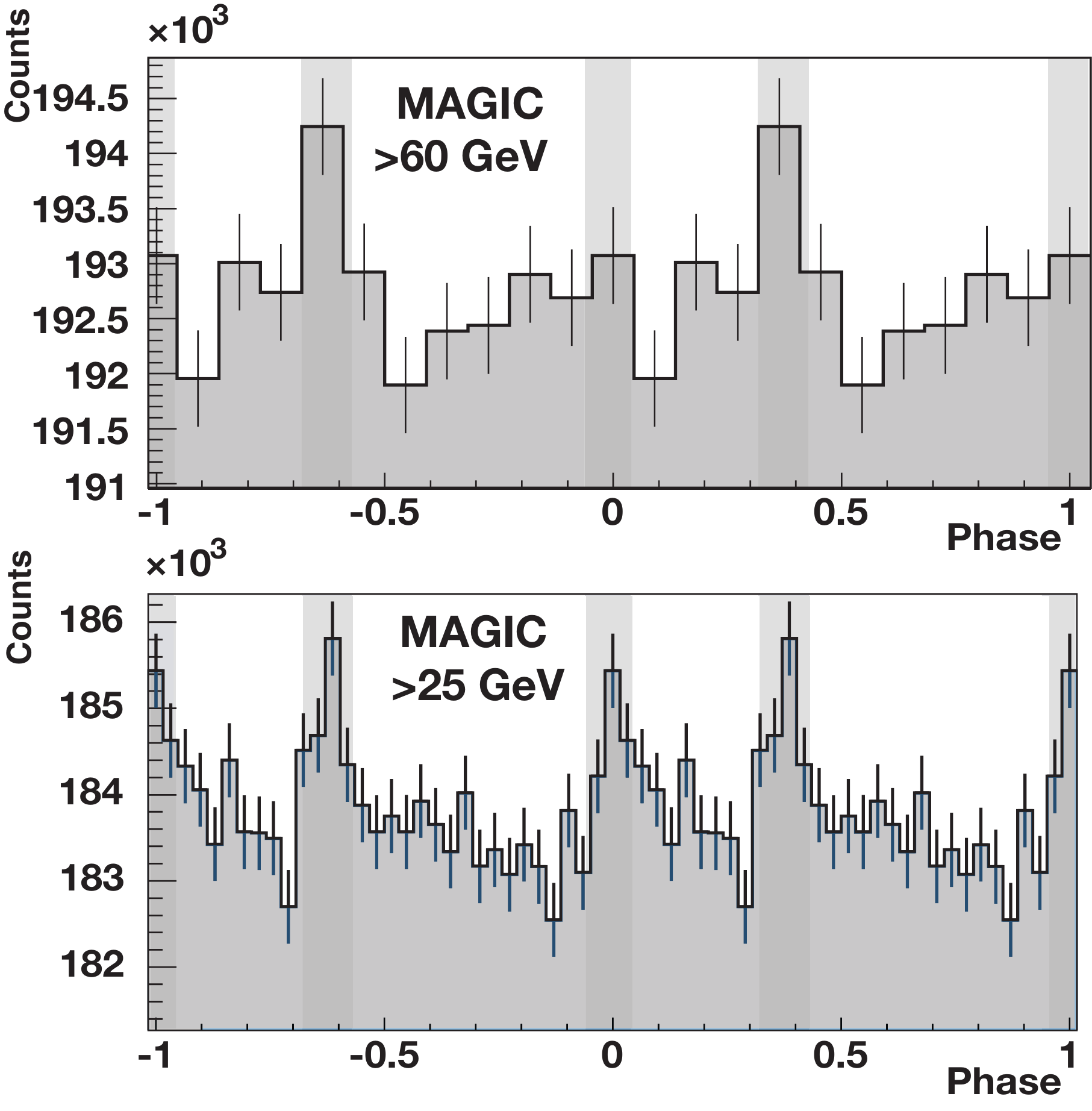}
\caption{First detected pulsed VHE gamma-ray emission of the Crab pulsar as measured by MAGIC \cite{Aliu2008}. A signal of 6.3 $\sigma$ significance, 8,300$\pm$1,300 pulsed events over a background of 6,106 events, has been detected. Figure courtesy the MAGIC collaboration.\label{Fig8.9.1}}
\end{figure}

Pulsars are one of the most interesting stellar objects. In the high-energy domain, satellite-borne gamma-ray detectors detected a few gamma-ray pulsars. The EGRET detector on board the \textit{Compton} gamma-ray satellite confirmed the observation of 7 high-significance pulsars in the MeV region while recently the follow-up satellite \textit{Fermi} added many more pulsars (6\% of all newly discovered stellar objects were pulsars \cite{Nolan2012}) and measured the spectra of the brightest ones up to 30/40 GeV). Ever since the discovery of VHE gamma-ray emission from the Crab nebula, groups have searched for pulsed emission from pulsars in the VHE domain, but up to 2009 without success. Trevor Weekes considers the finding of pulsed gamma-ray emission in the VHE domain as ``The Holy Grail'' of ground-based gamma-ray astronomy. In 2009, the MAGIC collaboration developed a new low threshold trigger, which could record data down to 26\,GeV, i.e., with considerable overlap with \textit{Fermi}-LAT data. Although \textit{Fermi}-LAT had predicted a cutoff of the pulsed gamma-ray emission at 12\,GeV for the Crab pulsar, MAGIC discovered pulsed emission from 26\,GeV upwards to nearly 100\,GeV (Fig. \ref{Fig8.9.1}) \cite{Aliu2008}. 

Two years later, the VERITAS team identified pulsed gamma-ray emission from the Crab pulsar using  data from 100 to 400\,GeV \cite{Aliu2011}. The spectrum followed a power
law up to 400 GeV which called for a new source mechanism 
than the standard outer-gap model. This was a physics breakthrough.
Some months later the MAGIC collaboration confirmed these results \cite{Aleksic2012}. These two measurements had opened the window of VHE pulsed gamma-ray studies.

\subsection{What to expect in the next decade: The next generation detectors for VHE gamma-astronomy.}
The first decade of the 21st century saw considerable progress in VHE gamma-ray astronomy. The third-generation Cherenkov telescopes achieved a sensitivity of $\approx 1\%$ of the Crab nebula flux and currently about one new source per month is discovered. Nevertheless, one sees a gradual shift from ``source hunting'' to the study of the underlying physics and to fundamental physics issues.  The recent successes have triggered ideas for quite a few new detectors with another large step in sensitivity increase and which should be realized in the coming years. There follows a very short overview of the new ideas.

\subsection{Technical progress in light sensors: FACT builds the first Cherenkov telescope with Geiger-mode avalanche photodiodes instead of PMTs}
Up to now the success of VHE gamma-ray astronomy was linked to the use of photomultipliers for the conversion of photons to photoelectrons, i.e. the so-called photoelectron detection efficiency (PDE). The currently best Cherenkov telescopes achieve an average PDE over the typical spectral range of $300-650$\,nm of only $\approx 15-18\%$ when folded by the Cherenkov spectrum. This low number dominated by the spectral sensitivity of PMTs clearly shows that there exists a large potential to improve the telescopes. Recently, the FACT collaboration tested new light sensors on the basis of silicon Geiger-mode avalanche photodiodes (G-apds) \cite{Anderhub2011}. These diodes have a peak PDE of $50-60\%$ and a rather flat detection efficiency across 300 to 700\,nm. When folded with the typical Cherenkov spectrum an averaged PDE of nearly a factor two improvement compared to the best PMTs should be achieved. The FACT collaboration built a full camera and installed it on a refurbished telescope from the old HEGRA experiment (the third telescope CT3 with a 9.4\,m$^2$ mirror area located at the Roque de los Muchachos observatory in the Canary Islands). The camera comprises 1,400 G-apds of 0.1$^\circ$ FOV each. Although G-apds are still in their development phase it is expected that such cameras will soon replace PMT cameras, because G-apds are estimated to eventually have up to a factor $2-3$ higher PDE in a few years.

\subsection{Towards a large VHE gamma-ray observatory: The CTA project}
\label{CTA}
\begin{figure}[h]
\centering
\includegraphics[width=.65\linewidth]{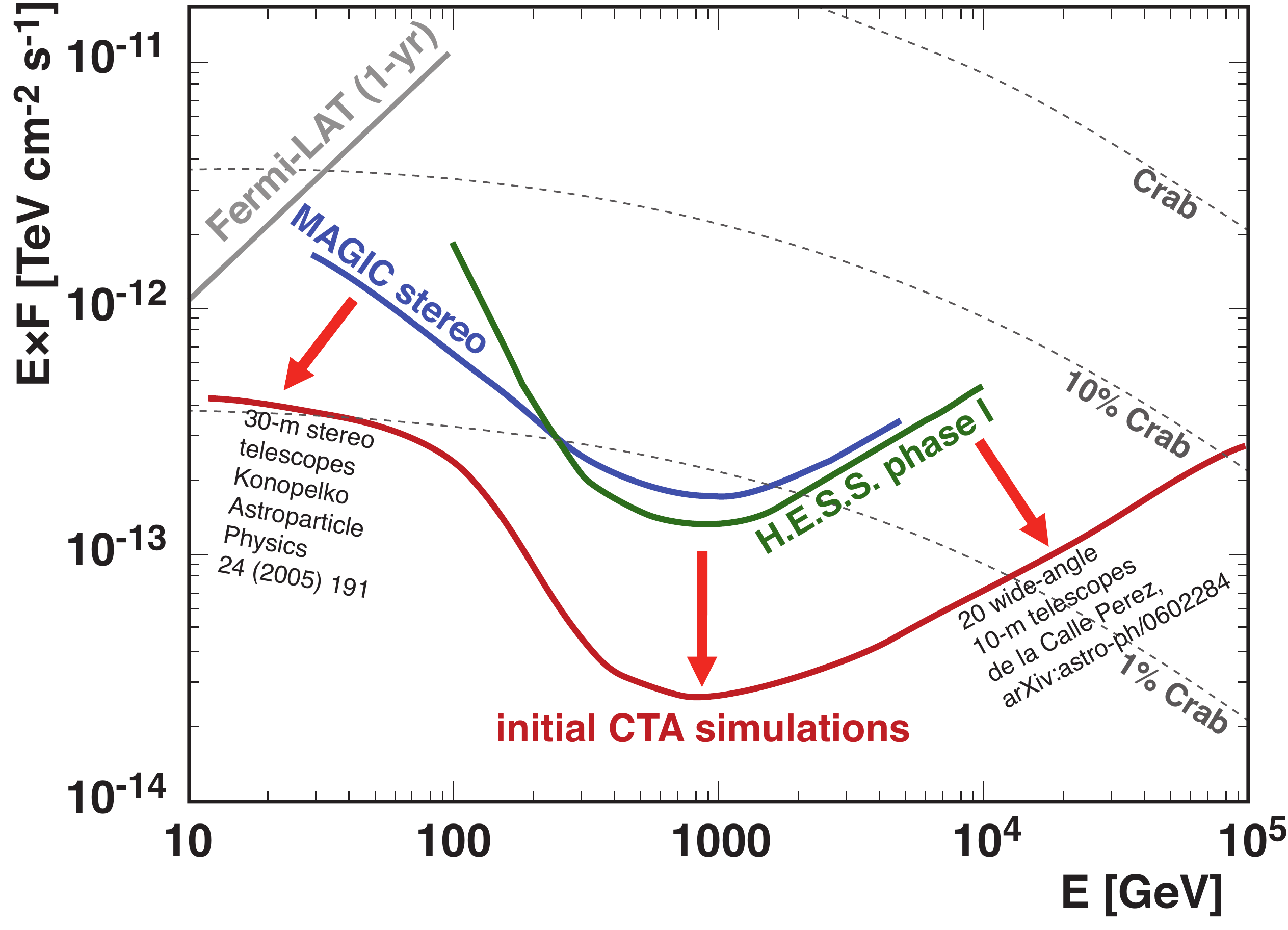}
\caption{Predicted sensitivity of CTA (in comparison to H.E.S.S., MAGIC, and \textit{Fermi}-LAT) and the Crab nebula flux.\label{Fig9.2.1}}
\end{figure}

Around 2007 it became evident that a further large increase in sensitivity could not be achieved by improving single telescopes but by considerably increasing the number of telescopes in an array configuration.  The idea for CTA (Cherenkov Telescope Array) was born. Building a detector covering the energy range of 20\,GeV to 100\,TeV requires a large number of three different sizes of telescopes (23\,m, 12\,m, and 3 to 5\,m diameter, respectively) in order to achieve a sensitivity 10 times higher compared to H.E.S.S., (see Fig. \ref{Fig9.2.1} for the predicted sensitivity) \cite{CTA}. The sites have not yet been selected. For covering the entire sky, it will be necessary to select one site in the Southern hemisphere and one in the Northern hemisphere. The energy range of CTA South will be extended to about 100\,TeV for the study of galactic sources while CTA North will need the two larger size telescopes types, because multi-TeV gamma rays from higher red-shift extragalactic sources are suppressed by the interaction with the low energy photon fields (see Sect. \ref{Sect:5}) and consequently no longer detectable. The initially European project is now enlarged to a worldwide collaboration approaching 900 members. CTA will start observations around 2015-2017. In their initial phase, the telescopes will be relatively conservative copies of current third-generation telescopes. 

\subsection{Other Projects: AGIS, MACE, HAWC, LHAASO}
Four other projects have passed the level of first ideas and are currently under detailed evaluation or in a first phase of construction. AGIS and MACE are Cherenkov telescopes, while HAWC is an extended air-shower (EAS) array at high altitude for achieving a low threshold. LHAASO is a facility that combines various air shower detector elements and Cherenkov telescopes.
 
\paragraph{AGIS} In 2008/9, US physicists proposed the Advanced Gamma-Ray Imaging System (AGIS), an array of 36 telescopes distributed over an 1~km$^2$ area \cite{Vandenbroucke2010}. The basic element is a dual-mirror telescope in an Schwarzschild-Coude\'e configuration. The primary mirror has an outer diameter of 9\,m. The Schwarzschild-Coud\'e configuration allows to use a rather compact camera with a wide FOV of 10$^\circ$. Later, it has been proposed to integrate these telescopes into the CTA project after successful prototyping and the AGIS collaboration has  become a member of CTA. Now, the Schwarzschild-Coude\'e telescope configuration is  also under consideration for some of the proposed designs of the small telescopes for CTA.

\paragraph{MACE} (Major Atmospheric Cerenkov Experiment) is predominantly pursued by physicists from India \cite{Koul2005}. The project comprises two large-diameter (21-m diameter) Cherenkov telescopes operated in stereo mode to be installed at the Hanle site at the Himalaya at 4,000\,m asl. The telescopes resemble in part the construction of the MAGIC telescopes but will have a lower threshold due to the high installation altitude and larger diameter. The first telescope should see first light in 2013.

\paragraph{HAWC} The HAWC (High Altitude Water Cherenkov) detector is a follow-up construction to the Milagro detector \cite{Salazar2009}. Instead of a single large pond the observatory comprises 900 densely packed water Cherenkov detectors of 5\,m height and 4.3\,m diameter. Again the top layer will be used to determine the shower energy and the angle of incidence while the bottom layer will detect muons and hadronic tracks in order to identify the hadronic background. Due to the installation on the flanks of the Sierra Negra volcano near Puebla, Mexico, at 4100 m asl, the detector will have a significantly lower threshold (100 GeV) compared to Milagro. The first detector elements have already been installed on site.

\paragraph{LHAASO} is a compound detector recently approved by the Chinese Academy of Sciences. LHAASO \cite{ZhenCao2009} plans to use the following elements:
\begin{itemize}
\item a 1~km$^2$ EAS array (KM2A),
\item 4 large water Cherenkov detector arrays (WCDAs),
\item a 5,000 m$^2$ shower core detector array (SCDA),
\item a wide-FOV Cherenkov/fluorescence telescope array (WFCA) and
\item two large imaging Cherenkov telescopes (LIACTs).
\end{itemize}
The observatory is a combination of a large EAS scintillator array, four large water Cherenkov ponds and a 5000\,m$^2$ shower core detector spread over 1\,km$^2$ interspaced by Cherenkov telescopes of different function and size.  The detector will be installed in Tibet near Yangbajing at around 4300\,m above sea level. Like HAWC, the EAS arrays, the Cherenkov ponds and the shower core detector will have a 24-hour uptime and a low threshold close to 100\,GeV. Two large Cherenkov telescopes should lower the threshold to 40\,GeV in gamma-ray energy. Other elements of the detector serve for general cosmic ray studies.

\section{A short summary of physics: What have we learned from VHE gamma-ray sources?}
The biggest success of ground-based gamma-ray astronomy, besides promoting VHE gamma-ray astronomy from the astronomy of a single source in 1989 by increasing the number of detected VHE gamma-ray sources to nearly 150 in 2012 is the diversity of source classes that could be established in this energy range.

A significant increase in the number of gamma-ray sources was due to the systematic scan of the Galactic plane performed by the H.E.S.S. collaboration from 2003 onwards. At almost the same pace, the extragalactic VHE gamma-ray sky became populated by VHE gamma-ray sources, dominantly blazars, due to systematic searches by the three VHE gamma-ray instruments H.E.S.S., MAGIC and VERITAS. 

\subsection{Supernova remnants}
A final stage of stellar evolution is reached when a star runs out of the fuel necessary for the fusion reactions that counteract the gravitational pressure. If the star is heavy enough, the collapse of the stellar core is followed by the ejection of the outer shells of the stellar material. Depending on the mass of the remaining object, a neutron star or a black hole is formed; the ejected material may interact with interstellar material. This expanding structure is called a supernova remnant. For a long time, supernova remnants have been suspected to be the sources of charged cosmic rays up to energies of at least $10^{15}$ eV. SNR generally are extended objects, and any VHE gamma-ray emission observed traces either, in case of hadronic origin, regions in which cosmic rays interact with target material, or, in case of leptonic origin, target electrons that exist in SNR. Showcase examples for detected and spatially resolved SNRs in gamma rays so far are the four objects RX\,J1713.7-3946 \cite{Aharonian2006}, RX\,J0852.0-4622 \cite{Lemoine-Gourmand2007}, RCW\,86 \cite{Aharonian2009}, and SN\,1006 \cite{Aharonian2010}. Generally, the VHE emission seems to resemble the X-ray morphology in these SNR, favoring a leptonic origin of the VHE emission, and particularly SN\,1006 and RX\,J1713.7-3946 are most certainly dominated by leptonic acceleration. On the other hand, an association of the gamma-ray emission with the presence of a molecular cloud (traced by CO density), which may serve as target material for hadronic gamma-ray production. Such an association is given in IC 443 \cite{ic443}, whereas in Tycho's supernova remnant, a combination of Fermi-LAT (GeV) and VERITAS spectra \cite{tycho} rule out leptonic acceleration models. The energy spectra from SNR are particularly hard, with a cutoff that sets in at about 20 TeV, indicating that the primary particles responsible for the gamma-ray emission must have had energies of some hundred TeV.

\subsection{Pulsars and pulsar-wind nebulae}
If a rotating neutron star remains in the system, it is referred to as a plerion or pulsar-wind nebula (PWN). The Crab nebula is a showcase example of a PWN. In such systems high energy electrons originating from the pulsar power the gamma-ray emission. PWN are the most commonly found type of galactic gamma-ray sources. Nonetheless,  not only the nebula itself may emit gamma radiation: As recently discovered \cite{Aliu2008}, the pulsar in the center of the Crab nebula emits pulsed VHE gamma radiation.

About one third of the sources found in scans of the galactic plane could not yet be associated with counterpart objects. For these, spectral and temporal properties of the TeV emission, and spatial co-location with known emission at other wavelengths are being investigated to learn about their nature.

\subsection{Compact objects and binary systems}
The source of high-energy particles in binary systems is the accretion of matter on one of the companions. Such systems provide vastly different conditions than the previously discussed objects, like high magnetic fields, high radiation densities, and high-energy photon fields. Due to this, particle acceleration and cooling timescales are short (typically in the order of the orbital periods of the systems). Compact objects (stellar-mass black holes or neutron stars) may also exhibit relativistic jet outflows. Such objects are then called microquasars in analogy to quasar-type active galactic nuclei. Well-known binary system TeV gamma-ray sources are PSR\,B1259-63 \cite{Aharonian2005a}, LS\,5039 \cite{Aharonian2005b}, LS\,I\,+61 303 \cite{Albert2006}, and HESS J0632+057 \cite{Maier2011}. 

\subsection{Stellar clusters and stellar winds}
Strong stellar winds, as they typically exist in star-forming regions and stellar clusters, may accelerate particles and lead to VHE gamma-ray production. Stellar winds seem natural candidate regions for VHE gamma-ray production as they also drive particle acceleration in binary systems and outflows in pulsar systems. Recently, TeV gamma-ray emission has been discovered in the young star system Westerlund 2 \cite{Aharonian2007,Aharonian2011}, and indications have been found in the Cyg\,OB2 star association.

\subsection{Unidentified sources}
The galactic plane scan revealed a substantial number of sources with no evident counterpart at any other wavelength -- about 20 such ``dark accelerators'' are now known. Some objects could later on be identified as PWN or SNR by catalog searches, by the revision of the likeliness of an association to a known object (e.g., HESS\,J1303-631/PSR\,J1301-6305) or by targeted follow-up observations (e.g., HESS\,J1813-178, \cite{Helfand2007}). However, for quite a few unidentified sources, such methods have failed to reveal their nature \cite{Aharonian2008d}.
Particularly a lack of X-ray emission may hint at an hadronic origin of the gamma-ray emission. Detailed studies of the (temporal, spectral, and morphological) features of these TeV-only emitters may help to identify the particle acceleration process at work and may also help answering the question whether these objects represent a source class of their own. However, as particle acceleration that leads to gamma-ray production generally requires certain rather characteristic parameters of the accelerator (like magnetic field strength, extension, densities), it may be difficult to establish a new class of TeV emitters.

In a certain sense, also the gamma-ray source at the center of our Galaxy is an unidentified TeV source \cite{Kosack2004,Aharonian2004}. Here, the difficulties come from source confusion, as the Galactic center region is a very busy one: Besides star-forming regions (Sgr B1, Sgr B2, Sgr D), the most prominent source towards the Galactic center is Sgr A, within which Sgr A* has been identified as possibly being a supermassive black hole. In addition, also a dark-matter annihilation signal could be expected from the center of our Galaxy. The gamma-ray energy spectrum determined from the Galactic center source is rather hard, favoring a PWN origin, and disfavoring a dark-matter origin. Dedicated searches for a dark-matter signal are reported, e.g., in \cite{Acero2011}.

\subsection{Extragalactic gamma-ray sources: Active galactic nuclei}
The second VHE gamma-ray source to be detected in 1992 was the active galactic nucleus Mkn 421. This source, like most of the well over twenty AGNs discovered as of today, is a blazar, which is a subclass of AGN with relativistically beamed emission towards the observer. Blazars have been detected at a redshift range of $z=0.031$ (Mkn 421; \cite{Punch1992}) up to $z=0.536$ (3C 279; \cite{3C279}) so far. Active galactic nuclei are powered by accretion of matter by supermassive black holes with some billion solar masses and show high variability down to timescale of minutes and below, indicating complex particle acceleration and cooling processes working within the jet acceleration regions. The most remarkable flaring activity so far has been observed in PKS 2155-304 \cite{2155high2} with flux intensities exceeding by an order of magnitude the otherwise mostly ``dormant'' emission \cite{2155low1,2155low2} and flux variations on timescales of minutes.

\begin{figure}[h]
\centering
\includegraphics[width=.95\linewidth]{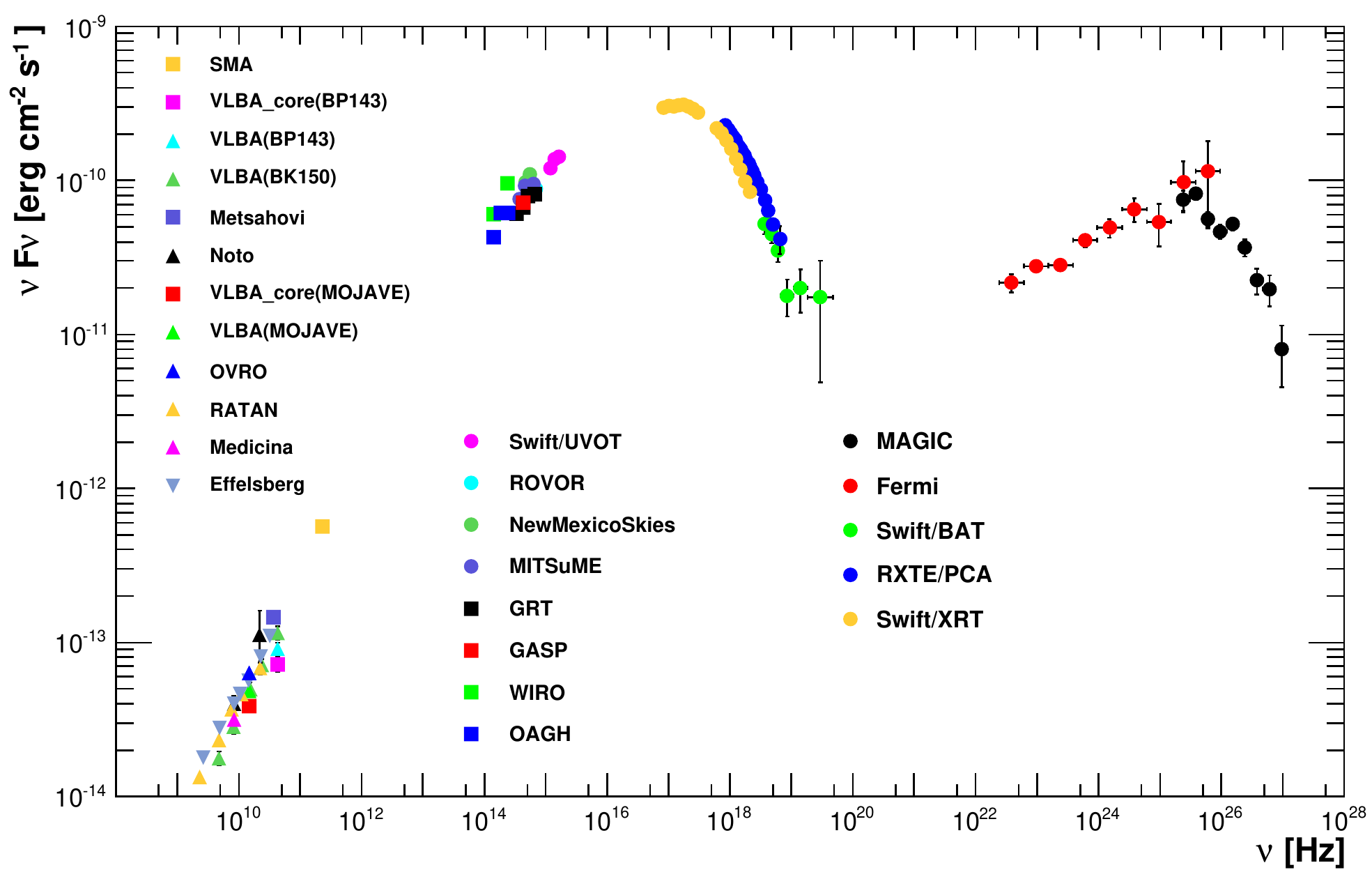}
\caption{A strictly-simultaneously measured spectral energy distribution of the blazar Mkn 421 \cite{421}. The low-energy peak is believed to represent synchrotron radiation off a population of relativistic electrons, while the origin of the second, high-energy peak is debated. It may be due to inverse-Comption radiation of the same electron and photon population (``self-synchrotron Compton'' emission), external photons scattering off the electrons (``external Compton'' emission), or it may be of hadronic origin. High-energy gamma-ray observations play a crucial role in discriminating possible scenarios due to their sensitivity to time variations and the spectral shape of the SED at around GeV/TeV energies.\label{fig421}}
\end{figure}

The TeV AGNs were for a long time dominated by so-called high-peaked BL Lac objects (Fig. \ref{fig421}), which are AGNs with a peak of their synchrotron emission in the X-ray range of the energy spectrum. In leptonic acceleration models the TeV emission is then interpreted as photons scattered off the same electron population that created the X-ray emission. Lately, some ``low-peaked'' BL Lac objects (with the synchrotron peak in the optical regime; e.g. BL\,Lac itself; W\,Comae) and flat-spectrum radio quasars with even lower X-ray peaks could be discovered (e.g., 3C\,279 \cite{3C279}, PKS\,1222+22 \cite{1222}).

Recently, also close-by radio galaxies like M\,87 and Centaurus A have also been identified as gamma-ray emitters. Those objects are close by and have jets misaligned to the line of sight. This allows spatial studies of the jets and the regions within them responsible for the particle acceleration, particularly by combining high-resolution radio observations and TeV light curves \cite{M871,M872,M873}.

\subsection{Starburst Galaxies}
Galaxies with entirely no activity in their central engine, like M\,82 and NGC\,253, could be identified as TeV emitters \cite{NGC253,M82}. In those objects, strong stellar winds created by high supernova activities are responsible for particle acceleration that leads to gamma-ray emission up to TeV energies.

\subsection{Galaxy Clusters}
In addition clusters of galaxies, which in some sense represent small ecosystems of the Universe itself, have been observed by ground-based Cherenkov instruments, and recently the central galaxy of the Perseus Cluster was detected in TeV gamma rays. The emission seen so far, however, is compatible with what is expected from the galaxy itself; no extended, inter-cluster emission could be claimed \cite{Aleksic1275}.

\subsection{Gamma-Ray Bursts}
Gamma-ray bursts (GRB) are transient extragalactic sources of high-energy gamma-ray emission that occur randomly and unpredictably. While gamma rays of energies of as high as 30 GeV could be detected by space-borne instruments, the specific difficulty in detecting them from ground is their presumably short time of activity. Fireball models describe the emission as being produced by relativistic shocks and predict ``prompt'' and ``delayed'' emission up to TeV energies both by leptonic and hadronic processes. Upon a trigger from space-borne, all-sky monitoring instruments, a ground-based detector has to slew very fast to the GRB direction. All ground-based instruments have GRB programs and while no GRB gamma rays have yet been detected, upper limits were reported, with observations starting up to 40 seconds after the GRB onset only. In one occasion, a (rather soft) GRB occurred while a ground-based instrument was accidentally pointing in its direction \cite{hessgrb}. 

\subsection{Astroparticle Physics and Fundamental Physics}
Besides studying of individual astrophysical objects, observations of very-high energy gamma rays are also used in the indirect search for dark matter. Some dark matter candidate particles, namely the lightest supersymmetric particles, the neutralinos, may decay into photons (which, however, is a disfavored channel) or into quark-antiquark pairs, which would undergo further reactions producing gamma rays. A recently found channel for the production of gamma rays in DM decay processes is the ``internal bremsstrahlung'' process, with a gamma-ray enhancement near the kinematical limit. For some regions of the dark-matter parameter space, those gamma rays will be in the GeV to TeV energy range and may be detectable from the directions of astrophysical objects in which the dark matter density is high. Such locations comprise the center of our Galaxy, intermediate-mass black holes, but also any object with a high mass-to-luminosity ratio, e.g. dwarf galaxies. Dark matter signatures have not been found yet, neither from observations of the center of our Galaxy \cite{Acero2011} nor from other candidate objects; the best exclusion limit so far comes from observations of the dwarf galaxy Segue\,1 \cite{segue}.

Another domain of fundamental physics in which ground-based gamma-ray astronomy can help is the search for violation of Lorentz Invariance, which is predicted, most notably, by theories of quantum gravity. Qualitatively speaking, the vacuum is considered to be interacting with traversing particles depending on their energy. Thus, observing delays of photon arrival times from strong AGN flares or pulsar emission constitute time of flight measurements. Any revealed relative delays of gamma rays with different energies, however, may also originate in the gamma-ray production mechanism. Thus, to demonstrate that Lorentz invariance is at work, an universal signature in many observations of objects at different distances needs to be found. Up to now, from strong flares in the AGN Mkn\,501 and PKS\,2155-304, only upper limits have been derived \cite{LIVH,LIVM} that reach few percent of the Planck energy scale, which is the natural scale expected at which quantum-gravity effects are expected to become apparent.

\section{The VHE sky map at the 98th year of Cosmic Ray studies}
\begin{figure}[h]
\centering
\includegraphics[width=.95\linewidth]{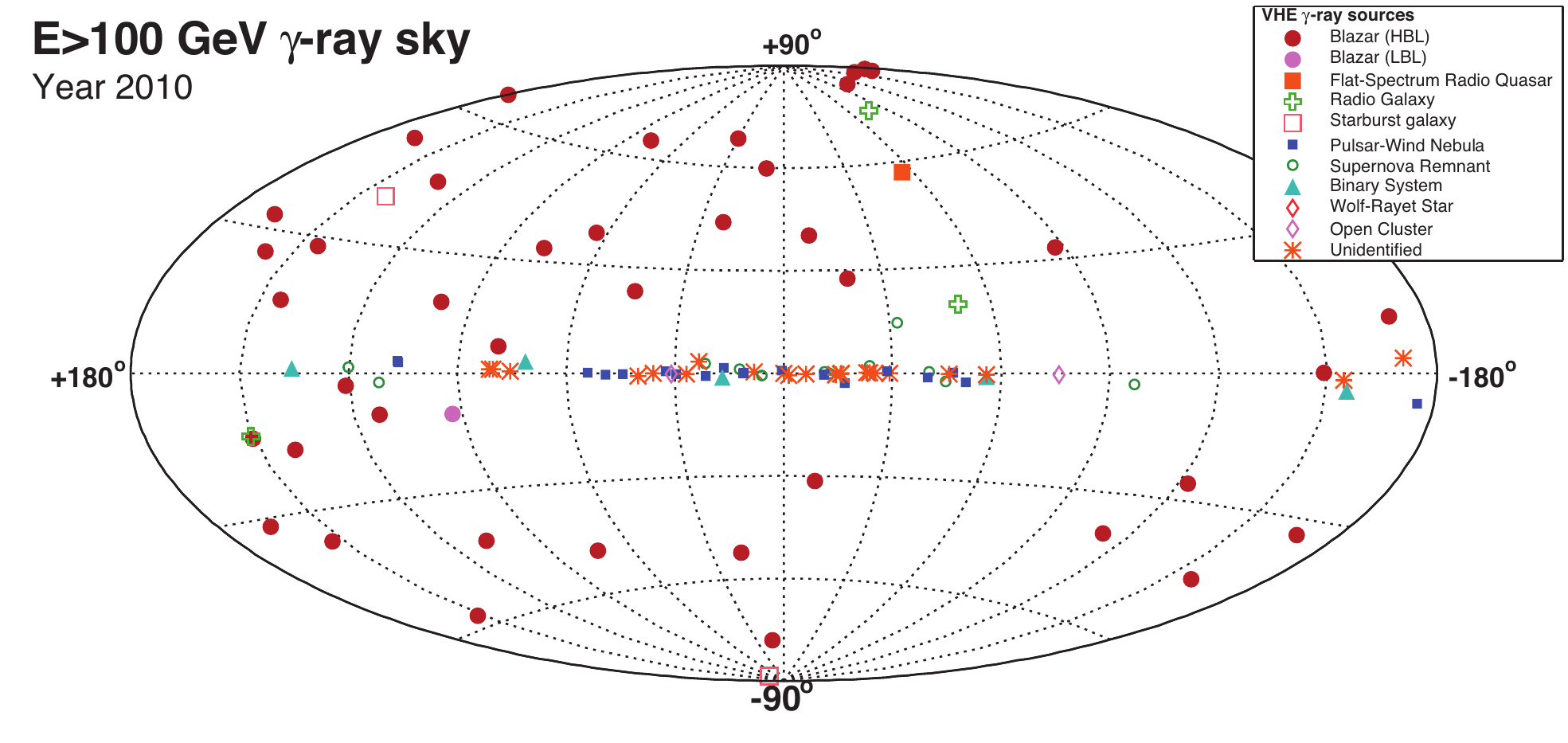}
\caption{The VHE ($E>100$\,GeV) sky map in the year 2010.\label{Fig11.1}}
\end{figure}

The first decade of the new millennium saw a large expansion of discoveries after the large Cherenkov observatories became fully operational. Nearly every month a new source was discovered. Fig.\,\ref{Fig11.1} shows the $E>$100\,GeV sky map in the year 2010 with over 110 sources. About 60\% of all sources are located in the galactic plane, while about 40\% of the sources are of extragalactic origin. The central part of the galactic plane is well visible from the H.E.S.S. observatory site while only the outer wings of the galactic plane are visible to the two northern Cherenkov observatories, MAGIC and VERITAS. Also, some sources are detected by the ``tail catcher'' detector Milagro.

Currently, the productivity of the three large telescope installations is high. In the 100th year of CR physics the number of 150 discovered sources is being approached. The two Northern installations have a sensitivity of about $0.8-1\%$ of the Crab nebula flux for a 5-$\sigma$ signal within 50\,h observation time, while H.E.S.S. has a sensitivity close to 0.7\% of the Crab nebula flux. Still, most of the extragalactic area has not been scanned. 

\vspace*{-1cm}


\begin{thebibliography}{}
%
\bibitem[Abdo 2007]{Abdo2007} Abdo, Aous A., et al. (Milagro collaboration). 2007.
TeV Gamma-Ray Sources from a Survey of the Galactic Plane with Milagro. \textit{Astrophysical Journal Letters} \textbf{664}: L91-L94.
\bibitem[Abdo 2011]{421} Abdo, Aous A., et al. 2011. Fermi-LAT Observations of Markarian 421: the Missing Piece of its Spectral Energy Distribution. \textit{Astrophysical Journal} \textbf{736}: 131-152.
\bibitem[Abramowski 2010]{2155low2} Abramowski, Attila, et al. (H.E.S.S. collaboration) 2010. VHE gamma-ray emission of PKS 2155-304: spectral and temporal variability. \textit{Astronomy \& Astrophysics} \textbf{520}: A83. 
\bibitem[Abramowski 2011a]{Aharonian2011} Abramowski, Attila, et al. (H.E.S.S. collaboration). 2011a. Revisiting the Westerlund 2 field with the H.E.S.S. Telescope Array. \textit{Astronomy \& Astrophysics} \textbf{525}: A46.
\bibitem[Abramowski 2011b]{LIVH} Abramowski, Attila, et al. (H.E.S.S. collaboration). 2011b. Search for Lorentz Invariance breaking with a likelihood fit of the PKS 2155-304 Flare Data Taken on MJD 53944. \textit{Astroparticle Physics} \textbf{34}: 738-747.
\bibitem[Abramowski 2011c]{Acero2011} Abramowski, Attila, et al. (H.E.S.S. collaboration). 2011c.
Search for a Dark Matter annihilation signal from the Galactic Center halo with H.E.S.S. \textit{Physical Review Letters} \textbf{106}: 161301.
\bibitem[Abramowski 2012]{M873} Abramowski, Attila, et al. (H.E.S.S., MAGIC, VERITAS collaboration). 2012. The 2010 very high energy gamma-ray flare \& 10 years of multi-wavelength observations of M 87. \textit{Astrophysical Journal} \textbf{746}: 151-169.
\bibitem[Acciari 2009a]{M871} Acciari, V. A., et al. (VERITAS, MAGIC, H.E.S.S. collaboration). 2009a. Radio Imaging of the Very-High-Energy Gamma-Ray Emission Region in the Central Engine of a Radio Galaxy. \textit{Science} \textbf{325}: 444-448. 
\bibitem[Acciari 2009b]{M82} Acciari, V. A., et al. (VERITAS collaboration). 2009b. A connection between star formation activity and cosmic rays in the starburst galaxy M82. \textit{Nature} \textbf{462}: 770-772. 
\bibitem[Acciari 2011]{tycho} Acciari, V. A., et al. (VERITAS collaboration). 2011. Discovery of TeV Gamma Ray Emission from Tycho's Supernova Remnant. \textit{Astrophysical Journal Letters} \textbf{730}: L20.
\bibitem[Acero 2009]{NGC253} Acero, F., et al. (H.E.S.S. collaboration). 2009. Detection of Gamma Rays from a Starburst Galaxy. \textit{Science} \textbf{326}: 1080-1082. 
\bibitem[Acero 2010]{Aharonian2010} Acero, F., et al. (H.E.S.S. collaboration). 2010. First detection of VHE $\gamma$-rays from SN 1006 by HESS. \textit{Astronomy \& Astrophysics} \textbf{516}: A62.
\bibitem[Acero 2012]{2155high2}Acero, F., et al. (H.E.S.S. collaboration). 2012. A multiwavelength view of the flaring state of PKS 2155-304 in 2006. \textit{Astronomy \& Astrophysics} \textbf{539}: A149.
\bibitem[Actis 2011]{CTA}Actis, M., et al. (CTA consortium). 2011. Design concepts for the Cherenkov Telescope Array CTA: an advanced facility for ground-based high-energy gamma-ray astronomy. \textit{Experimental Astronomy} \textbf{32}: 193-316.
\bibitem[Aharonian 1989]{Aharonian1989} Aharonian, Felix, et al. 1989. Cerenkov Imaging TeV Gamma-ray Telescope. In: \textit{Very High Energy Gamma Ray Astronomy: proceedings of the International Workshop, Crimea, USSR, April 17-21, 1989}: p. 36.
\bibitem[Aharonian 1991]{Aharonian1991} Aharonian, Felix A., Akhperjanian, A. G., Kankanian, A. S., Mirzoyan, R. G., Stepanian, A. A., M\"uller, N., Samorski, M., Stamm, W., Bott-Bodenhausen, M., Lorenz, E., Sawallisch, P. 1991. A System of Air Cherenkov Telescopes in the HEGRA Array. In: \textit{Proceedings of the 22nd International Cosmic Ray Conference, Dublin, Ireland} \textbf{2}: 615-618.
\bibitem[Aharonian 1999]{hegra501} Aharonian, Felix, et al. (HEGRA collaboration). 1999. The temporal characteristics of the TeV gamma-radiation from MKN 501 in 1997. I. Data from the stereoscopic imaging atmospheric Cherenkov telescope system of HEGRA. \textit{Astronomy \& Astrophysics} \textbf{342}: 69-86.
\bibitem[Aharonian 2004]{Aharonian2004} Aharonian, Felix, et al. (H.E.S.S. collaboration). 2004. Very high energy gamma rays from the direction of Sagittarius A*. \textit{Astronomy \& Astrophysics} \textbf{425}: L13-L17.
\bibitem[Aharonian 2005a]{Aharonian2005a} Aharonian, Felix, et al. (H.E.S.S. collaboration). 2005a. Discovery of the binary pulsar PSR B1259-63 in very-high-energy gamma rays around periastron with HESS. \textit{Astronomy \& Astrophysics} \textbf{442}: 1-10.
\bibitem[Aharonian 2005b]{Aharonian2005b} Aharonian, Felix, et al. (H.E.S.S. collaboration). 2005b. Discovery of Very High Energy Gamma Rays Associated with an X-ray Binary. \textit{Science} \textbf{309}: 746-749.
\bibitem[Aharonian 2006a]{Aharonian2006} Aharonian, Felix, et al. (H.E.S.S. collaboration). 2006a. The H.E.S.S. survey of the Inner Galaxy in very high-energy gamma-rays. \textit{Astrophysical Journal} \textbf{636}: 777-797.
\bibitem[Aharonian 2006b]{ls5039} Aharonian, Felix, et al. (H.E.S.S. collaboration). 2006b. 3.9 day orbital modulation in the TeV gamma-ray flux and spectrum from the X-ray binary LS 5039. \textit{Astronomy \& Astrophysics} \textbf{460}: 743-749.
\bibitem[Aharonian 2007a]{Aharonian2007} Aharonian, Felix, et al. (H.E.S.S. collaboration). 2007a. Detection of extended very-high-energy gamma-ray emission towards the young stellar cluster Westerlund 2. \textit{Astronomy \& Astrophysics} \textbf{467}: 1075-1080.
\bibitem[Aharonian 2008]{Aharonian2008d} Aharonian, Felix, et al. (H.E.S.S. collaboration). 2008. HESS very-high-energy gamma-ray sources without identified counterparts. \textit{Astronomy \& Astrophysics} \textbf{477}: 353-363.
\bibitem[Aharonian 2009a]{hessgrb} Aharonian, Felix, et al. (H.E.S.S collaboration). 2009a. H.E.S.S. observations of the prompt and afterglow phases of GRB 060602B. \textit{Astrophysical Journal} \textbf{690}: 1068-1073.
\bibitem[Aharonian 2009b]{Aharonian2009} Aharonian, Felix, et al. (H.E.S.S. collaboration). 2009b. Discovery of Gamma-Ray Emission from the Shell-Type Supernova Remnant RCW 86 with HESS. \textit{Astrophysical Journal} \textbf{692}: 1500-1505.
\bibitem[Aharonian 2009c]{2155low1} Aharonian, Felix, et al. (H.E.S.S. collaboration). 2009c. Simultaneous observations of PKS 2155-304 with H.E.S.S., Fermi, RXTE and ATOM: spectral energy distributions and variability in a low state. \textit{Astrophysical Journal} \textbf{696}: L150.
\bibitem[Aharonian 2009d]{b1259} Aharonian, Felix, et al. (H.E.S.S. collaboration). 2009d. Very high energy gamma-ray observations of the binary PSR B1259-63/SS2883 around the 2007 Periastron
H.E.S.S. collaboration. \textit{Astronomy \& Astrophysics} \textbf{507}: 389-396.
\bibitem[Aiso 1997]{Aiso1997} Aiso, S., et al. 1997. The Detection of TeV Gamma Rays from Crab using the Telescope Array Prototype. In \textit{Proceedings of the 25th International Cosmic Ray Conference, Durban, South Africa} \textbf{3}: 177-180.
\bibitem[Akerlof 1991]{Akerlof1991} Akerlof, C.W., et al. 1991. Granite, a new very high energy gamma-ray telescope. \textit{Nuclear Physics B -- Proceedings Supplements} \textbf{14}: 237-243.
\bibitem[Albert 2006]{Albert2006} Albert, Jord\'\i, et al. (MAGIC collaboration). 2006. Variable Very-High-Energy Gamma-Ray Emission from the Microquasar LS I +61 303. \textit{Science} \textbf{23}: 1771-1773.
\bibitem[Albert 2007]{ic443} Albert, Jord\'\i, et al. (MAGIC collaboration). 2007. Discovery of VHE Gamma Radiation from IC443 with the MAGIC Telescope. \textit{Astrophysical Journal Letters} \textbf{664}: L87-L90.
\bibitem[Albert 2008a]{3C279} Albert, Jord\'\i, et al. (MAGIC collaboration). 2008a. Very high energy gamma rays from a distant Quasar: How transparent is the Universe? 2008a. \textit{Science} \textbf{320}: 1752-1754.
\bibitem[Albert 2008b]{LIVM} Albert, Jord\'\i, et al. (MAGIC collaboration). 2008b. Probing quantum gravity using photons from a flare of the active galactic nucleus Markarian 501 observed by the MAGIC telescope. \textit{Physics Letters B} \textbf{668}: 253-257.
\bibitem[Albert 2009]{lsi} Albert, Jord\'\i, et al. (MAGIC collaboration). 2009. Periodic Very High Energy $\gamma$-Ray Emission From LS I+61$^\circ$ 303 Observed With The MAGIC Telescope. \textit{Astrophysical Journal} \textbf{693}: 303-310.
\bibitem[Aleksi\'c 2011]{1222} Aleksi\'c, Jelena, et al. (MAGIC collaboration). 2011. MAGIC discovery of VHE Emission from the FSRQ PKS 1222+21. \textit{Astrophysical Journal} \textbf{730}: L8.
\bibitem[Aleksi\'c 2012a]{Aleksic1275} Aleksi\'c, Jelena, et al. (MAGIC collaboration). 2012a. Detection of very-high energy $\gamma$-ray emission from NGC 1275 by the MAGIC telescopes. \textit{Astronomy \& Astrophysics} \textbf{539}: L2. 
\bibitem[Aleksi\'c 2012b]{Aleksic2012} Aleksi\'c, Jelena, et al. (MAGIC collaboration). 2012b. Phase-resolved energy spectra of the Crab pulsar in the range of 50-400GeV measured with the MAGIC telescopes. \textit{Astronomy \& Astrophysics} \textbf{540}: A69.
\bibitem[Aliu 2008]{Aliu2008} Aliu, Ester, et al. (MAGIC collaboration). 2008. Observation of Pulsed $\gamma$-Rays Above 25 GeV From the Crab Pulsar with MAGIC. \textit{Science} \textbf{322}: 1221-1224.
\bibitem[Aliu 2011]{Aliu2011} Aliu, Ester, et al. (VERITAS collaboration). 2011. Detection of Pulsed Gamma Rays Above 100 GeV from the Crab Pulsar. \textit{Science} \textbf{334}: 69-72.
\bibitem[Aliu 2012]{segue} Aliu, Ester, et al. (VERITAS collaboration). 2012. VERITAS Deep Observations of the Dwarf Spheroidal Galaxy Segue 1. \textit{Physical Review} \textbf{D85}: 062001.
\bibitem[Anderhub 2011]{Anderhub2011} Anderhub, Hans, et al. (FACT collaboration). 2011. A G-APD based Camera for Imaging Atmospheric Cherenkov Telescopes. \textit{Nuclear Instruments and Methods in Physics Research Section A: Accelerators, Spectrometers, Detectors and Associated Equipment} \textbf{628}: 107-110.
\bibitem[Auger 1939]{Auger1937} Auger, P., Ehrenfest, P., Maze, R., Daudin, J., Fr\'eon, Robley A. 1939. Extensive Cosmic-Ray Showers. \textit{Reviews of Modern Physics} \textbf{11}: 288-291.
\bibitem[Augustin 1974]{Augustin1974} Augustin, J.-E., et al. 1974. Discovery of a Narrow Resonance in e$^+$e$^-$ Annihilation. \textit{Physical Review Letters} \textbf{33}: 1406-1408.
\bibitem[Bagge 1977]{Bagge1977} Bagge, E. R., Samorski, M., Stamm, W. 1977. A new Air Shower Experiment at Kiel. In: \textit{Proceedings of the 15th International Cosmic Ray Conference, Plovdiv, Bulgaria} \textbf{12}: 24-29.
\bibitem[Baillon 1992]{Baillon1992} Baillon, P., et al. 1992. Detection of very high energy gamma rays from the Crab source. In: \textit{
AIP Conference Proceedings} \textbf{272}: 1218-1221.
\bibitem[Baixeras 2003]{Baixeras2003} Baixeras, Carmen, et al. (MAGIC collaboration). 2003. The MAGIC Telescope. \textit{Nuclear Physics B (Proceedings Supplement)} \textbf{114}: 247-252.
\bibitem[Barrau 1998]{Barrau1998} Barrau, A., et al. (CAT collaboration). 1998. The CAT imaging telescope for very-high-energy gamma-ray astronomy. \textit{Nuclear Instruments and Methods in Physics Research Section A: Accelerators, Spectrometers, Detectors and Associated Equipment} \textbf{416}: 278-292.
\bibitem[Bastieri 1999]{Bastieri1999} Bastieri, Denis, et al. (CLUE collaboration). 1999. The CLUE experiment running with 8 telescopes; observations of gamma sources and runs on Moon. In \textit{AIP Conference Proceedings} \textbf{515}: 436-440.
\bibitem[Blackett 1948]{Blackett1948} Blackett, P. M. S. 1948. A possible contribution to the night sky from the Cerenkov radiation emitted by cosmic rays. In: The Emission Spectra of the Night Sky and Aurorae, Papers read at an International Conference held in London, July, 1947. London: The Physical Society, 1948., p.\,34.
\bibitem[Borione 1997]{Borione1997} Borione, A., et al. 1997. High statistics search for ultrahigh energy $\gamma$-ray emission from Cygnus X-3 and Hercules X-1. \textit{Physical Review} \textbf{D55}: 1714-1731.
\bibitem[Bowden 1991]{Bowden1991} Bowden, C. C. G., et al. 1991. The University of Durham Mark V Composite Gamma Ray Telescope. In \textit{Proceedings of the 22nd International Cosmic Ray Conference, Dublin, Ireland} \textbf{2}: 626-629.
\bibitem[Bradbury 1997]{Bradbury1997} Bradbury, Stella, et al. (HEGRA collaboration). 1997. Detection of gamma-rays above 1.5 TeV from Mkn 501. \textit{Astronomy \& Astrophysics Letters} \textbf{320}: 5-8.
\bibitem[Budnev 2005]{Budnev2005} Budnev, N.M., et al. 2005. The Tunka Experiment: Towards a 1-km$^2$ Cherenkov EAS Array in the Tunka Valley. In \textit{Proceedings of the 29th International Cosmic Ray Conference, Pune, India} \textbf{8}: 255-259.
\bibitem[Cao 2011]{ZhenCao2009} Cao, Zhen, et al., (LHAASO collaboration). 2011. The ARGO-YBJ Experiment Progresses and Future Extension. \textit{International Journal of Modern Physics D} \textbf{20}: 1713-1721.
\bibitem[Cassidi 1997]{Cassidi1997} Cassidi, M., et al. 1997. CASA-BLANCA: A Large Non-imaging Cherenkov Detector at CASA-MIA. In \textit{Proceedings of the 25th International Cosmic Ray Conference, Durban, South Africa} \textbf{5}: 189-192.
\bibitem[Catanese 1999]{Catanese1999} Catanese, Michael, Weekes, Trevor C. 1999. Very High Energy Gamma-Ray Astronomy. \textit{Publications of the Astronomical Society of the Pacific} \textbf{111}: 1193-1332.
\bibitem[Cherenkov 1934]{Cherenkov1934} Cherenkov, Pavel A. 1934. Visible emission of clean liquids by action of $\gamma$ radiation. \textit{Doklady Akademii Nauk SSSR} \textbf{2}: 451.
\bibitem[Chi 1992]{Chi1992}Chi, X, Wdowczyk, J, Wolfendale, A. W. 1992. The clustering of the arrival directions of the highest-energy cosmic rays. \textit{Journal of Physics G: Nuclear Particle Physics} \textbf{18}: 1867-1868; further references therein.
\bibitem[Chudakov 1961]{Chudakov1961} Chudakov, A. E., Zatsepin G. 1961.  On the methods of searching for local sources of high-energy photons. \textit {Journal of Experimental and Theoretical Physics} \textbf{41}: 655.
\bibitem[Chudakov 1963]{Chudakov1963} Chudakov, A. E., Dadykin, V. L., Zatsepin, V. I., Nesterova, N. M. 1963. On the high energy photons from local sources. In \textit{Proceedings of the 8th International Cosmic Ray Conference, Jaipur, India} \textbf{4}: 199-203.
\bibitem[Chudakov 1965]{Chudakov1965} Chudakov, A. E. et al. 1965. A serach for photons with energy
$10^{13}$ eV from local sources of cosmic radiation. \textit{Proceedings of Lebedev Institute} \textbf{26}: 118-141.  
English Translation: Consultants Bureau, 99 (1965).
\bibitem[Cocconi 1959]{Cocconi1959} Cocconi, G. 1959. An air shower telescope and the detection of 10$^{12}$ eV photon sources. In \textit{Proceedings of the 2nd International Cosmic Ray Conference, Moscow, Russia} \textbf{2}: 309-312.
\bibitem[Enomoto 2006]{Enomoto2006} Enomoto, R., et al. 2006. A Search for Sub-TeV Gamma Rays from the Vela Pulsar Region with CANGAROO-III. \textit{Astrophysical Journal} \textbf{638}: 397-408.
\bibitem[Fazio 1968]{Fazio1968}Fazio, G. G., Helmken, H. F., Rieke, G. H., Weekes, T. C. 1968. An experiment to search for discrete sources of cosmic gamma rays in the 10$^{11}$ to 10$^{12}$ eV region. \textit{Canadian Journal of Physics} \textbf{46}: S451-S456.
\bibitem[Fleury 1992]{Fleury1992} Fleury, Patrick, Vacanty, Giuseppe, editors. 1992. \textit{Conference Proceedings: Towards a Major Atmospheric Cherenkov Detector for TeV Astroparticle Physics}. Editions Fronti\`eres, ISB N 2-86332-126-9.
\bibitem[Fomin 1991]{Fomin1991} Fomin, V. P., et al. 1991. Comparison of the Imaging Gamma-Ray Telescopes at the Crimea and Whipple Observatories. In: \textit{Proceedings of the 22nd International Cosmic Ray Conference, Dublin, Ireland} \textbf{2}: 603-606.
\bibitem[Gaidos 1996]{Gaidos1996} Gaidos, J. A., et al. (Whipple collaboration). 1996. Extremely rapid bursts of TeV photons from the active galaxy Markarian 421. \textit{Nature} \textbf{383}: 319-320.
\bibitem[Galbraith 1953]{Galbraith1953} Galbraith, W., Jelley, J.V. 1953. Light Pulses from the Night Sky associated with Cosmic Rays. \textit{Nature} \textbf{171}: 349-350.
\bibitem[Gingrich 2005]{Gingrich2005} Gingrich D.M., et al., 2005. The STACEE ground-based gamma-ray detector. \textit{IEEE Transactions on Nuclear Science} \textbf{52}: 2977-2985.
\bibitem[Goret 1991]{Goret1991} Goret, P., et al. 1991. ASGAT: A Fast Timing VHE Gamma-Ray Telescope. In: \textit{Proceedings of the 22nd International Cosmic Ray Conference, Dublin, Ireland} \textbf{2}: 630-633.
\bibitem[Gould 1966]{Gould1966} Gould, Robert J., Schr\'eder, Gerald. 1966. Opacity of the Universe to High-Energy Photons. \textit{Physical Review Letters} \textbf{16}: 252-254.
\bibitem[Grindlay 1975]{Grindlay1975} Grindlay, J. E., Helmken, H. F., Brown, R. H., Davis, J., Allen, L. R. 1975. Results of a Southern-Hemisphere search for gamma-ray sources at energies of at least 300 GeV. \textit{Astrophysical Journal} \textbf{201}: 82-89.
\bibitem[Harris 2011]{M872} Harris, Dan E., Massaro, F., Cheung, C. C., Horns, D., Raue, M., Stawarz, \L., Wagner, S., Colin, P., Mazin, D., Wagner, R., Beilicke, M., LeBohec, S., Hui, M., Mukherjee, R. 2011. An Experiment to Locate the Site of TeV Flaring in M87. \textit{Astrophysical Journal} \textbf{743}: 177. 
\bibitem[Helfand 2007]{Helfand2007} Helfand, D. J., Gotthelf, E.V., Halpern, J.P., Camilo, F., Semler, D.R., et al. 2007. Discovery of the Putative Pulsar and Wind Nebula Associated with the TeV Gamma-Ray Source HESS J1813-178. \textit{The Astrophysical Journal} \textbf{665}: 1297-303.
\bibitem[Hess 1912]{hess1912} Hess, Victor Franz. 1912. \"Uber Beobachtungen der durchdringenden Strahlung bei sieben Freiballonfahrten. \textit{Physikalische Zeitschrift} \textbf{13}: 1084-1091.
\bibitem[Hillas 1985]{Hillas1985} Hillas, A. Michael. 1985. Cerenkov light images of EAS produced by primary gamma rays and by nuclei. In: \textit{Proceedings of the 18th International Cosmic Ray Conference, La Jolla, USA} \textbf{3}: 445-449.
\bibitem[Hofmann 2001]{Hofmann2001} Hofmann, Werner. 2001. Status of the H.E.S.S. project. In: \textit{Proceedings of the 27th International Cosmic Ray Conference, Hamburg, Germany} \textbf{7}: 2785-7288.
\bibitem[Holder 2006]{Holder2006} Holder, Jamie, et al. (VERITAS collaboration). 2006. The First VERITAS Telescope. \textit{Astroparticle Physics} \textbf{25}: 391-401.
\bibitem[Huang 2009]{Huang2009} Huang, J., et al. (Tibet-AS collaboration). 2009. The complex EAS hybrid arrays in Tibet. \textit{Nuclear Physics: Proceedings Supplement} \textbf{196}: 147-152.
\bibitem[Kampert 2012]{Kampert} Kampert, Karl-Heinz, Watson, Alan A. 2012. Extensive air showers and ultra high-energy cosmic rays: a historical review. \textit{European Physical Journal H} \textbf{37}: 359-412.
\bibitem[Karle 1995]{Karle1995} Karle, Albrecht, et al. (AIROBICC collaboration). 1995. Design and performance of the angle integrating Cerenkov array AIROBICC. \textit{Astroparticle Physics} \textbf{3}: 321-347.
\bibitem[Kifune 1992]{Kifune1989} Kifune, Tadashi. 1992. The Energy Threshold of Imaging Cerenkov Technique and 3.8m Telecope of CANGAROO. In \textit{Conference Proceedings: Towards a Major Atmospheric Cherenkov Detector for TeV Astroparticle Physics} (ref. \cite{Fleury1992}): 229-237.
\bibitem[Kranich 2002]{Kranich2002} Kranich, Daniel. 2002. Temporal and spectral characteristics of the active galactic nucleus Mkn 501 during a phase of high activity in the TeV range. Ph.D. thesis, Technische Universit\"at M\"unchen.
\bibitem[Kolh\"orster 1913]{Kolhoerster1913} Kolh\"orster, Werner. 1913. Messungen der durchdringenden Strahlung in Freiballons in gr\"o\ss eren H\"ohen. \textit{Physikalische Zeitschrift} \textbf{14}: 1153-1156.
\bibitem[Kosack 2004]{Kosack2004} Kosack, Karl, et al. 2004. TeV Gamma-Ray Observations of the Galactic Center. \textit{Astrophysical Journal Letters} \textbf{608}: L97-L100.
\bibitem[Koul 2005]{Koul2005} Koul, R., et al. 2005. The Himalayan Gamma Ray Observatory at Hanle. In \textit{Proceedings of the 29th International Cosmic Ray Conference, Pune, India} \textbf{5}: 243-246.
\bibitem[Joshi 2000]{tactic501} Joshi, U., et al. (TACTIC collaboration). 2000. Coordinated TeV gamma-ray and optical polarization study of BL Lac object Mkn 501. \textit{Bulletin of the Astronomical Society of India} \textbf{28}: 409-411.
\bibitem[Lloyd-Evans 1983]{Lloyd-Evans1983} Lloyd-Evans, J., Coy, R. N., Lambert, A., Lapikens, J., Patel, M., Reid, R. J. O., Watson, A. A. 1983. Observation of gamma rays with greater than 1000 TeV energy from Cygnus X-3. \textit{Nature} \textbf{305}: 784-787.
\bibitem[Lorenz 2006]{lorenznima} Lorenz, Eckart. 2006. High-energy astroparticle physics. \textit{Nuclear Instruments and Methods in Physics Research Section A: Accelerators, Spectrometers, Detectors and Associated Equipment} \textbf{567}: 1-11.
\bibitem[Lemoine-Gourmand 2007]{Lemoine-Gourmand2007} Lemoine-Gourmard, Marianne, et al. (H.E.S.S. collaboration). 2007. HESS observations of the supernova remnant RX J0852.0-4622: shell-type morphology and spectrum of a widely extended VHE gamma-ray source. In: \textit{Proceedings of the 30th International Cosmic Ray Conference, Merida, Mexico} \textbf{2}: 667-670.
\bibitem[Maier 2011]{Maier2011} Maier, Gernot, Skilton, Joanna, (VERITAS and H.E.S.S. collaborations). 2011. VHE Observations of the Binary Candidate HESS J0632+057 with H.E.S.S. and VERITAS. In: \textit{Proceedings of the 32th International Cosmic Ray Conference, Beijing, China}. Available at arXiv:1111.2155 [astro-ph].
\bibitem[Marshak 1985]{Marshak1985} Marshak, M. L., et al. 1985. 
Evidence for muon production by particles from Cygnus X-3. \textit{Physical Review Letters} \textbf{54}: 2079-2082.
\bibitem[Merck 1991]{Merck1991} Merck, Martin, et al. (HEGRA collaboration). 1991. Search for Steady and Sporadic Emission of Neutral Radiation Above 50 TeV with the HEGRA Array. In: \textit{Proceedings of the 22nd International Cosmic Ray Conference, Dublin, Ireland} \textbf{1}: 261-264.
\bibitem[Merck 1993]{Merck1993} Merck, Martin. 1993. Suche nach Quellen untrahochenergetischer kosmischer Strahlung mit dem HEGRA-Detektor.
Ph.D. thesis, Ludwig-Maximilians-Universit\"at M\"unchen.
\bibitem[Millikan 1928]{Millikan1925} Millikan, R. A, Cameron, G. H. 1928. New results on cosmic rays. \textit{Nature} supplement \textbf{121}: 19-26.
\bibitem[Mirabel 2006]{Mirabel2006}Mirabel, I. Felix. 2006. Very Energetic Gamma-Rays from Microquasars and Binary Pulsars. \textit{Science} \textbf{312}: 1759-1760.
\bibitem[Morrison 1958]{Morrison1958} Morrison, P. 1958. On gamma-ray astronomy. \textit{Il Nuovo Cimento} \textbf{7}: 858-865.
\bibitem[Nikolski 1989]{Nikolski1989} Nikolsky, S. I., Sinitsyna, V. G. 1989. Investigation of Gamma-sources by Mirror Telescopes. In: \textit{Very High Energy Gamma Ray Astronomy: proceedings of the International Workshop, Crimea, USSR, April 17-21, 1989}: p. 11.
\bibitem[Nolan 2012]{Nolan2012} Nolan, Patrick L., et al. (Fermi-LAT collaboration). 2012. Fermi Large-Area Telescope Second Source Catalog. \textit{Astronomical Journal Supplement Series} \textbf{199}: 31-77
\bibitem[Par\'e 2002]{Pare2002} Par\'e, E., et al. 2002. CELESTE: an atmospheric Cherenkov telescope for high energy gamma astrophysics. \textit{Nuclear Instruments and Methods in Physics Research Section A: Accelerators, Spectrometers, Detectors and Associated Equipment} \textbf{490}: 71-89.
\bibitem[Parsignault 1976]{Parsignauld1976} Parsignault, D. R., Schreier, E., Grindlay, J., Gursky, H. 1976. On the stability of the period of Cygnus X-3. \textit{Astrophysical Journal Letters} \textbf{209}: L73-L75.
\bibitem[Pfeffermann 1996]{Pfeffermann1996} Pfeffermann, E., Aschenbach, R., 1996. ROSAT observation of a new supernova remnant in the constellation Scorpius. In: \textit{Proceedings of International Conference on X-ray Astronomy and Astrophysics: R\"ontgenstrahlung from the Universe}: p. 267-268.
\bibitem[Plaga 2000]{Plaga2000} Plaga, Rainer. 2000. In \textit{Proceedings of 17th European Cosmic Ray Symposium, \L\'od\'z, Poland.}
\bibitem[Penzias 1965]{Penzias1965} Penzias, Arno, Wilson R. W. 1965. A Measurement Of Excess Antenna Temperature At 4080 Mc/s. \textit{Astrophysical Journal Letters} \textbf{142}: 419-421.
\bibitem[Punch 1992]{Punch1992} Punch, Michael, et al. (Whipple collaboration). 1992. Detection of TeV photons from the active galaxy Markarian 421. \textit{Nature} \textbf{358}: 477-478.
\bibitem[Quinn 1996]{Quinn1996} Quinn, John, et al. (Whipple collaboration). 1996. Detection of Gamma Rays with $E > 300$~GeV from Markarian 501. \textit{Astrophysical Journal Letters} \textbf{456}: L83-86.
\bibitem[Quinn 1999]{whipple501} Quinn, John, et al. (Whipple collaboration). 1999. The Flux Variability of Markarian 501 in Very High Energy Gamma Rays. \textit{Astrophysical Journal} \textbf{518}: 693-698.
\bibitem[Remillard 1997]{Remillard1997} Remillard, R. A., Levine, M. L. 1997. The RXTE All Sky Monitor: First Year of Performance. In: \textit{Proceedings All Sky X-ray Observations in the Next Decade, RIKEN, Japan}: p. 29.
\bibitem[Salazar 2009]{Salazar2009} Salazar, H. 2009. The HAWC observatory and its synergies at Sierra Negra Volcano. In: \textit{Proceedings of the 31st International Cosmic Ray Conference, \L\'od\'z, Poland} 
\bibitem[Samorski 1983]{Samorski1983} Samorski, Manfred, Stamm, W. 1983. Detection of $2\times10^{15}$ to $2\times10^{16}$ eV gamma-rays from Cygnus X-3. \textit{Astrophysical Journal Letters} \textbf{268}: L17-21.
\bibitem[Sinnis 2009]{Sinnis2009} Sinnis, Gus. 2009. Cosmic-Ray Physics with the Milagro Gamma-Ray Observatory. \textit{Journal of the Physical Society of Japan Supplement A} \textbf{78}: 84-87.
\bibitem[Smith 2006]{Smith2006} Smith, Davud. A., et al. 2006. Mrk 421, Mrk 501, and 1ES 1426+428 at 100 GeV with the CELESTE Cherenkov telescope. \textit{Astronomy \& Astrophysics} \textbf{459}: 453-464.
\bibitem[Spiering 2012]{Spiering} Spiering, Christian. 2012. Towards High-Energy Neutrino Astronomy.  \textit{European Physical Journal H} \textbf{37}: 515-565.
\bibitem[Stepanian 1983]{Stepanian1983} Stepanian, Arnold A., Fomin, Valery P., Vladimirsky, B. M. 1983. A method to distinguish the gamma-ray Cherenkov flashes from proton component of cosmic rays. \textit{Izv. Krimskoi astrofiz. obs.} \textbf{66}: 234-240.
\bibitem[T\"umer 1999]{Tumer1999} T\"umer, T., Bhattacharya, D., Mohideen, U., Rieben, R., Souchkov, V., Tom, H., Zweerink J. 1999. Solar Two Gamma-Ray Observatory. \textit{Astroparticle Physics} \textbf{11}: 271-273.
\bibitem[Urban 1991]{artemis} Urban, M., et al. (ARTEMIS and Whipple collaborations). 1991. ARTEMIS: Anti-Matter Research Through the Earth Moon Ion Spectrometer. In: \textit{Proceedings of the 22nd International Cosmic Ray Conference, Dublin, Ireland} \textbf{2}: 189-192.
\bibitem[Urry 1995]{UrryPadovani} Urry, C. Megan, Padovani, Paolo. 1995. Unified Schemes for Radio-Loud Active Galactic Nuclei. \textit{Publications of the Astronomical Society of the Pacific} \textbf{107}: 803-845.
\bibitem[Vandenbroucke 2010]{Vandenbroucke2010} Vandenbroucke, Justin. 2010. AGIS: A Next-generation TeV Gamma-ray Observatory. \textit{Bulletin of the American Astronomical Society} \textbf{41}: 909.
\bibitem[Vassiliev 2000]{Vassiliev2000} Vassiliev, Vladimir V. 2000. Extragalactic background light absorption signal in the TeV gamma-ray spectra of blazars. \textit{Astroparticle Physics} \textbf{12}: 217-238.
\bibitem[Vladimirski 1989]{Vladimirski1989} Vladimirsky, B. M., Zyskin, Yu. L., Neshpor, Yu. I., Stepanian, A. A., Fomin, V. P., Shitov, V. G. 1989. Cerenkov Gamma-telescope GT-48 of the Crimean Astrophysical Observatory of the USSR Academy of Sciences. In: \textit{Very High Energy Gamma Ray Astronomy: proceedings of the International Workshop, Crimea, USSR, April 17-21, 1989}: p. 11.
\bibitem[Wagner 2008]{Wagner2008} Wagner, Robert. 2008. Synoptic studies of seventeen blazars detected in very high-energy gamma-rays. \textit{Monthly Notices of the Royal Astronomical Society} \textbf{385}: 119-135.
\bibitem[Watson 1985]{Watson1985} Watson, Alan A. 1985. 
High-energy astrophysics: Is Cygnus X-3 a source of gamma rays or of new particles? \textit{Nature} \textbf{315}: 454-455.
\bibitem[Weekes 1977]{Weekes1977} Weekes, Trevor C., Turver, K. E. 1977. Gamma-ray astronomy from 10-100 GeV: A new approach. in: \textit{ESA Recent Advances in Gamma-Ray Astronomy}: 279-286.
\bibitem[Weekes 1983]{Weekes1981} Weekes, Trevor C. 1983. A Fast Large Aperture Camera for Very High Energy Gamma-Ray Astronomy. In: \textit{Proceedings of the 17th International Cosmic Ray Conference, Paris, France} \textbf{8}: 34-37.
\bibitem[Weekes 1989]{Weekes1989} Weekes, Trevor C., et al. (Whipple collaboration). 1989. Observation of TeV gamma rays from the Crab nebula using the atmospheric Cerenkov imaging technique. \textit{Astrophysical Journal} \textbf{342}: 379-395.
\bibitem[Weekes 2003]{Weekes2003} Weekes, Trevor C. 2003. \textit{Very High Energy Gamma-Ray Astronomy}. Institute of Physics, Bristol.
\bibitem[Winston 1970]{Winston} Winston, Roland. 1970. Light Collection within the Framework of Geometrical Optics,. \textit{Journal of the Optical Society of America} \textbf{60}: 245-247.
\end{thebibliography}
\end{document}